\documentclass{jpp}
\usepackage{graphicx}
\usepackage{tensor}
\usepackage{bm}
\usepackage{enumitem}
\newlist{myitemize}{itemize}{3}
\setlist[myitemize,1]{label=\textbullet,leftmargin=2em,rightmargin=2em,itemindent=0pt,labelsep=5pt,labelwidth=2em}
\setlist[myitemize,2]{label=$\rightarrow$,leftmargin=1em}
\setlist[myitemize,3]{label=$\diamond$}

\newlist{myenumerate}{enumerate}{1}
\setlist[myenumerate,1]{leftmargin=3em,rightmargin=3em,itemindent=0pt,labelsep=5pt,labelwidth=2em}

\usepackage[utf8]{inputenc}
\usepackage[T1]{fontenc}
\usepackage{amsmath}

\newcommand{\diff}[2]{\frac{\mathrm{d}#1}{\mathrm{d}#2}}	
\renewcommand{\tt}{\tilde{\theta}}							
\renewcommand{\d}{\textrm{d}}								

\renewcommand{\div}{\boldsymbol{\nabla}\bcdot}							

\title{An approximate analytic solution to the coupled problems of coronal heating
  and solar-wind acceleration}

\author{Benjamin D.\ G.\ Chandran}

\affiliation{Department of Physics and Astronomy, University of New
    Hampshire, Durham, New Hampshire 03824,  USA }

\begin{document}

\maketitle

\begin{abstract}
  Between the base of the solar corona at $r=r_{\rm b}$ and the
  Alfv\'en critical point at $r=r_{\rm A}$, where $r$ is heliocentric
  distance, the solar-wind density decreases by a
  factor~$\gtrsim 10^5$, but the plasma temperature varies by a factor
  of only a few. In this paper, I show that such quasi-isothermal
  evolution out to~$r=r_{\rm A}$ is a generic property of outflows
  powered by reflection-driven Alfv\'en-wave (AW) turbulence, in which
  outward-propagating AWs partially reflect, and counter-propagating
  AWs interact to produce a cascade of fluctuation energy to small
  scales, which leads to turbulent heating. Approximating the
  sub-Alfv\'enic region as isothermal, I first present a brief, simplified
  calculation showing that in a solar or stellar wind powered by AW
  turbulence with minimal conductive losses,
  $\dot{M} \simeq P_{\rm AW}(r_{\rm b})/v_{\rm esc}^2$,
  $U_{\infty} \simeq v_{\rm esc}$, and
  $T\simeq m_{\rm p} v_{\rm esc}^2/[8 k_{\rm B} \ln(v_{\rm esc}/\delta
  v_{\rm b})]$, where $\dot{M}$ is the mass outflow rate, $U_{\infty}$
  is the asymptotic wind speed, $T$ is the coronal temperature,
  $v_{\rm esc}$ is the escape velocity of the Sun, $\delta v_{\rm b}$
  is the fluctuating velocity at~$r_{\rm b}$, $P_{\rm AW}$ is the
  power carried by outward-propagating AWs, $k_{\rm B}$ is the
  Boltzmann constant, and $m_{\rm p}$ is the proton mass. I then
  develop a more detailed model of the transition region, corona, and
  solar wind that accounts for the heat flux~$q_{\rm b}$ from the
  coronal base into the transition region and momentum deposition by
  AWs. I solve analytically for~$q_{\rm b}$ by balancing conductive
  heating against internal-energy losses from radiation, $p\,\d V$
  work, and advection within the transition region. The density
  at~$r_{\rm b}$ is determined by balancing turbulent heating and
  radiative cooling at~$r_{\rm b}$. I solve the equations of the model
  analytically in two different parameter regimes. In one of these regimes,
  the leading-order analytic solution reproduces the results of the aforementioned
  simplified calculation of~$\dot{M}$, $U_\infty$, and~$T$. Analytic
  and numerical solutions to the model equations match a number of
  observations.
\end{abstract}

\vspace{0.2cm} 
\section{Introduction}
\label{sec:intro}
\vspace{0.2cm} 

Pioneering works by \cite{parker58, parker65},
\cite{hartle68}, and \cite{durney72} modelled the solar wind as a
steady-state, spherical outflow powered by the outward conduction of
heat from the base of the corona. These models succeeded in producing
a supersonic wind, but were unable to explain the large outflow
velocities measured in fast-solar-wind streams near Earth. They also
had little predictive power for the mass outflow rate from the
Sun,~$\dot{M}$, because they specified the temperature of the
coronal base as a boundary condition, and~$\dot{M}$ is highly
sensitive to the coronal temperature~\citep{hansteen95}.

A possible solution to these problems was proposed almost as soon as
the difficulties became apparent, namely that the solar wind is
powered by an Alfv\'en-wave (AW) energy flux (Parker 1965, p.~686;
Hollweg~1973, 1978).  \nocite{parker65,hollweg73b,hollweg78c} This
idea received strong support from the discovery of large-amplitude AWs
in the interplanetary medium that propagate away from the Sun in the
local plasma frame~\citep{belcher71} as well as the remote observation
of AW-like motions in the low solar atmosphere carrying an energy flux
sufficient to power the solar wind~\citep{depontieu07}. A leading
paradigm for how AWs energise the solar wind is based on
reflection-driven AW turbulence. As AWs propagate away from the Sun,
they undergo partial reflection because of the radial variation in the
Alfv\'en speed \citep{heinemann80,velli93}. Counter-propagating AWs
then interact nonlinearly \citep{iroshnikov63, kraichnan65}, causing
wave energy to cascade from large wavelengths to small wavelengths and
dissipate, thereby heating the ambient plasma
\citep{velli89,matthaeus99b,cranmer05,verdini07,perez13,vanballegooijen16,vanballegooijen17}.
The presence of turbulence in the interplanetary medium is confirmed
by spacecraft measurements~\citep{coleman68,matthaeus82,
  tumarsch95,smith01,bruno05,horbury08,wicks10,chen20}, and
numerical solar-wind models based on reflection-driven AW turbulence
have proven successful at explaining a number of observations of
the solar wind and corona
\citep{cranmer07,verdini10,chandran11,vanderholst14,usmanov14}.
Three-dimensional compressible magnetohydrodynamic (MHD)
simulations have also shown that the
solar wind can be self-consistently generated by an AW energy flux
\citep{shoda19}.

Although the aforementioned numerical models reproduce many of the
observed properties of the solar wind \citep[see
also][]{riley11,gressl14}, analytic formulae that determine $\dot{M}$
and the outflow speed far from the Sun ($U_\infty$) remain elusive. A
number of studies have obtained a single equation that constrains the
two unknowns~$\dot{M}$ and~$U_{\infty}$. For example,
\cite{sandbaek94} pointed out that if the energy flux far from the Sun
is mostly in the form of bulk-flow kinetic energy, and if the energy
flux at the coronal base is dominated by the flux of gravitational
potential energy, heat, and some additional form of mechanical energy
(from, e.g., AWs), then energy conservation implies that
\begin{equation}
  \dot{M} = \frac{\dot{E}_{\rm m0} + \dot{E}_{q0}}{\frac{1}{2}(v_{\rm esc}^2 +
    U_\infty^2)},
  \label{eq:sandbaek94}
\end{equation}
where $\dot{E}_{\rm m0}$ is the mechanical-energy input into the solar
wind at the corona base, $\dot{E}_{q0}$ is the energy input at the
coronal base from thermal conduction (which is negative in models that
include the lower solar atmosphere),
\begin{equation}
  v_{\rm esc} = \left(\frac{2 G M_{\odot}}{R_{\odot}}\right)^{1/2} =
  617.7 \mbox{ km} \mbox{ s}^{-1}
  \label{eq:defvesc}
\end{equation}
is the escape velocity of the Sun, $G$ is the gravitational
constant, $M_\odot$ is the solar mass, and~$R_{\odot}$ is the solar
radius. \cite{schwadron03} derived a variant of~(\ref{eq:sandbaek94})
that explicitly relates $\dot{E}_{q0}$ to the altitude of the
coronal-temperature maximum.  \cite{hansteen95} and \cite{hansteen97}
showed that $\dot{E}_{q0} \ll \dot{E}_{\rm m0}$, and  \cite{hansteen12}
made use of this finding to further refine~(\ref{eq:sandbaek94}),
obtaining
\begin{equation}
  \dot{M} = \frac{\dot{E}_{\rm m0}}{\frac{1}{2}(v_{\rm esc}^2 +
    U_\infty^2)}.
  \label{eq:sandbaek94b}
\end{equation}
Another important result was obtained by \cite{leer80}, who found that heating
inside the sonic critical point enhances~$\dot{M}$ but has little
effect on~$U_{\infty}$, whereas heating beyond the sonic critical
point increases~$U_{\infty}$ but has little effect on~$\dot{M}$.

The main goal of the present paper is to obtain approximate analytic
solutions for $\dot{M}$, $U_{\infty}$, the temperature of the corona,
the heat flux from the coronal base into the lower solar atmosphere,
and the plasma density at the coronal base under the assumption that
AW turbulence is the primary energisation mechanism of the solar wind.
Section~\ref{sec:heuristic} takes a first step towards this goal
by presenting a simplified approximate calculation of~$\dot{M}$,
$U_{\infty}$, and the coronal temperature.  Section~\ref{sec:analytic}
develops a more detailed solar-wind model that accounts for physical
processes neglected in \S~\ref{sec:heuristic}. Approximate
analytic solutions to the equations of this model are presented in
\S~\ref{sec:analytic}, and numerical solutions are presented in
\S~\ref{sec:numerical}. Section~\ref{sec:conclusion} discusses
and summarises the main results of the paper.

\section{Heuristic calculation of~$\dot{M}$, $U_{\infty}$,
  and the coronal temperature 
in AW-driven winds  with minimal conductive losses}
\label{sec:heuristic} 

Approximate expressions for~$\dot{M}$, $U_\infty$, and the coronal
temperature can be quickly obtained by modelling the solar wind as a
spherically symmetric, steady-state outflow and assuming that: (1) AW
turbulence is the dominant heating mechanism; (2) solar rotation can
be neglected, so that the magnetic field~$\bm{B}$ and flow velocity
are aligned; (3) $\bm{B}\propto r^{-2}\bm{\hat{r}}$, where
$\bm{\hat{r}}$ is the radial unit vector (i.e., a split monopole, with
$B_r>0$ in one hemisphere and $B_r<0$ in the other); (4) momentum
deposition by AWs can be neglected between the coronal base
and sonic critical point; and (5) $p\,\d V$ work is the dominant sink of
internal energy in the sub-Alfv\'enic region of the solar wind, in
which the solar-wind outflow velocity~$U$ is smaller than the Alfv\'en
speed
\begin{equation}
  v_{\rm A} = \frac{B}{\sqrt{4\upi \rho}},
  \label{eq:defvA}
\end{equation} 
where $\rho$ is the mass density.
Assumptions~(3)--(5) are relaxed in the next section.

In steady state, given assumption~(2) above, mass and flux
conservation imply (see \S~\ref{sec:flux_mass}) that
\begin{equation}
  v_{\rm Ab} = y_{\rm b} U_{\rm b},
  \label{eq:vAbybUb}
\end{equation} 
where
\begin{equation}
  y_{\rm b} \equiv \left[\frac{\rho_{\rm b}}{\rho(r_{\rm A})}\right]^{1/2},
  \label{eq:defyb}
\end{equation}
$\rho_{\rm b} = \rho(r_{\rm b})$, $r_{\rm b}$ is the radius of the
coronal base, which in this
section (but not the next) is simply set equal to~$R_{\odot}$,
$r_{\rm A}$ is the Alfv\'en critical point at which~$U=v_{\rm A}$,
$v_{\rm Ab} = v_{\rm A}(r_{\rm b})$, and $U_{\rm b} = U(r_{\rm b})$.
The mass outflow rate can thus be written in the form
\begin{equation}
  \dot{M} = 4\upi R_{\odot}^2 \rho_{\rm b} U_{\rm b} = 4\upi R_{\odot}^2
    \rho_{\rm b} v_{\rm Ab} y_{\rm b}^{-1}.
  \label{eq:defMdot}
\end{equation}

As shown in \S~\ref{sec:AW}, heating by
reflection-driven AW turbulence causes the sub-Alfv\'enic part of the
solar wind at $r< r_{\rm A}$ to become quasi-isothermal, in the sense that
\begin{equation}
\frac{1}{c_{\rm s}^2}   \left|\diff{ c_{\rm s}^2}{r}\right| \ll \frac{1}{\rho}
  \left| \diff{\rho}{r} \right|,
  \label{eq:QIT1}
\end{equation}
where
\begin{equation}
c_{\rm s}^2 \equiv \frac{p}{\rho} = \frac{2 k_{\rm B} T}{m_{\rm p}}
\label{eq:defcs} 
\end{equation} 
is the square of the isothermal sound speed, 
$k_{\rm B}$ is the Boltzmann constant, $T$ is the temperature,
and $m_{\rm p}$ is the proton mass. This argument is
consistent with observations and models of the solar wind, which
suggest that the temperature varies by a factor of only a few
between~$r_{\rm b}$ and $r_{\rm A}$, whereas~$\rho$ varies by a
factor of approximately~$ 10^5$ \citep[see, e.g.,][]{cranmer07}.  When the
sub-Alfv\'enic solar wind is approximated as an isothermal plasma, the
momentum equation at $r<r_{\rm A}$ can be written in the form
\begin{equation}
  U \diff{U}{r} = -\frac{c_{\rm s}^2}{\rho} \diff{\rho}{r} -
  \frac{v_{\rm esc}^2 R_{\odot}}{2 r^2}.
  \label{eq:momentum_heur}
\end{equation}
As $\dot{M} = 4\upi r^2 U \rho$
is independent of~$r$,
$U^{-1} \d U/\d r = - \rho^{-1} \d \rho / \d r - 2/r$.
Upon substituting this relation into~(\ref{eq:momentum_heur}) and
rearranging terms, one obtains
\begin{equation}
   \frac{(c_{\rm s}^2 - U^2)}{\rho}\diff{\rho}{r} =
  \frac{2 U^2}{r} - \frac{v_{\rm esc}^2 R_{\odot}}{2r^2}.
  \label{eq:crit_pt_heur}
\end{equation} 
In order for~(\ref{eq:crit_pt_heur}) to have a smooth transonic
solution, the right-hand side of~(\ref{eq:crit_pt_heur}) must vanish
at the radius~$r_{\rm c}$ at which~$U=c_{\rm s}$, so that $\d \rho/\d r$ remains finite. This
leads to the two critical-point conditions
\begin{equation}
  U_{\rm c} = c_{\rm s} \qquad \frac{r_{\rm c}}{R_{\odot}} =
  \frac{v_{\rm esc}^2}{4 c_{\rm s}^2},
  \label{eq:crit_pt_conds_heur}
\end{equation}
where $U_{\rm c} = U(r_{\rm c})$. Integrating~(\ref{eq:momentum_heur}),
one obtains the Bernoulli integral
\begin{equation}
  \frac{1}{2} U^2 + c_{\rm s}^2 \ln\left(\frac{\rho}{\rho_{\rm
        b}}\right) - \frac{v_{\rm esc}^2 R_{\odot}}{2 r} = \mbox{
    constant} = - \frac{v_{\rm esc}^2}{2},
\label{eq:Bernoulli_heur} 
\end{equation}
where the second equality in~(\ref{eq:Bernoulli_heur}) results from
evaluating the left-hand side of~(\ref{eq:Bernoulli_heur})
at~$r=r_{\rm b}$ and dropping the $U_{\rm b}^2/2$ term, which is $ \ll v_{\rm
  esc}^2/2$. After evaluating the left-hand side
of~(\ref{eq:Bernoulli_heur}) at $r=r_{\rm c}$ and
using~(\ref{eq:crit_pt_conds_heur}) to rewrite~$U_{\rm c}$ and
$r_{\rm c}$ in terms of~$c_{\rm s}^2$, one obtains
\begin{equation}
\ln\left( \frac{  \rho_{\rm c}}{\rho_{\rm b}}\right) = - \frac{v_{\rm
      esc}^2}{2 c_{\rm s}^2} + \frac{3}{2},
  \label{eq:lnrhoc}
\end{equation}
where $\rho_{\rm c} = \rho(r_{\rm c})$.
Upon setting $\dot{M} = 4\upi r_{\rm c}^2 \rho_{\rm c} U_{\rm c}$ and
using~(\ref{eq:crit_pt_conds_heur}) and~(\ref{eq:lnrhoc}) to rewrite
$r_{\rm c}$, $\rho_{\rm c}$, and~$U_{\rm c}$ in terms of~$c_{\rm s}$,
one obtains~\citep{hansteen12}
\begin{equation}
  \dot{M} = \frac{\upi R_{\odot}^2 v_{\rm esc}^4 \rho_{\rm b}}{4 c_{\rm
      s}^3} \exp\left( - \frac{v_{\rm esc}^2}{2 c_{\rm s}^2} +
    \frac{3}{2}\right).
  \label{eq:Mdot_heur2}
\end{equation}
The exponential appearing on the right-hand side
of~(\ref{eq:Mdot_heur2}) reflects the fact that, at $r< r_{\rm c}$,
the flow is subsonic and the density drops off approximately
as in a static atmosphere
\citep{hansteen12}.
Equating the right-hand sides of (\ref{eq:defMdot})
and~(\ref{eq:Mdot_heur2}), one finds that
\begin{equation}
  \ln y_{\rm b} = \frac{v_{\rm esc}^2}{2 c_{\rm s}^2} - \ln c_1 \simeq
  \frac{v_{\rm esc}^2}{2 c_{\rm s}^2},
  \label{eq:lnyx}
\end{equation}
where
$c_1 \equiv e^{3/2} v_{\rm esc}^4/(16 c_{\rm s}^3 v_{\rm
  Ab})$.

When radiative cooling and thermal conduction are neglected, the
internal-energy equation becomes
\begin{equation}
  -c_{\rm s}^2 U \diff{\rho}{r} + \frac{\rho U}{\gamma-1} \diff{c_{\rm
      s}^2}{r} = Q,
  \label{eq:IE_heur}
\end{equation}
where~$Q$ is the turbulent heating rate, and~$\gamma$ is the ratio of
specific heats.
In the quasi-isothermal approximation, the
second term on the left-hand side of~(\ref{eq:IE_heur}) can be
neglected. Multiplying~(\ref{eq:IE_heur}) by~$4\upi r^2$ and
integrating over the quasi-isothermal sub-Alfv\'enic region, one obtains
\begin{equation}
  -c_{\rm s}^2 \dot{M} \int_{r_{\rm b}}^{r_{\rm A}} \frac{1}{\rho}
  \diff{\rho}{r} \d r = 4\upi \int_{\rm r_{\rm b}}^{r_{\rm A}} r^2 Q(r)
  \d r \simeq P_{\rm AW}(r_{\rm b}),
  \label{eq:IE_heur1}
\end{equation}
where the approximate equality assumes that the volume-integrated turbulent heating rate
between $r_{\rm b}$ and~$r_{\rm A}$ is comparable to~$P_{\rm
  AW}(r_{\rm b})$, the power carried by
outward-propagating AWs at the coronal base, consistent with direct
numerical simulations~\citep{perez21}. As $\ln (\rho_{\rm
  b}/\rho(r_{\rm A})) = 2 \ln y_{\rm b}$, equation~(\ref{eq:IE_heur1}) yields
\begin{equation}
  \dot{M} \simeq \frac{P_{\rm AW}(r_{\rm b})}{2 c_{\rm s}^2 \ln y_{\rm
      b}} \simeq \frac{P_{\rm AW}(r_{\rm b})}{v_{\rm esc}^2},
  \label{eq:Mdot_heur3}
\end{equation}
where the second relation in~(\ref{eq:Mdot_heur3}) follows
from~(\ref{eq:lnyx}).

Equations~(\ref{eq:IE_heur1}) and (\ref{eq:Mdot_heur3}) can be
understood as follows. Within the quasi-isothermal sub-Alfv\'enic
region, each time the density of a parcel of plasma decreases by a
factor of~$e$, its thermal energy must be replaced via heating to
offset internal-energy losses from~$p\,\d V$ work. The quantity
$c_{\rm s}^2 \ln (\rho_{\rm b}/\rho(r_{\rm A})) = 2 c_{\rm s}^2 \ln
y_{\rm b} \simeq v_{\rm esc}^2$ is thus the heating cost per unit mass
for plasma to transit the sub-Alfv\'enic region. The reason that the
product $c_{\rm s}^2 \ln (\rho_{\rm b}/\rho(r_{\rm A}))$ is
approximately constant is that increasing~$c_{\rm s}^2$ leads to an
exponential increase in~$\dot{M}$ and the solar-wind density and an
exponential reduction of~$\rho_{\rm b}/\rho(r_{\rm A})$, leaving
$c_{\rm s}^2 \ln(\rho_{\rm b}/\rho(r_{\rm A}))$ approximately
unchanged.  Equations~(\ref{eq:IE_heur1}) and~(\ref{eq:Mdot_heur3})
state that~$\dot{M}$ is the net heating power within the
sub-Alfv\'enic region (which is taken to
be~$\simeq P_{\rm AW}(r_{\rm b})$) divided by the heating cost per
unit mass.

Assuming that the AW-energy flux and gravitational-potential-energy
flux are the dominant mechanical-energy fluxes at the coronal base
and that the kinetic-energy flux is the dominant mechanical-energy
flux at large~$r$, and equating the mechanical luminosities
at~$r=r_{\rm b}$ and at large~$r$, one obtains
\begin{equation}
P_{\rm AW}(r_{\rm b}) -  \frac{1}{2}\dot{M} v_{\rm esc}^2 =
\frac{1}{2} \dot{M} U_{\infty}^2.
\label{eq:Uinf_heur1}
\end{equation} 
Substituting~(\ref{eq:Mdot_heur3}) into~(\ref{eq:Uinf_heur1}) yields
\begin{equation}
  U_{\infty} \simeq v_{\rm esc}.
  \label{eq:Uinf_heur2}
\end{equation}
Reflection-driven AW turbulence thus changes the energy per unit mass
from $\simeq - v_{\rm esc}^2/2$ at the coronal base to~$\simeq 
v_{\rm esc}^2/2$ far from the~Sun.

In the absence of super-radial expansion,
$P_{\rm AW}(r_{\rm b}) = 4 \upi R_{\odot}^2 v_{\rm Ab} \rho_{\rm
  b} (\delta v_{\rm b})^2$, where~$\delta v_{\rm b}$ is
the root-mean-square (r.m.s.) amplitude of the fluctuating velocity at
the coronal base. This relation,  in conjunction
with~(\ref{eq:defMdot}) and~(\ref{eq:Mdot_heur3}), implies that $y_{\rm b} \simeq v_{\rm esc}^2 /
(\delta v_{\rm b})^2$, which leads via~(\ref{eq:defcs}) and
(\ref{eq:lnyx}) to the approximate value of the coronal temperature,
\begin{equation}
T \simeq \frac{m_{\rm p} v_{\rm esc}^2}{8 k_{\rm B} \ln(v_{\rm
    esc}/\delta v_{\rm b})}.
\label{eq:Theuristic} 
\end{equation} 

A number of factors can cause $\dot{M}$,~$U_{\infty}$, and the coronal
temperature to deviate from the estimates in
(\ref{eq:Mdot_heur3})~(\ref{eq:Uinf_heur2}),
and~(\ref{eq:Theuristic}).  For example, some of the AW power survives
out to~$r_{\rm A}$, which reduces the total turbulent heating in the
sub-Alfv\'enic region appearing on the right-hand side
of~(\ref{eq:IE_heur1}), which, in turn, reduces~$\dot{M}$. The heat that
is conducted from the corona into the transition region adds a
negative sink term to the right-hand sides of~(\ref{eq:IE_heur})
and~(\ref{eq:IE_heur1}), which likewise acts to reduce~$\dot{M}$.  On
the other hand, the AW pressure force helps drive plasma away from the
Sun at the critical point, which acts to increase~$\dot{M}$. These
effects, as well as super-radial expansion of the magnetic field, are
included in the more detailed solar-wind model developed in the next
section.

\section{Steady-state model of the transition region, corona, and
  solar wind}
\label{sec:analytic} 

The lowest layer of the solar atmosphere is the
chromosphere, which extends $2000-3000$ km above the
photosphere with a temperature~$T$ ranging from several
thousand~$K$ to around~$10^4$~K. Bounding the chromosphere from above is
the transition region, a narrow layer approximately $ 100$~km thick, in which
the density~$\rho$ drops (and $T$ increases) by approximately two
orders of magnitude. Above the transition region lies the corona,
which extends out a few solar radii from the Sun and has a temperature
of around~$10^6$~K. The corona contains both closed magnetic loops,
which connect back to the Sun at both ends, and open magnetic-field
lines that connect the solar surface to the distant interplanetary
medium. Regions of the corona permeated by open magnetic-field lines
have lower densities than closed-field-line regions and are referred to as coronal holes.
In the analysis to follow, $r_{\rm b}$ denotes the
radius of the coronal base just above the transition region in
Sun-centred spherical coordinates~$(r, \theta, \phi)$, which is taken
to be
\begin{equation}
r_{\rm b} = 1.005 R_{\odot}.
\label{eq:defrb} 
\end{equation} 

\begin{figure}
\begin{center}
\includegraphics[width=12cm]{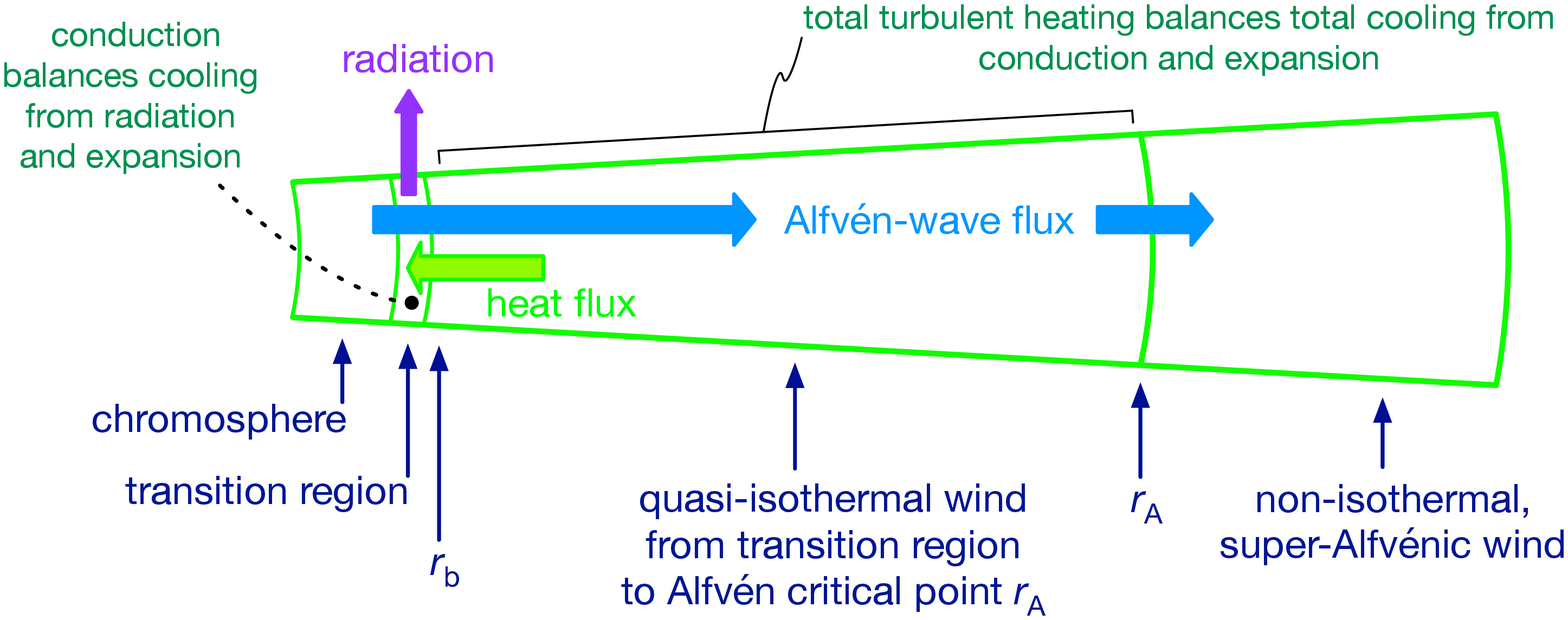}
\end{center}
\caption{Schematic overview of model.
\label{fig:schematic_diagram} }
\end{figure}

The steady-state model developed in this section describes the
outflowing plasma within an open magnetic
flux tube from the transition region to
beyond the Alfv\'en critical point~$r_{\rm A}$, at which the plasma
outflow velocity~$U$ equals $v_{\rm A}$.
Figure~\ref{fig:schematic_diagram} provides a schematic overview of
the model, which determines five unknowns --- the density at the
coronal base~$\rho_{\rm b}$, the temperature
of the quasi-isothermal sub-Alfv\'enic region, the mass outflow rate~$\dot{M}$,
the asymptotic outflow velocity~$U_\infty$, and the flux of heat from the coronal
base into the transition region~$q_{\rm b}$ --- through the following five steps:

\vspace{0.2cm} 
\begin{myenumerate}
  \item[1.] balancing turbulent heating
    at $r=r_{\rm b}$ against radiative cooling at $r=r_{\rm b}$;

\vspace{0.2cm} 
    \item[2.] balancing the total turbulent heating between $r_{\rm b}$ and
$r_{\rm A}$ against the two primary sinks of internal energy in this
region, $p\,\d V$ work and the flux of heat into the transition region;

\vspace{0.2cm} 
\item[3.] balancing, within the transition region, conductive heating
  against internal-energy losses from 
  $p\,\d V$ work, advection, and radiation;

\vspace{0.2cm} 
  \item[4.] equating the mass outflow
rate at $r=r_{\rm b}$ with the mass outflow rate at the wave-modified
sonic critical point $r=r_{\rm c}$; and

\vspace{0.2cm} 
\item[5.] equating the wave-modified Bernoulli integral at $r=r_{\rm b}$ and
$r=r_{\rm c}$.
\end{myenumerate}
\vspace{0.2cm} 

These five steps are detailed in \S~\ref{sec:coronal_base}--\ref{sec:crit_point}.  Section~\ref{sec:flux_mass} reviews
some identities that follow from the conservation of mass and magnetic
flux. Sections~\ref{sec:AW} and~\ref{sec:photosphere} review previous
results on reflection-driven AW turbulence in the solar wind and
present the simplified model of reflection-driven AW turbulence that
is used in this paper. A glossary is given in Table~\ref{tab:Glossary}. 

\begin{table}
\label{tab:t1} 
\begin{center}
\begin{tabular}{ccc}
  Quantity &  Meaning & First use \\
           &  & \\
  $\dot{M}$ & mass outflow rate & (\ref{eq:sandbaek94}) \\
  $U_\infty$ & asymptotic wind speed & (\ref{eq:sandbaek94}) \\
  $v_{\rm esc}$ & escape velocity at photosphere & (\ref{eq:defvesc}) \\
  $R_{\odot}$ & solar radius & (\ref{eq:defvesc}) \\
  $v_{\rm A}$ & Alfv\'en speed & (\ref{eq:defvA}) \\
  $\rho$ & plasma density & (\ref{eq:defvA}) \\
  $U$ & solar-wind outflow velocity & (\ref{eq:vAbybUb}) \\
  $y_{\rm b}$ & $[\rho(r_{\rm b})/\rho(r_{\rm A})]^{1/2}$ & (\ref{eq:vAbybUb}) \\
  $r_{\rm b}$ & radius of coronal base & (\ref{eq:defyb}) \\
  $r_{\rm A}$ & radius of Alfv\'en critical point & (\ref{eq:defyb}) \\
  $c_{\rm s}$ & isothermal sound speed & (\ref{eq:defcs}) \\
  $T$ & temperature &  (\ref{eq:defcs}) \\
  $m_{\rm p}$ & proton mass & (\ref{eq:defcs}) \\
  $k_{\rm B}$ & Boltzmann constant & (\ref{eq:defcs}) \\
  $r_{\rm c}$ & radius of wave-modified sonic critical point & (\ref{eq:crit_pt_conds_heur}) \\
  $\gamma$ & ratio of specific heats (5/3) & (\ref{eq:IE_heur}) \\
  $Q$ & turbulent heating rate & (\ref{eq:IE_heur}) \\
  $P_{\rm AW}$ & power of outward-propagating AWs & (\ref{eq:IE_heur1})
  \\
  $\delta v_{\rm b}$ & r.m.s. amplitude of fluctuating velocity at $r_{\rm
                       b}$ & (\ref{eq:Theuristic}) \\
  $\eta(r)$ & local super-radial expansion factor & (\ref{eq:Bprofile}) \\
  $\bar{ B}$ & field strength at $r=R_{\odot}$ in absence of
                    super-radial expansion & (\ref{eq:Bprofile}) \\
  $\psi$ & $(R_{\odot}/r_{\rm b})^2\simeq 0.99$ & (\ref{eq:defBb})\\
  $A(r)$ & cross-sectional area of the outflow & (\ref{eq:Mdot}) \\
  $y(r)$ & $ [ \rho(r)/\rho(r_{\rm A})]^{1/2}$ &  (\ref{eq:defy}) \\
  $z_{\pm}$ & r.m.s. amplitude of $\bm{z}^\pm$ &  (\ref{eq:defEtot}) \\
  $\bm{z}^+$, $\bm{z}^-$ & Elsasser variables &    (\ref{eq:time_avg}) \\
  $ q$ & heat flux & (\ref{eq:defq}) \\
  $l_{\rm b}$ & AW dissipation length scale at $r_{\rm b}$ & (\ref{eq:Qb}) \\
  $ \sigma$ & dimensionless coefficient in turbulent heating rate &
                                                                    (\ref{eq:zminus2}) \\
$\chi_{\rm H}$ & fraction of~$P_{\rm AW}(r_{\rm b})$ that dissipates
                 between $r_{\rm b}$ and~$r_{\rm A}$ & (\ref{eq:H2}) \\
  $f_{\rm chr}$ & AW transmission coefficient, $P_{\rm AW}(r_{\rm b}) / P_{\rm AW \odot}(R_{\odot})$ &
                                                     (\ref{eq:chichr})
  \\
$\rho_\odot$ & plasma density at the photosphere, $\simeq 10^{17}
               m_{\rm p} \mbox{ cm}^{-3}$ & (\ref{eq:chichr}) \\
  $\delta v_{\odot}$ & r.m.s. amplitude of fluctuating velocity at
                       $R_{\odot}$ & (\ref{eq:chichr}) \\
  $\delta v_{\odot \rm eff}$ & $f_{\rm chr}^{1/2} \delta v_{\odot}$ &
                                                                      (\ref{eq:dvbdvsun}) \\
 $\Lambda(T)$ & optically thin radiative loss function & (\ref{eq:defR})
  \\
  $c_{\rm R}$ & numerical constant in approximate formula
                for~$\Lambda(T)$  & (\ref{eq:Lambda_approx}) \\
  $\overline{ c_{\rm s}}$ & constant value of~$c_{\rm s}$ in the
                            sub-Alfv\'enic region & (\ref{eq:rhob}) \\
  $B_{\rm ref}$ & reference value for magnetic-field-strength,~$\simeq
                  118.8 \mbox{ G}$ & (\ref{eq:rhob2}) \\
  $B_\ast$ & $\bar{B} / B_{\rm ref}$ &  (\ref{eq:rhob2}) \\
$  \tilde{\rho}_\odot$ &  $ 4\upi \rho_{\odot} v_{\rm esc}^2 / B_{\rm
                         ref}^2  \simeq 5.7 \times 10^5$ & (\ref{eq:rhob2}) \\
  $\xi$  & the parameter combination $[\epsilon_{\odot} \eta_{\rm b}/(B_\ast^3 \tilde{l}_{\rm
           b})]^{1/4}$ & (\ref{eq:rhob2}) \\
  $x$ & dimensionless temperature $(\overline{ c_{\rm s}} / v_{\rm
        esc})^2$ & (\ref{eq:defx}) \\
  $\epsilon_{\odot}$ & $ (\delta v_{\odot\rm eff} / v_{\rm esc})^2$ &                                                                      (\ref{eq:defepssun})  \\
  $\tilde{l}_{\rm b} $ & $l_{\rm b} / R_{\odot}$ & (\ref{eq:defxi}) \\
  $\epsilon$ & $(\delta v_{\rm b} / v_{\rm esc})^2$ & (\ref{eq:defeps}) \\
  $w$ & dimensionless heat flux &  (\ref{eq:defw}) \\
  $\gamma_B(r)$ & $-(r/2B) \d B/ \d r$ &  (\ref{eq:crit_pt1}) \\
           &&\\
\end{tabular}
\caption{Glossary \label{tab:Glossary} }
\end{center}
\end{table}

\subsection{Flux and mass conservation}
\label{sec:flux_mass} 

To simplify the analysis, solar rotation is neglected, and
the magnetic field is taken to be radial, except in the corona, where
open magnetic-field lines fan out to fill the space above closed
magnetic loops at lower altitudes. Mathematically,
\begin{equation}
  B(r) = \frac{\bar{ B} \eta(r) R_{\odot}^2}{r^2},
  \label{eq:Bprofile}
\end{equation} 
where $\eta(r)$ is the local super-radial expansion factor, which
approaches~1 when~$r/R_{\odot} \gg 1$, and
$\bar{ B}$ is the magnetic-field strength that would arise at the
photosphere in the absence of super-radial expansion (i.e., if
$\eta(r)$ were unity everywhere.) Given~(\ref{eq:Bprofile}), the magnetic-field strength at the
coronal base is
\begin{equation}
B_{\rm b} \equiv B(r_{\rm b}) = \bar{ B} \eta_{\rm b} \psi,
\label{eq:defBb} 
\end{equation} 
where here and in the following a `b' subscript indicates that the subscripted
quantity is evaluated at~$r=r_{\rm b}$, and
\begin{equation}
  \psi \equiv \frac{R_{\odot}^2}{r_{\rm b}^2} = 0.9901.
  \label{eq:defpsi}
\end{equation}

Because rotation is neglected, the steady-state solar-wind outflow
velocity is aligned with the background magnetic field \citep{mestel61}:
\begin{equation}
\bm{v} = U \frac{\bm{B}}{B}.
\label{eq:defU} 
\end{equation}
The density and velocity satisfy the steady-state continuity equation,
\begin{equation}
  \div(\rho \bm{v}) = 0.
  \label{eq:cont0}
\end{equation} 
It follows from~(\ref{eq:defU}), (\ref{eq:cont0}), and $\div\bm{B}=0$ that
$ \bm{B} \cdot \boldsymbol{\nabla}( \rho U / B) = 0$, and, hence,
\begin{equation} 
  \frac{\rho U}{B} = \mbox{ constant}.
  \label{eq:cont1}
\end{equation}   
Equation~(\ref{eq:cont1}) is equivalent to
\begin{equation}
\dot{M} \equiv  A(r) \rho(r) U(r) = \mbox{ constant},
  \label{eq:Mdot}
\end{equation} 
where
\begin{equation}
A(r) = \frac{4 \upi r^2}{\eta(r)}
\label{eq:AB} 
\end{equation} 
is the cross-sectional area of the flow, which satisfies
$A(r) = \mbox{ constant}/ B(r)$, as required by flux conservation.
Equations~(\ref{eq:Mdot}) and~(\ref{eq:AB}) follow the convention of
expressing the mass outflow rate as the total solar mass-loss rate
that would arise if the local mass flux at some~$r$ characterised the
entire outflow at that~$r$, even though the actual solar wind is
comprised of different wind streams with different properties. All of
the results of this paper can be applied to an individual
flux tube accounting for some fraction~$\nu$ of the total flow area
by multiplying the right-hand side of~(\ref{eq:AB}) by~$\nu$.

It follows from~(\ref{eq:defvA}) and~(\ref{eq:cont1}) that
$\rho^{1/2} U/v_{\rm A}$ is a constant. Because
$U(r_{\rm A}) = v_{\rm A}(r_{\rm A})$,
this constant must
be~$\rho_{\rm A}^{1/2}$, where
\begin{equation}
\rho_{\rm A} \equiv \rho(r_{\rm A}).
\label{eq:defrhoA} 
\end{equation} 
Thus,
\begin{equation}
v_{\rm A} = y U,
\label{eq:vAyU} 
\end{equation} 
where
\begin{equation}
  y \equiv \left(\frac{\rho}{\rho_{\rm A}}\right)^{1/2}.
  \label{eq:defy}
\end{equation}
With the aid of~(\ref{eq:defBb}), (\ref{eq:defpsi}), (\ref{eq:AB})
and~(\ref{eq:vAyU}), equation~(\ref{eq:Mdot}) can be rewritten as
\begin{equation}
  \dot{M} = A_{\rm b} \rho_{\rm b} U_{\rm b} =
\frac{4 \upi R_{\odot}^2}{\psi \eta_{\rm b}}  \times
\rho_{\rm b} \times
  \frac{v_{\rm Ab}}{y_{\rm b}}
  = \frac{R_{\odot}^2 \bar{ B} (4 \upi \rho_{\rm b})^{1/2}}{y_{\rm
      b}} = \bar{ B} R_{\odot}^2 (4 \upi \rho_{\rm
    A})^{1/2}.
  \label{eq:Mdot2}
\end{equation}
Thus, $\dot{M}$ is determined uniquely by the value of~$\rho_{\rm A}$
and the
single-hemisphere open magnetic flux,~$2 \upi R_{\odot}^2 \bar{ B} $.

\subsection{Reflection-driven AW turbulence}
\label{sec:AW} 

\cite{dmitruk02} (hereafter D02) developed an analytic model of
reflection-driven AW turbulence in the solar corona valid in the limit
of small~$L_\perp$, where $L_\perp$ is the correlation length of the
AW fluctuations measured in the plane perpendicular to the background
magnetic field.  \cite{chandran09c} (hereafter CH09) generalised this
model by accounting for the solar-wind outflow
velocity.  Section~\ref{sec:CH09} summarises the main results of the
CH09 model, and \S~\ref{sec:isothermal} uses the CH09 model to
show that heating by reflection-driven AW turbulence causes the
sub-Alfv\'enic region of the solar wind (at $r<r_{\rm A}$, where
$U< v_{\rm A}$) to become approximately isothermal.
Section~\ref{sec:modified} presents a modified version of the
CH09 model that is easier to work with analytically, which is used
to incorporate AW turbulence into the solar-wind model developed in this
section.

\subsubsection{The \cite{chandran09c} model of reflection-driven AW turbulence}
\label{sec:CH09} 

In classical mechanics, a simple harmonic oscillator with frequency~$\omega$ and
energy~$E$ possesses an adiabatic invariant~$E/\omega$. If the
parameters of the oscillator vary on a time scale~$t_0$ satisfying $t_0 \gg \omega^{-1}$
(e.g., if the length of a pendulum is slowly varied), then~$E/\omega$ is
almost exactly conserved. As~$(\omega t_0)^{-1} \rightarrow 0$,
changes in~$E/\omega$ vanish faster than any power of~$(\omega t_0)^{-1}$ \citep{lan60}.

An AW is like a space-filling collection of harmonic oscillators, and
the wave action is analogous to the harmonic oscillator's adiabatic
invariant. The wave action per unit volume per unit~$\omega$ is
${\cal E}_\omega / \omega^\prime$, where $\omega$ is the AW frequency
in an inertial frame centred on the Sun, $\omega^\prime$ is the AW
frequency in the local plasma frame, and ${\cal E}_\omega$ is the AW
energy per unit volume per unit~$\omega$.  In the
Wentzel-Kramers-Brillouin (WKB) limit, in which the wave period is
much shorter than the time scale on which the plasma parameters vary
appreciably and the wave length is much shorter than the length scales
over which the background plasma varies appreciably, the wave action
satisfies the conservation law 
\begin{equation}
\frac{\partial }{\partial t} \left(\frac{{\cal
      E}_\omega}{\omega^\prime}\right) + \div \left(\frac{\bm{c} {\cal
        E}_\omega}{\omega^\prime}\right) = 0,
    \label{eq:actioncons} 
\end{equation}
where~$\bm{c}$ is the group velocity of the waves 
\citep{bretherton68,dewar70}.
For outward-propagating AWs in the solar wind,
$\omega = k_r(U+v_{\rm A})$, $\omega^\prime = k_r v_{\rm A}$, and
$\bm{c} = (U+v_{\rm A})\bm{\hat{r}}$, where $\bm{\hat{r}}$ is the
radial unit vector and~$k_r$ is the radial component of the wave
vector. In a steady-state solar
wind, $\omega$ depends on neither position nor time.
Upon multiplying~({\ref{eq:actioncons}) by $\omega$, integrating
  over~$\omega$, and assuming a steady state, one obtains
\begin{equation} 
    \div \left[ \frac{\bm{\hat{r}} (U+v_{\rm A})^2 {\cal E}_{\rm
          tot}}{v_{\rm A}}\right] = 0,
    \label{eq:divGamma}
\end{equation} 
where
\begin{equation}
  {\cal E}_{\rm tot} = \int {\cal E}_\omega \d \omega = \frac{1}{4} \rho z_+^2
  \label{eq:defEtot}
\end{equation}
is the total AW energy density,
  \begin{equation}
    z_\pm \equiv \langle |\bm{z}^\pm|^2 \rangle^{1/2},
    \label{eq:time_avg}
  \end{equation}
$\langle \dots \rangle$ indicates a time average,
\begin{equation}
  \bm{z}^\pm = \delta \bm{v} \mp \frac{\delta \bm{B}}{\sqrt{4\upi
      \rho}}
  \label{eq:Elsasser}
\end{equation}
are the Elsasser variables, $\delta \bm{v}$ and $\delta \bm{B}$ are
the fluctuating velocity and magnetic field, and $\bm{z}^+$
($\bm{z}^-$) corresponds to AW fluctuations propagating
away from (toward) the Sun.\footnote{The use of $\mp$ on the
  right-hand side of~(\ref{eq:Elsasser}) instead of~$\pm$ implies
  that~$\bm{z}^+$ fluctuations propagate parallel to the background
  magnetic field, and~$\bm{z}^-$ fluctuations propagate anti-parallel
  to the background magnetic field. The identification of~$\bm{z}^+$
  with outward-propagating AWs thus corresponds to the
  case~$B_r>0$. If $B_r<0$, the same analysis goes through by replacing
  the $\mp$ on the right-hand side of~(\ref{eq:Elsasser}) with~$\pm$.}
In (\ref{eq:time_avg}) and the following, a $\pm$ sign is used as a subscript
(as opposed to a superscript) when the subscripted quantity is an r.m.s.\
value. As $\div (\rho U \bm{\hat{r}}) = 0$, (\ref{eq:divGamma}) can
be rewritten as
\begin{equation}
\diff{}{r}g^2 = 0,
  \label{eq:constgsq}
\end{equation}
where
\begin{equation}
  g^2 = \frac{(U+v_{\rm A})^2 z_+^2}{Uv_{\rm A}} =  \frac{(1+y)^2 z_+^2}{y}
  \label{eq:defg}
\end{equation}
is the wave-action flux per unit mass flux
per unit~$\omega$ times~$4 \omega$ integrated over~$\omega$, or, for
brevity, the `wave action flux per unit mass flux'.

When the finite radial wavelength of the AWs is taken into account, 
the radial gradient
in~$v_{\rm A}$ causes partial non-WKB reflection of $\bm{z}^+$
fluctuations, leading to the production of~$\bm{z}^-$
fluctuations. Counter-propagating AWs then interact,
causing fluctuation energy to cascade to small scales and
dissipate. In the CH09 model, this loss of fluctuation energy causes
$g^2$ to decrease with radius according to the equation\footnote{CH09
  derived~(\ref{eq:dgdr}) starting from the MHD equations; the alternative derivation presented here is
  given for brevity.}
\begin{equation}
  (U+v_{\rm A})  \diff{}{r}g^2 = - \frac{z_-}{L_\perp} g^2.
  \label{eq:dgdr}
\end{equation}
Equation~(\ref{eq:dgdr}) states that in a reference frame that follows
an outward-propagating AW, the wave action flux per unit mass flux
decays on the eddy turnover time scale~$L_\perp / z_-$. The reason
that only~$z_-$ (and not~$z_+$) appears in this eddy turnover time
scale is that ${\bm z}^+$ fluctuations are not sheared or distorted by
other~$\bm{z}^+$ fluctuations, but they are sheared and distorted by
counter-propagating~$\bm{z}^-$ fluctuations
\citep{iroshnikov63,kraichnan65}.  The radial decay of~$g^2$ is
accompanied by turbulent heating at the rate
\begin{equation}
  Q = \frac{\rho z_+^2 z_-}{4 L_\perp},
  \label{eq:defQ}
\end{equation}
which is the energy density of the outward-propagating AWs~$\rho z_+^2
/ 4$ divided by their eddy turnover time scale~$L_\perp / z_-$.
Equation~(\ref{eq:defQ}) drops a term $\rho z_-^2 z_+ / (4 L_\perp)$
that is normally included in the turbulent heating rate because of
CH09's assumption that
\begin{equation}
z_- \ll   z_+ ,
  \label{eq:zplusdom}
\end{equation}
an inequality that holds in the small-$L_\perp$ limit, as can be seen later in (\ref{eq:zminusCH09}).
  
Following D02, CH09 determined~$z_- $ by
balancing the rate at which $z_-$ is produced through reflections
against the rate at which $z_-$ cascades to small scales through nonlinear
interactions in the small-$L_\perp$ limit, obtaining
\begin{equation}
\frac{(U+v_{\rm A})}{v_{\rm A}}\left|
  \diff{v_{\rm A}}{r}\right|  z_+
=   \frac{z_+}{L_\perp} z_-,
\end{equation}
or, equivalently,
\begin{equation}
  \frac{z_-}{L_\perp} = \frac{(U+v_{\rm A})}{v_{\rm A}}\left|
    \diff{v_{\rm A}}{r}\right|.
\label{eq:zminusCH09} 
\end{equation}
Upon substituting~(\ref{eq:zminusCH09}) into~(\ref{eq:dgdr}), assuming
that $v_{\rm A}(r)$ has a single maximum at~$r_{\rm m} > r_{\rm b}$,
and solving for~$g(r)$,  CH09 found that
\begin{equation}
  g^2 = g_{\rm b}^2 h(r)
  \label{eq:CH09gsq},
\end{equation}
where
\begin{equation}
  h(r) = \left\{ \begin{array}{ll}
v_{\rm Ab}/v_{\rm A}(r) & \hspace{0.3cm} \mbox{ for $r_{\rm b} < r <
                          r_{\rm m}$} \\
                   v_{\rm Ab} v_{\rm A}(r) / v_{\rm Am}^2 &
                                                            \hspace{0.3cm}
                                                            \mbox{ if
                                                            $r>r_{\rm m}$}
                   \end{array}\right.,
\label{eq:defh} 
\end{equation}
where $v_{\rm Am} = v_{\rm A}(r_{\rm m})$ is the maximum value
of~$v_{\rm A}$. Conceptually, 
(\ref{eq:CH09gsq}) and (\ref{eq:defh})
state that an appreciable fraction of the local AW action flux per unit
mass flux dissipates within
each Alfv\'en-speed scale height, which causes~$z_+^2(r)$ to drop
below the WKB value that would be predicted from (\ref{eq:defg}) with constant~$g^2$.

Upon substituting (\ref{eq:CH09gsq}) into~(\ref{eq:defQ}) and making use~(\ref{eq:defg}),
CH09 obtained
\begin{equation}
Q = \rho (\delta v_{\rm b})^2 \left(\frac{U+v_{\rm A}}{v_{\rm
      A}}\right) \left|\diff{v_{\rm A}}{r}\right|
\left(\frac{y}{y_{\rm b}}\right)\left(\frac{1 + y_{\rm
      b}}{1+y}\right)^2 h(r),
\label{eq:QCH09} 
\end{equation} 
where
\begin{equation}
  (\delta v)^2 \equiv \langle | \delta \bm{ v}|^2 \rangle =
  \frac{z_+^2}{4}.
  \label{eq:defdv}
\end{equation}
The second equality in~(\ref{eq:defdv}) follows from~(\ref{eq:zplusdom}).
As
\begin{equation}
y_{\rm b} \gg 1,
\label{eq:yblim} 
\end{equation} 
equation~(\ref{eq:QCH09}) can be rewritten to a good approximation in the
following simplified form with the aid of~(\ref{eq:vAyU}):
\begin{equation}
Q = \rho (\delta v_{\rm b})^2 y_{\rm b}\left(\frac{U }{U+v_{\rm A} }\right) \left| \diff{v_{\rm A}}{r}\right|
h(r).
\label{eq:QCH09b} 
\end{equation}

\subsubsection{Approximate isothermality between the transition region
  and Alfv\'en critical point}
\label{sec:isothermal}

In steady state, the plasma internal-energy equation takes the form
\begin{equation} 
\div\left[\bm{v} \left(\frac{p}{\gamma-1}\right)\right] = -p \div \bm{v} -
\div\bm{q} + Q - R,
\label{eq:internal_energy}
\end{equation} 
where~$p$ is the pressure, $p/(\gamma-1)$ is
the internal-energy density,
 $-p
\div\bm{v}$ is the rate at which $p\,\d V$ work is done on the plasma 
per unit volume,
$Q$ is the rate of turbulent heating per unit volume, $\bm{q}$ is the
heat flux, which is written in the form\footnote{As in
  \S~\ref{sec:CH09},
  $\bm{B}$ is taken to point radially outwards. If $\bm{B}$ in fact
  points toward the Sun, the same analysis goes through if one 
  introduces a minus sign in front of the right-hand side of~(\ref{eq:defq}).}
\begin{equation}
  \bm{q} = q_r \frac{\bm{B}}{B},
  \label{eq:defq}
\end{equation}
and~$R$ is the rate of radiative cooling per unit volume. 
In the corona and solar wind, the density is sufficiently small that
radiative cooling can be neglected, and, to a good approximation,
(\ref{eq:internal_energy})
can be rewritten with the aid of~(\ref{eq:cont0}) as
\begin{equation} 
- c_{\rm s}^2 U \frac{\d \rho}{\d r} + \frac{\rho U}{\gamma-1} 
\diff{}{r} c_{\rm s}^2 = - \frac{1}{A}
\diff{}{r}\left(A q_r\right) + Q .
\label{eq:internal_energy2} 
\end{equation} 

A generic consequence of heating by
reflection-driven AW turbulence is that, when other forms of heating
(including conduction)
are subdominant, the flow becomes quasi-isothermal at $r_{\rm b} < r<r_{\rm A}$,
meaning that 
\begin{equation}
\left| \frac{1}{c_{\rm s}^2} \diff{c_{\rm s}^2}{r}\right| \ll 
\left| \frac{1}{\rho} \diff{\rho}{r}\right|,
\label{eq:iso_ineq} 
\end{equation} 
whereas~(\ref{eq:iso_ineq}) is not satisfied at~$r>r_{\rm A}$.  This can be
demonstrated by (1) neglecting conductive heating
in~(\ref{eq:internal_energy2}), (2) assuming that~(\ref{eq:iso_ineq})
is satisfied so that the second term on the left-hand side
of~(\ref{eq:internal_energy2}) can be neglected, and (3)
solving~(\ref{eq:internal_energy2}) for~$c_{\rm s}^2$.  If the
resulting expression for~$c_{\rm s}(r)^2$ satisfies~(\ref{eq:iso_ineq}),
then the neglect of the second term on the left-hand side
of~(\ref{eq:internal_energy2}) is self-consistent, and this
expression for~$c_{\rm s}(r)^2$ is a reasonable approximation for the full
solution of~(\ref{eq:internal_energy2}).  On the other hand, if the
resulting value of~$c_{\rm s}^2(r)$ does not satisfy~(\ref{eq:iso_ineq}),
then~(\ref{eq:internal_energy2}) does not possess a quasi-isothermal
solution.

Carrying out this procedure and equating the first term on the
left-hand side of~(\ref{eq:internal_energy2}) with the turbulent
heating term on the right-hand side (which is given
by~(\ref{eq:QCH09b})), one obtains
\begin{equation}
c_{\rm s}^2 = y_{\rm b} (\delta v_{\rm b})^2 \left(\frac{ L_\rho}{L_{v_{\rm A}}}\right)
\left(\frac{v_{\rm
      A}}{U+v_{\rm A}}\right)h(r),
\label{eq:isotherm1} 
\end{equation} 
where 
\begin{equation}
L_\rho = \rho\left|\diff{\rho}{r}\right|^{-1} 
\qquad
L_{v_{\rm A}} = v_{\rm A}\left|\diff{v_{\rm A}}{r}\right|^{-1} 
\label{eq:scale_heights} 
\end{equation} 
are the density and Alfv\'en-speed scale heights. These scale heights
are generally some constant of order unity times~$r$, with
$L_\rho/L_{\rm v_{\rm A}}$ increasing by a factor of a few between the
low corona and~$r_{\rm A}$. On the other hand, $h(r)$ decreases by a
factor of a few between the low corona and~$r_{\rm A}$, so that the
product $(L_\rho/L_{\rm v_{\rm A}}) h(r)$ varies quite weakly
with~$r$.\footnote{An exception to this statement arises in the
  vicinity of the Alfv\'en speed maximum~$r_{\rm m}$, where the
  right-hand side of~(\ref{eq:isotherm1}) vanishes. This vanishing is
  an artifact of the CH09 model, which determines the amplitude of
  inward-propagating AWs through a purely local balancing of wave
  reflections against cascade and dissipation.  In numerical
  simulations of reflection-driven AW turbulence, in which
  inward-propagating AWs travel some distance before cascading and
  dissipating, the heating profile varies smoothly with~$r$ without
  any strong reduction at~$r=r_{\rm m}$ \citep{perez13}.} The $v_{\rm A}/(U+v_{\rm A})$
term in~(\ref{eq:isotherm1}) likewise exhibits little variation in the
sub-Alfv\'enic region, ranging from~$\simeq 1$ at~$r=r_{\rm b}$
to~$0.5$ at~$r=r_{\rm A}$. On the other hand, $\rho$ varies by a
factor of~$\sim 10^5$ between the coronal base and~$r_{\rm A}$
\citep[see, e.g.,][]{cranmer07}. Thus, the approximate
solution for~$c_{\rm s}^2$ in (\ref{eq:isotherm1})
satisfies~(\ref{eq:iso_ineq}) at $r_{\rm b} < r < r_{\rm A}$, and AW
heating indeed causes the sub-Alfv\'enic region to become
quasi-isothermal.  In contrast, at $r>r_{\rm A}$, $U$ asymptotes
toward a constant value,
$h \propto v_{\rm A} \propto \rho^{1/2} U \propto \rho^{1/2}$,
$v_{\rm A}/(U+v_{\rm A}) \propto \rho^{1/2}$, and the right-hand side
of~(\ref{eq:isotherm1}) becomes proportional to~$\rho$,
contradicting~(\ref{eq:iso_ineq}).

\subsubsection{A modified version of the \cite{chandran09c} model}
\label{sec:modified} 

There are two difficulties with incorporating the CH09 model
into an analytic solar-wind model that includes the region immediately
above the transition region. First, (\ref{eq:zminusCH09}) is consistent
with~(\ref{eq:zplusdom}) only if
\begin{equation}
\frac{v_{\rm A} L_\perp}{z_+} \ll \frac{v_{\rm A}}{|\d v_{\rm A}/ \d r|}.
\label{eq:CH09ineq}
\end{equation} 
The left-hand side of~(\ref{eq:CH09ineq}) is the characteristic
distance a $\bm{z}^-$ fluctuation 
at scale~$L_\perp$ (measured perpendicular to the background magnetic
field) propagates along the magnetic field before cascading and
dissipating, and the right-hand side is the Alfv\'en-speed scale
height.
Just above the transition region, the magnetic-field strength
has a scale height of order~$10^{-2} R_{\odot}$
\citep{hackenberg00,cranmer07}, the $ v_{\rm A}$ scale height has a
similar value, and (\ref{eq:CH09ineq}) is not satisfied \citep[see,
e.g., equation (3.13-a) and table 3
of][]{chandran19}. On the other hand, beyond the low corona, the
$v_{\rm A}$ scale height grows to values~$\sim r$ that are 
$\gg L_\perp$.  Thus, the CH09 model should be applied only beyond some
minimum heliocentric distance, and some other approximation is needed
to treat the region immediately above the transition region. 

The second difficulty with incorporating the CH09 model into an
analytic solar-wind model is that the function $h(r)$
in~(\ref{eq:defh}) depends on the number of local extrema in the
Alfv\'en-speed profile and the locations of these extrema. In models
that account for the rapid increase in~$B(r)$ as~$r$ drops
below~$\simeq 1.01 R_{\odot}$, $v_{\rm A}$ typically has two local
extrema within the corona: a local minimum just above the coronal base and
a local maximum a few tenths of a solar radius above the coronal base.
(See, e.g., figure~3 of \cite{cranmer05} or figure~9 of
\cite{cranmer07}.)  In the context of the analytic solar-wind model
developed in the following, each local extremum introduces two new unknowns: the
location of the extremum, and either the density or flow speed at the
extremum.

In this paper, the first of the two aforementioned difficulties is handled by
replacing the CH09 result for $Q(r)$ at $r=r_{\rm b}$ with the expression
\begin{equation}
  Q_{\rm b}  =\frac{\rho_{\rm b} (\delta v_{\rm b})^2 v_{\rm Ab}
  }{l_{\rm b}},
  \label{eq:Qb}
\end{equation}
where $l_{\rm b}$ is a free parameter, which is set equal to~$0.3
R_{\odot}$ in the numerical examples presented in \S~\ref{sec:numerical}.
 The second difficulty is
resolved by replacing (\ref{eq:zminusCH09}) with the expression
\begin{equation}
  \frac{z_-}{L_\perp} = - \sigma (U + v_{\rm A}) \diff{}{r} \ln(1+y)
\label{eq:zminus2} 
\end{equation}
at $r_{\rm b} < r < r_{\rm A}$, where $\sigma$ is a free parameter,
whose value is approximately $ 0.1 - 0.5$ in the numerical examples in
\S~\ref{sec:numerical}. Whereas
$z_- / L_\perp = (U+v_{\rm A})/L_{v_{\rm A}}$ in
(\ref{eq:zminusCH09}), equation~(\ref{eq:zminus2}) takes  $z_- / L_\perp$ to be 
$\simeq (U+v_{\rm A})/L_{\rho}$ times a
free parameter.\footnote{More precisely,
  $(\d / \d r) \ln(1+y) = (1+y)^{-1} \d y/ \d r$, which is
  $\simeq - 1/(2 L_\rho)$ when $y \gg 1$ and $\simeq - 1/(4 L_\rho)$ near
  $r=r_{\rm A}$ where $y = 1$.}
 Because $\rho$ (and, hence,~$y$) is a
monotonically decreasing function of~$r$, the minus sign on the
right-hand side of~(\ref{eq:zminus2}) ensures that~$z_- > 0$. The
right-hand side of (\ref{eq:zminus2}) is taken to be proportional to
$\ln(1+y)$ instead of $\ln y$ simply to make some of the expressions
encountered later on easier to integrate analytically.  Another
motivation for using~(\ref{eq:zminus2}) instead
of~(\ref{eq:zminusCH09}) is that the parameter-free CH09 result
in~(\ref{eq:zminusCH09}) overestimates the rate at which AWs cascade
and dissipate in the solar wind \citep{chandran09c}, and the
introduction of a free parameter in~(\ref{eq:zminus2}) in principle
makes it possible to correct for this.

Upon substituting~(\ref{eq:zminus2}) into~(\ref{eq:dgdr}), solving
for~$g^2(r)$, and rewriting~$g^2$ in terms of~$z_+^2$
using~(\ref{eq:defg}),
one finds that
\begin{equation}
  z_+^2 = z_{+{\rm b}}^2 \frac{y}{y_{\rm b}} \left(\frac{1+y_{\rm b}}{1+y}
  \right)^{2-\sigma} .
    \label{eq:zplusy}
\end{equation}
Equations~(\ref{eq:defQ}), (\ref{eq:defdv}), (\ref{eq:zminus2}), and
(\ref{eq:zplusy}) can then be used 
to show that
\begin{equation}
  H = \chi_{\rm H}  P_{\rm AW}(r_{\rm b}),
  \label{eq:H2}
\end{equation} 
where
\begin{equation}
  H = \int_{r_{\rm b}}^{r_{\rm A}} Q(r) A(r) \d r 
  \label{eq:defH}
\end{equation} 
is the total turbulent heating power between~$r_{\rm b}$ and~$r_{\rm A}$,
\begin{equation}
  P_{\rm AW}(r) =   \rho (\delta v)^2 (U + v_{\rm A}) A
  \label{eq:defLAW}
\end{equation}
is the power (energy flux times area) of outward-propagating AWs at radius~$r$, and
\begin{equation}
  \chi_{\rm H} = 1 - 2^{\sigma -1}
  \left(\frac{2-\sigma}{1-\sigma}\right)
  \frac{(1+y_{\rm b})^{1-\sigma}}{y_{\rm b}}
  + \frac{1}{y_{\rm b}(1-\sigma)}
  \label{eq:defchiH}
\end{equation}
is the fraction of~$P_{\rm AW}(r_{\rm b})$ that is dissipated
between~$r_{\rm b}$ and~$r_{\rm A}$.
    
\subsection{Relating the AW amplitudes at the coronal base and  photosphere}
\label{sec:photosphere}

In the absence of wave reflection and dissipation, $P_{\rm AW}$
(defined in~(\ref{eq:defLAW})) would have almost exactly the same value at
$r_{\rm b}$ and~$R_{\odot}$.\footnote{When~$U$ is non-zero, it is $g^2$ in~(\ref{eq:defg})
  that is invariant in the absence of dissipation and reflection,
  not~$P_{\rm AW}$. However, between~$r=R_{\odot}$ and~$r=r_{\rm b}$,
  $U\ll v_{\rm A}$, and~$P_{\rm AW}$ is to an excellent approximation
  proportional to~$g^2$.}
However, the steep Alfv\'en-speed
gradient in the transition region leads to strong AW reflection, and a
vigorous energy cascade in the chromosphere leads to substantial AW
dissipation \citep{vanballegooijen11, chandran19}. This reduces
$P_{\rm AW}(r_{\rm b}) / P_{\rm AW }(R_{\odot})$ to some
value~$f_{\rm chr} < 1$, where~$f_{\rm chr}$ is an effective AW
transmission coefficient for the chromosphere and transition
region. Equivalently, because~$U\ll v_{\rm A}$ at $r\leq r_{\rm b}$,
\begin{equation}
P_{\rm AW}(r_{\rm b}) =   \rho_{\rm b} \delta v_{\rm b}^2 v_{\rm Ab}
A_{\rm b} = f_{\rm chr} \rho_{\odot} \delta v_{\odot}^2 v_{\rm A \odot} A_{\odot},
\label{eq:chichr}
\end{equation} 
where a $\odot$ subscript indicates that the subscripted quantity is
evaluated at the photosphere. For the numerical
calculations in \S~\ref{sec:numerical}, I set
\begin{equation}
  \rho_\odot = 10^{17} m_{\rm p} \mbox{ cm}^{-3}.
  \label{eq:rhoodotval}
\end{equation}
As $B A = \mbox{ constant}$ and $v_{\rm A} = B /(4\upi \rho)^{1/2}$,
equation~(\ref{eq:chichr}) can be rewritten as
\begin{equation}
  \delta v_{\rm b}^2 = \delta v_{\odot\rm eff}^2
  \left(\frac{\rho_{\odot}}{\rho_{\rm b}}\right)^{1/2} ,
  \label{eq:dvbdvsun}
\end{equation}
where
\begin{equation}
\delta v_{\odot\rm eff} =  f_{\rm chr}^{1/2} \delta v_{\odot}.
  \label{eq:dvsuneff}
\end{equation}
The difference between $\delta v_{\odot}$ and
$\delta v_{\odot \rm eff}$ is that $\delta v_{\odot}$ gives rise to
the fluctuating velocity~$\delta v_{\rm b}$ at $r=r_{\rm b}$ when
reflection and dissipation are accounted for, whereas
$\delta v_{\odot \rm eff}$ would give rise to the same value
of~$\delta v_{\rm b}$ via WKB propagation (i.e., without reflection or
dissipation).

The value of~$\delta v_{\odot \rm eff}$ can be constrained in
different ways. For example, observations
fix~$\delta v_{\odot}$ at a value
of~$\simeq 1 \mbox{ km} \mbox{ s}^{-1}$ \citep{richardson50}, and
numerical simulations suggest that $f_{\rm chr}$ is
$\simeq 0.04 - 0.08$ \citep{vanballegooijen11,chandran19}.
Alternatively, if the solar wind is assumed to be powered primarily by
an AW energy flux, then
$\delta v_{\odot \rm eff}$ can be inferred directly from measurements
of the mass flux and energy flux far from the Sun. This latter method is
used to determine $\delta v_{\odot \rm eff}$ in some of the numerical
solutions presented in \S~\ref{sec:numerical} and is described in
more detail in \S~\ref{sec:PAWb}.  It is worth noting that in
contrast to $\delta v_{\odot}$, $f_{\rm chr}$,
and~$\delta v_{\odot \rm eff}$, which are plausibly independent of the
properties of the coronal plasma and coronal magnetic field,
$\delta v_{\rm b}$ depends upon~$\rho_{\rm b}$, which varies between
different flux tubes with different super-radial expansion
factors~$\eta_{\rm b}$, as shown later in~(\ref{eq:rhob}).

\subsection{Balancing turbulent heating and radiative cooling at the
  coronal base}
\label{sec:coronal_base} 

Within the corona and transition region, the plasma is optically thin,
and the rate of radiative cooling is given by
\begin{equation}
R = \left(\frac{\rho}{m_{\rm p}}\right)^2 \Lambda(T),
\label{eq:defR} 
\end{equation} 
where~$m_{\rm p}$ is the proton mass, and~$\Lambda(T)$ is the
optically thin radiative loss function.  Figure~\ref{fig:Lambda_T}
shows three different approximations to~$\Lambda(T)$.
The dashed lines correspond to
equation~A1 of \cite{rosner78}.  The solid line is a plot of
\begin{equation}
\Lambda(T) = c_{\rm R} T^{-1/2},
\label{eq:Lambda_approx} 
\end{equation} 
where
\begin{equation}
c_{\rm R} = 1.549 \times 10^{-19}
 \mbox{ erg} \mbox{ cm}^3 \mbox{ s}^{-1} \mbox{ K}^{1/2}.
\label{eq:defcR} 
\end{equation} 
Equation~(\ref{eq:Lambda_approx}) is the temperature derivative of
equation~A3 of \cite{rosner78}.  The dotted line in
figure~\ref{fig:Lambda_T} is a piecewise-continuous linear
approximation to the low-temperature range of~$\Lambda(T)$ in figure~1
of \cite{cranmer07}, which is included to illustrate that optically
thin radiative cooling becomes extremely weak at $T\lesssim
10^4$~K. 

\begin{figure}
\begin{center}
\includegraphics[width=7cm]{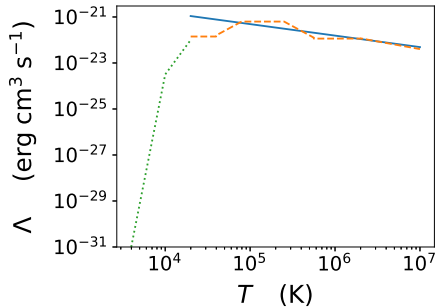}  
\end{center}
\caption{Three approximations to the optically thin radiative loss
  function~$\Lambda(T)$; see the text for further discussion.
  \label{fig:Lambda_T} }
\end{figure}

Throughout most of the corona, $\rho$ is sufficiently small that
radiative cooling is negligible. However, as~$r$ decreases within the
corona, $\rho$ increases by orders of magnitude, which causes $R/Q$ to increase,
because $R\propto \rho^2$. There is thus some radius~$r_{\rm b}$ at
which
\begin{equation}
R(r_{\rm b})  = Q(r_{\rm b}),
\label{eq:rbeq}
\end{equation} 
which marks the transition between the corona, in which radiative
cooling is negligible, and the low solar atmosphere, in which
radiation is thermodynamically important. In this paper, the
radius~$r_{\rm b}$ is identified as the base of the corona, as already
noted in~(\ref{eq:defrb}).  As $r$ decreases below~$r_{\rm b}$,
$\rho(r)$ increases above~$ \rho_{\rm b}$, $R/Q$ increases to
values~$\gg 1$, and the temperature gradient length scale must
decrease so that conductive heating can balance radiative cooling (as
well as internal-energy losses from advection and~$p\,\d V$ work). This
shortening of the temperature gradient length scale gives rise to the
transition region, which is discussed further in \S~\ref{sec:qb}.
To facilitate an analytic solution, I neglect turbulent heating
at~$r<r_{\rm b}$ and radiative cooling at~$r>r_{\rm b}$.

With the aid of~(\ref{eq:Qb}), (\ref{eq:chichr}), and~(\ref{eq:defR}), 
equation~(\ref{eq:rbeq}) can be written in the form
\begin{equation}
\frac{\rho_{\rm b}^2}{m_{\rm p}^2} \Lambda(T_{\rm b}) = 
\frac{\rho_{\rm b} (\delta v_{\rm b})^2 v_{\rm Ab}
  }{l_{\rm b}}
= \frac{\rho_{\odot} (\delta v_{\odot\rm eff})^2 v_{\rm A \odot} B_{\rm b}
  }{l_{\rm b} B_{\odot}}.
\label{eq:bal1} 
\end{equation} 
Solving for~$\rho_{\rm b}$, one finds via~(\ref{eq:Lambda_approx}) that
\begin{equation}
 \rho_{\rm b} = \left(\frac{\delta v_{\odot\rm eff}^2
     \psi \bar{ B} \eta_{\rm b} m_{\rm
       p}^{5/2} \overline{ c_{\rm s}}}{c_{\rm R} l_{\rm
       b}}\right)^{1/2}
\left(\frac{\rho_{\odot}}{8 \upi k_{\rm B}}\right)^{1/4},
 \label{eq:rhob}
\end{equation}
where
\begin{equation}
  \overline{ c_{\rm s}} = \mbox{ constant}
  \label{eq:defcsbar}
\end{equation}
is the sound speed (\ref{eq:defcs}) in the sub-Alfv\'enic region
($r_{\rm b} < r < r_{\rm A}$), which is approximated as a constant for
the reasons discussed in \S~\ref{sec:isothermal}.
Equation~(\ref{eq:rhob}) can be rewritten as
\begin{equation}
  \rho_{\rm b} = \frac{B_{\rm ref}^2}{4 \upi v_{\rm esc}^2}
  \tilde{\rho}_{\odot}^{1/4} B_\ast^2  \xi^2 \psi^{1/2} x^{1/4},
  \label{eq:rhob2}
\end{equation}
where
\begin{equation}
  x = \frac{\overline{ c_{\rm s}}^2}{v_{\rm esc}^2}
  \label{eq:defx}
\end{equation}
is the dimensionless temperature of the sub-Alfv\'enic region,
\begin{equation}
  B_{\rm ref} = \left(\frac{4 \upi m_{\rm p}^{5/2} v_{\rm esc}^6}{c_{\rm R}
      R_{\odot} \sqrt{2 k_{\rm B }}}\right)^{1/2} = 118.8 \mbox{ G}
\label{eq:defBref} 
\end{equation}
is the magnetic-field strength for which $v_{\rm A} = v_{\rm esc}$
when the radiative cooling time $\rho \overline{ c_{\rm s}}^2 / R$,
free-fall time $R_{\odot}/v_{\rm esc}$, and sound-crossing
time~$R_{\odot}/\overline{ c_{\rm s}}$ are all equal,
\begin{equation}
  \epsilon_{\odot} = \frac{(\delta v_{\odot\rm eff})^2}{v_{\rm
      esc}^2},
  \label{eq:defepssun}
\end{equation}
\begin{equation} 
  \tilde{\rho}_{\odot} = \frac{4 \upi v_{\rm esc}^2 \rho_{\odot}
  }{B_{\rm ref}^2},
  \label{eq:defrhosuntilde}
  \end{equation} 
(for reference, $\tilde{\rho}_\odot = 5.7 \times 10^5$
given~(\ref{eq:rhoodotval})), and
\begin{equation}
  \xi = \left(\frac{\epsilon_{\odot} \eta_{\rm b}}{B_\ast^3
      \tilde{l}_{\rm b}}\right)^{1/4} \qquad   B_\ast =
  \frac{\bar{ B}}{B_{\rm ref}}
  \qquad \tilde{l}_{\rm b} =
  \frac{l_{\rm b}}{R_{\odot}}.
  \label{eq:defxi}
\end{equation}
Equation~(\ref{eq:rhob2}) can  used to express $v_{\rm Ab}$ in the
form
\begin{equation}
  v_{\rm Ab} = v_{\rm esc} \psi^{3/4} \xi^{-1} \eta_{\rm b} (x\tilde{\rho}_\odot)^{-1/8},
  \label{eq:vAb}
\end{equation}
and (\ref{eq:dvbdvsun}), (\ref{eq:rhob2}), and (\ref{eq:defepssun}) can
be used to write
\begin{equation}
\epsilon \equiv \frac{(\delta   v_{\rm b})^2}{v_{\rm esc}^2} =
\epsilon_{\odot} \tilde{\rho}_\odot^{3/8} B_\ast^{-1} \xi^{-1}
\psi^{-1/4} x^{-1/8}.
\label{eq:defeps}
\end{equation}

\subsection{Internal-energy equilibrium within the sub-Alfv\'enic region}
\label{sec:Teff} 

In the quasi-isothermal approximation~(\ref{eq:iso_ineq}),
the second term on the left-hand side of (\ref{eq:internal_energy2})
is negligible, and the volume integral of~(\ref{eq:internal_energy2})
between $r=r_{\rm b}$ and $r=r_{\rm A}$ yields
\begin{equation}
- \dot{M} \overline{ c_{\rm s}}^2 \int_{r_{\rm b}}^{r_{\rm A}} \frac{1}{\rho}\diff{\rho}{r} \d r =
- \left(A q_{r}\right)\bigg|^{r_{\rm A}}_{r_{\rm b}} + \chi_{\rm H} \dot{M}
\delta v_{\rm b}^2 (1 + y_{\rm b}),
\label{eq:vol_int} 
\end{equation} 
where (\ref{eq:H2})--(\ref{eq:defLAW}) have been used to
rewrite the volume integral of the turbulent heating rate,
equation~(\ref{eq:Mdot}) has been used to pull a constant factor of
$\dot{M} = \rho A U $ out of the integral on the left-hand side, and
(\ref{eq:vAyU}) has been used to write
$U_{\rm b} + v_{\rm Ab} = U_{\rm b} (1 + y_{\rm b})$.

The heat flux at $r=r_{\rm A}$ is a small fraction of the total energy
flux in AW-driven solar-wind models and is thus neglected
in~(\ref{eq:vol_int}).  Upon evaluating the integral on the left-hand
side of~(\ref{eq:vol_int}), 
noting that $q_{r\rm b} = - |q_{r \rm b}| \equiv - q_{\rm b}$, 
making use of~(\ref{eq:defy}), and rearranging terms, one obtains
\begin{equation}
\chi_{\rm H} \dot{M}\delta v_{\rm b}^2 (1 +
y_{\rm b})  = 2 \dot{M}\overline{ c_{\rm s}}^2 \ln y_{\rm b} + A_{\rm
  b} q_{\rm b}.
\label{eq:Eq3} 
\end{equation} 
The left-hand side of~(\ref{eq:Eq3}) is the total turbulent heating
rate within the sub-Alfv\'enic region, which represents the source of
internal energy between~$r_{\rm b}$ and~$r_{\rm A}$. The two terms on
the right-hand side of~(\ref{eq:Eq3}) are the dominant sinks of
internal energy in the sub-Alfv\'enic region: $p\,\d V$ work and
thermal conduction into the transition region. Dividing~(\ref{eq:Eq3})
by $\dot{M} v_{\rm esc}^2$ leads to
\begin{equation}
 \epsilon \chi_{\rm H} (1 +
y_{\rm b})  = 2 x \ln y_{\rm b} + \frac{ q_{\rm b}}{\rho_{\rm b}
  U_{\rm b} v_{\rm esc}^2}.
\label{eq:IEbal} 
\end{equation}

\subsection{The flux of heat from the corona into the transition region}
\label{sec:qb} 

The temperature structure within the~transition region can be
determined using a method similar to the methods of \cite{rosner78}
and \cite{schwadron03}.  The Knudsen number~$N_{\rm K}$ (the electron
Coulomb mean free path~$\lambda_{\rm mfp}$ divided by the temperature
gradient scale length~$l_T$) is approximately~$ 10^{-3}$ in the low
corona, approximately~$ 10^{-3}$ at the upper end of the transition
region, and approximately~$10^{-6}$ at the lower end of the transition
region.\footnote{These estimates follow from setting
  $\rho \sim 10^8 m_{\rm p} \mbox{ cm}^{-3}$, $T \sim 10^6 \mbox{ K}$,
  and~$l_T \sim R_{\odot}$ in the low corona,
  $\rho \sim 10^9 m_{\rm p} \mbox{ cm}^{-3}$,
  $T \sim 3\times 10^5 \mbox{ K}$, and~$l_T \sim 0.01 R_{\odot}$ at
  the upper end of the transition region, and
  $\rho \sim 10^{11} m_{\rm p} \mbox{ cm}^{-3}$,
  $T \sim 10^4 \mbox{ K}$, and~$l_T \sim 100 \mbox{ km}$ at the lower
  end of the transition region.}  Because~$N_{\rm K} \ll 1$, the
radial component of the heat flux in the transition region is well
approximated by the \cite{spitzer53} formula,
\begin{equation}
q_r = -\alpha T^{5/2} \diff{T}{r},
\label{eq:spitzer_harm} 
\end{equation} 
where
\begin{equation}
\alpha =\frac{1.84 \times 10^{-5} \mbox{ erg} \mbox{ cm}^{-1} \mbox{
    s}^{-1} \mbox{ K}^{-7/2}}{\ln \Lambda_{\rm Coul}},
\label{eq:defsigma} 
\end{equation} 
and $\ln \Lambda_{\rm coul}$ is the Coulomb logarithm. In the
numerical examples of \S~\ref{sec:numerical},
\begin{equation}
\ln \Lambda_{\rm Coul} = 18.1,
\label{eq:lnLambda} 
\end{equation} 
the value for electron-electron collisions in a
proton-electron plasma with $\rho/m_{\rm p} = 10^{9} \mbox{ cm}^{-3}$
and $T = 10^6 \mbox{ K}$ \citep{huba13}.
The magnetic field near~$r=r_{\rm  b}$ is taken to be
approximately radial, and the width of the
transition region ($\sim 10^2 \mbox{ km}$) is 
so narrow that $p$, $B$, and~$A$ are treated as constants within
the transition region, with
\begin{equation}
\div \bm{q} = \diff{q_r}{r}.
\label{eq:divq} 
\end{equation} 
As mentioned previously, turbulent heating is neglected at
$r< r_{\rm b}$. Within the transition
region, equation~(\ref{eq:internal_energy}) thus becomes
\begin{equation}
  \left(\frac{\gamma}{\gamma-1}\right) p \div \bm{v} = - \diff{q_{\rm r}}{r} -
  \frac{c_{\rm R} p^2}{4 k_{\rm B}^2 T^{5/2}},
  \label{eq:internal_energy_TR}
\end{equation}
where $\rho$ has been expressed in terms of~$p$ and~$T$ using~(\ref{eq:defcs}).
The velocity divergence in~(\ref{eq:internal_energy_TR}) can be
expressed in terms of the heat flux via
\begin{equation}
  \div \bm{v} = - \frac{\bm{v}}{\rho} \bcdot \boldsymbol{\nabla} \rho =
  \frac{\bm{v}}{T} \bcdot \boldsymbol{\nabla} T =  \frac{U}{T}
  \left(-\frac{q_r}{\alpha T^{5/2}}\right).
    \label{eq:divv}
\end{equation}
where the first equality follows from (\ref{eq:cont0}).
With the aid of~(\ref{eq:divv}),
and using (\ref{eq:spitzer_harm}) to write
\begin{equation}
  \diff{q_r}{r} = - \diff{q_r}{T} \frac{q_r}{\alpha T^{5/2}}
  \label{eq:dqrdr},
\end{equation}
one can rewrite~(\ref{eq:internal_energy_TR}) in the form
\begin{equation}
  -a_2 q_r = q_r \diff{q_r}{T} - a_1,
  \label{eq:IE_TR2}
\end{equation}
where
\begin{equation}
  a_1 = \frac{\alpha p^2 c_{\rm R}}{4 k_{\rm B}^2} \qquad a_2 =
  \left(\frac{\gamma}{\gamma-1}\right) \frac{p U}{T}
  \label{eq:a1a2}
\end{equation} 
are both constants. Upon defining
\begin{equation}
  w = - \frac{a_2}{a_1} q_{\rm r},
  \label{eq:defw}
\end{equation}
which is positive, one can rewrite~(\ref{eq:IE_TR2}) as
\begin{equation}
  \frac{a_2^2}{a_1} \d T = \frac{w }{1 + w} \d w,
  \label{eq:w_eq}
\end{equation}
which can be integrated from the chromospheric values of~$T$ and~$w$,
denoted $T_{\rm chr}$ and $w_{\rm chr}$, to the values of~$T$ and~$w$
at $r=r_{\rm b}$, denoted $T_{\rm b}$ and~$w_{\rm b}$. As $T_{\rm
  chr} \ll T_{\rm b}$ and the chromospheric heat flux is negligible
($q$ scaling as $T^{7/2}$ divided by the temperature-gradient scale length), the chromospheric terms are
dropped, and the integral becomes
\begin{equation}
  w_{\rm b} - \ln(1+w_{\rm b}) = a_3,
  \label{eq:wb}
\end{equation}
where 
\begin{equation}
  a_3 = \frac{a_2^2 T_{\rm b}}{a_1} =
 \frac{2}{I_1^2} \left(\frac{\gamma}{\gamma-1}\right)^2 \left(\frac{U_{\rm
      b}}{\overline{ c_{\rm s}}}\right)^2,
  \label{eq:defa3}
\end{equation}
\begin{equation}
  I_1 = \left(\frac{\alpha c_{\rm R} m_{\rm p}}{4 k_{\rm
        B}^3}\right)^{1/2} = 0.1581,
  \label{eq:defI1}
\end{equation}
and the numerical value on the right-hand side of~(\ref{eq:defI1}) is
calculated using~(\ref{eq:lnLambda}). Equation~(\ref{eq:wb})  can be
solved in terms of the lower branch of the Lambert~$W$
function,~$W_{-1}$. This solution, in conjunction
with~(\ref{eq:defw}), yields
\begin{equation}
  q_{\rm b} = -q_{r\rm b} = - \frac{a_1}{a_2}\left[ 1 +
    W_{-1}\left(-e^{-(1+a_3)}\right)\right],
  \label{eq:lambertqb}
\end{equation}
which is positive since $\displaystyle W_{-1}\left(- e^{-(1+a_3)}\right)<-1 $.

If $U_{\rm b}/ \overline{ c_{\rm s}}$ is sufficiently small, then $a_3
\ll 1$. In this low-Mach-number limit, (\ref{eq:lambertqb}) becomes, to leading order in~$a_3$, 
\begin{equation}
  q_{\rm b} = I_1 \rho_{\rm b} \overline{ c_{\rm s}}^3.
  \label{eq:qb_approx}
\end{equation}
Equation~(\ref{eq:qb_approx}) is equivalent to equation~(3.15) of
\cite{rosner78} when thermal conduction is the only source of heating
in the transition region, i.e., when $f_{\rm H}$ is set equal to~zero in
their equation~(3.15).  Equation~(\ref{eq:lambertqb}) follows
\cite{schwadron03} in generalising the work of~\cite{rosner78} to
include internal-energy losses
from $p\,\d V$~work and advection. However, the analysis leading to~(\ref{eq:lambertqb})
differs from that of \cite{schwadron03} in that (\ref{eq:lambertqb})
completely neglects turbulent heating at $r<r_{\rm b}$ and does not
assume that $\int q(T) \d T \propto T$.\footnote{\cite{schwadron03}
  integrate the internal-energy equation from the upper chromosphere
  all the way out to the coronal temperature maximum, where the heat
  flux vanishes. The value of~$q_{\rm b}$ in~(\ref{eq:lambertqb})
  corresponds to $r=r_{\rm b}$, at which turbulent heating and
  radiative cooling balance, which, in the low-Mach-number limit,
  corresponds to the maximum of the heat flux. This heat-flux maximum
  is lower down in the solar atmosphere than the temperature maximum.}

\subsection{Constraints associated with the wave-modified sonic critical point}
\label{sec:crit_point} 

In the presence of mostly outward-propagating AWs, a radial background
magnetic field, and a radial outflow, the momentum equation in the
approximately isothermal sub-Alfv\'enic region takes the form
\begin{equation}
  \rho U \diff{U}{r} = - \overline{ c_{\rm s}}^2\diff{\rho}{r} - \diff{}{r}\left(\frac{\rho
      z_+^2}{8}\right) - \frac{G M_{\odot} \rho}{r^2},
  \label{eq:momentum1}
  \end{equation} 
where $\rho z_+^2 / 8$ is the AW pressure, which is one-half the AW
energy density~\citep{dewar70}. Equation~(\ref{eq:cont1}) implies that
\begin{equation}
  \frac{1}{\rho}\diff{\rho}{r} + \frac{1}{U} \diff{U}{r} -
  \frac{1}{B}\diff{B}{r} = 0.
  \label{eq:cont3}
\end{equation}
Equations~(\ref{eq:zplusy})  and (\ref{eq:cont3}) can be used to
rewrite~(\ref{eq:momentum1}) in the form
\begin{equation}
  \frac{1}{\rho}\diff{\rho}{r} \left\{
    -U^2 + \overline{ c_{\rm s}}^2 + \frac{(\delta v_{\rm b})^2
      (1+y_{\rm b})^{2-\sigma}[y^2(1+\sigma) + 3y]}{4 y_{\rm
        b}(1+y)^{3-\sigma}}\right\} = \frac{2}{r}\left(\gamma_B U^2 -
    \frac{v_{\rm esc}^2 R_{\odot}}{4 r}\right),
  \label{eq:crit_pt1}
\end{equation} 
where
\begin{equation}
  \gamma_B \equiv - \frac{r}{2B} \diff{B}{r} = 1 - \frac{r}{2\eta}
  \diff{\eta}{r},
  \label{eq:defgammaB}
\end{equation}
and $\eta$ is defined in~(\ref{eq:Bprofile}). In order for~(\ref{eq:crit_pt1})
to possess a transonic-wind solution for~$U$, the quantity in braces on the
left-hand side must be positive near the Sun and negative far from the
Sun; i.e., it must pass through zero at some radius~$r_{\rm c}$ (the
wave-modified
sonic critical point).  In order for $\d \rho/ \d r$ to
remain finite at~$r=r_{\rm c}$, the right-hand side
of~(\ref{eq:crit_pt1}) must also vanish at~$r_{\rm c}$. Together,
these conditions yield the constraints
\begin{equation} 
  r_{\rm c}  =  \frac{v_{\rm esc}^2 R_{\odot}}{4 \gamma_{B\rm c} U_{\rm
                  c}^2} \label{eq:defrc} 
\end{equation} 
and
\begin{equation} 
  U_{\rm c}^2  =  \overline{ c_{\rm s}}^2 + \frac{(\delta v_{\rm b})^2
      (1+y_{\rm b})^{2-\sigma}[y_{\rm c}^2(1+\sigma) + 3y_{\rm c}]}{4 y_{\rm
                    b}(1+y_{\rm c})^{3-\sigma}}
                    \label{eq:defUc},
\end{equation} 
where, here and in the following, a `c' subscript indicates that the
subscripted quantity is evaluated at~$r=r_{\rm c}$.
An implicit assumption underlying~(\ref{eq:defrc})
and~(\ref{eq:defUc}) is that
\begin{equation}
  r_{\rm c} < r_{\rm A},
  \label{eq:rcrA}
\end{equation}
so that the quasi-isothermal approximation applies at $r_{\rm
  c}$. 

Mass and flux conservation (i.e., (\ref{eq:cont1})) imply that
\begin{equation}
\frac{\rho_{\rm c} U_{\rm c}}{B_{\rm c}} = \frac{\rho_{\rm  b} U_{\rm
    b}}{B_{\rm b}} = \frac{\rho_{\rm b} v_{\rm Ab}}{B_{\rm b} y_{\rm
    b}},
\label{eq:mass_cons_c1}
\end{equation}  
where the second equality in~(\ref{eq:mass_cons_c1}) follows from~(\ref{eq:vAyU}).
Equation~(\ref{eq:Bprofile}) implies that
\begin{equation}
  \frac{B_{\rm b}}{B_{\rm c}} = \frac{r_{\rm c}^2}{R_{\odot}^2} \psi
  \frac{\eta_{\rm b}}{\eta_{\rm c}}
= \frac{v_{\rm esc}^4}{16 \gamma_{B\rm c}^2 U_{\rm c}^4} \psi
  \frac{\eta_{\rm b}}{\eta_{\rm c}},
  \label{eq:BbBc}
\end{equation}
where the second equality in~(\ref{eq:BbBc}) follows from~(\ref{eq:defrc}).
Upon substituting~(\ref{eq:BbBc}) into~(\ref{eq:mass_cons_c1}) and
evaluating~$v_{\rm Ab}$ using~(\ref{eq:vAb}), one obtains
\begin{equation}
U_{\rm c}^2 = v_{\rm esc}^2 \left[\frac{y_{\rm c}^2
  \psi^{1/4} \xi (x \tilde{\rho}_{\odot})^{1/8}}{16 \gamma_{B\rm c}^2
  \eta_{\rm c} y_{\rm b}}\right]^{2/3}.
\label{eq:Uc3}
\end{equation} 
Substituting~(\ref{eq:Uc3}) into~(\ref{eq:defUc}) yields
\begin{equation}
  \left[\frac{y_{\rm c}^2 \psi^{1/4} \xi (x
      \tilde{\rho}_{\odot})^{1/8}}{16 \gamma_{B\rm c}^2 \eta_{\rm c}
      y_{\rm b}}\right]^{2/3} - x - \frac{\epsilon (1+y_{\rm
      b})^{2-\sigma}[y_{\rm c}^2(1+\sigma) + 3
    y_{\rm c}]}{4 y_{\rm b}(1+y_{\rm c})^{3-\sigma}} = 0.
  \label{eq:crit_point_cond}
\end{equation}

The integral over~$r$ of $\rho^{-1}$ times the momentum equation
(\ref{eq:momentum1}) yields the Bernoulli integral,
\begin{equation} 
\frac{U^2}{2} + \overline{ c_{\rm s}}^2 \ln\left(\frac{\rho}{\rho_{\rm
      b}}\right) - \frac{v_{\rm esc}^2 R_{\odot}}{2 r} - \frac{(\delta
  v_{\rm b})^2 (1+y_{\rm b})^{2-\sigma}}{2y_{\rm b}} \left[
  \left(\frac{1+\sigma}{1-\sigma}\right) (1+y)^{\sigma - 1} +
  (1+y)^{\sigma -2}\right] = \Gamma,
\label{eq:Bernoulli1}
\end{equation}
where $\Gamma$ is independent of~$r$. Evaluating~(\ref{eq:Bernoulli1})
at $r=r_{\rm b}$ leads to the equality
\begin{equation}
  \Gamma = \frac{U_{\rm b}^2}{2} - \frac{v_{\rm esc}^2 \psi^{1/2}}{2}
  - \frac{(\delta v_{\rm b})^2}{2}\left[ \frac{1+\sigma}{1-\sigma} +
    \frac{2}{y_{\rm b}(1-\sigma)}\right].
  \label{eq:Gamma}
\end{equation}
Evaluating~(\ref{eq:Bernoulli1}) at $r=r_{\rm c}$, rewriting~$r_{\rm
  c}$ and~$U_{\rm c}$ using~(\ref{eq:defrc}) and~(\ref{eq:defUc}),
rewriting~$\Gamma$ using~(\ref{eq:Gamma}), and
multiplying the resulting equation by~$2/v_{\rm esc}^2$ yields
\[
\psi^{1/2} + x\left[4 \ln\left( \frac{y_{\rm c}}{y_{\rm b}}\right) + 1 - 4
  \gamma_{B\rm c}\right] + \frac{\epsilon (1 + y_{\rm b})^{2-\sigma}
  \Phi}{4 (1-\sigma)y_{\rm b} (1+y_{\rm c})^{3-\sigma}}
\]
\begin{equation}
  - \frac{\eta_{\rm
    b}^2 \psi^{3/2}}{y_{\rm b}^2 \xi^2 (x \tilde{\rho}_{\odot})^{1/4}}
+ \epsilon \left[\frac{1+\sigma}{1-\sigma} + \frac{2}{y_{\rm
      b}(1-\sigma)}\right] = 0 ,
  \label{eq:Bernoulli2}
\end{equation}
where
\begin{equation}
  \Phi \equiv y_{\rm c}^2[(1-4\gamma_{B\rm c})(1-\sigma^2) - 4(1+\sigma)]
  + y_{\rm c}[3(1-4\gamma_{B\rm c})(1-\sigma) - 12 - 4\sigma] - 8.
  \label{eq:defPhi}
  \end{equation} 

\subsection{Mathematical structure of the model and approximate
  analytic solutions}
\label{sec:knowns} 

The various quantities appearing in the model equations can be divided
into five groups: (1) quantities that are determined observationally
($R_{\odot}$, $M_{\odot}$, $v_{\rm esc}$, $\rho_{\odot}$,
$\tilde{\rho}_{\odot}$, $\psi$, $\delta v_{\odot \rm eff}$,
$\bar{ B}$); (2) free parameters ($\sigma$, $l_{\rm b}$); (3) the
super-radial expansion factor $\eta(r)$, which takes on different
values in different magnetic flux tubes and in different models for
the solar magnetic field; (4) the three principal unknowns,
$y_{\rm b}$, $y_{\rm c}$, and $x$; and (5) additional unknowns that
can be determined once $y_{\rm b}$, $y_{\rm c}$, and $x$ are found
($\dot{M}$, $\chi_{\rm H}$, $U(r)$, $\rho_{\rm b}$, $q_{\rm b}$,
$v_{\rm Ab}$, $r_{\rm c}$, $r_{\rm A}$).  The three principal unknowns
$y_{\rm b}$, $y_{\rm c}$, and $x$ are determined by solving the three
simultaneous equations (\ref{eq:IEbal}), (\ref{eq:crit_point_cond}),
and~(\ref{eq:Bernoulli2}), where it must be remembered that~$\epsilon$
is itself a function of~$x$ via~(\ref{eq:defeps}).  The additional
unknowns $\dot{M}$, $\chi_{\rm H}$, $\rho_{\rm b}$, $v_{\rm Ab}$,
$q_{\rm b}$, and $r_{\rm c}$ then follow immediately
from~(\ref{eq:Mdot2}), (\ref{eq:defchiH}), (\ref{eq:rhob2}),
(\ref{eq:vAb}), (\ref{eq:lambertqb}), and (\ref{eq:defrc}),
respectively. For example,
\begin{equation}
  \dot{M} = \frac{R_\odot^2 \bar{ B}^2}{v_{\rm esc}} y_{\rm
    b}^{-1} (x \tilde{\rho}_\odot)^{1/8} \xi \psi^{1/4} .
  \label{eq:Mdotgeneral}
\end{equation}
The procedures for determining~$r_{\rm A}$ and~$U(r)$
involve a few more steps, which are described in
Appendix~\ref{ap:U_of_r}.

Two approximate analytic solutions to (\ref{eq:IEbal}), 
(\ref{eq:crit_point_cond}) and~(\ref{eq:Bernoulli2}), valid in two
different parameter regimes, are derived in
Appendix~\ref{ap:analytic}. Both solutions 
rely on the approximations
\begin{equation}
 y_{\rm b} \gg 1 \qquad \psi = 1 \qquad \epsilon \ll 1
  \qquad \eta_{\rm c} = \gamma_{B \rm c} = 1.
  \label{eq:approxs0}
\end{equation}
The last equality in~(\ref{eq:approxs0}) amounts to taking the
magnetic-field lines to be purely radial at $r\geq r_{\rm c}$. The first of the two approximate analytic solutions is
valid in the conduction-dominated regime, in which the dominant sink
of internal energy in the sub-Alfv\'enic region is the
flux of heat from the corona into the transition region, rather than
$p\,\d V$ work. As discussed further in the following, this regime  arises only for
values of~$\delta v_{\odot \rm eff}$ much smaller than the solar
value. This limit is thus not directly relevant to solar-wind observations.
 To leading order in~$\epsilon_\odot$, the mass outflow
rate~$\dot{M}^{\rm (cond)}$  and asymptotic flow velocity
$U_{\infty}^{\rm (cond)}$ in this parameter regime are given by
\begin{equation}
\dot{M}^{\rm (cond)}=\frac{R_{\odot}^2 B_{\rm ref}^2}{v_{\rm esc}}
\frac{1}{I_1^{1/14} I_2}
\left[
  \epsilon_\odot^{14-4\sigma} (\eta_{\rm b} B_\ast)^{-4\sigma}
  \tilde{l}_{\rm b}^{3\sigma}\tilde{\rho}_\odot^{7-2\sigma}
  \right]^{1/(7-7\sigma)},
  \label{eq:Mdotcond}
\end{equation}
where $I_2$ is a numerical constant given in~(\ref{eq:defI2}), and
\[
U_\infty^{\rm (cond)} = v_{\rm esc}\Bigg[
    \frac{2^{9-4\sigma} \zeta}
    {(1+\sigma)^2 I_1^{\sigma/14}}
    \left( \frac{1-\sigma}{7-3\sigma}\right)^{(7-3\sigma)/2}
\]
\begin{equation}
\epsilon_\odot^{-(7-2\sigma)/7} B_\ast^{(7-5\sigma)/7}
\eta_{\rm b}^{2\sigma/7} \tilde{\rho}_\odot^{(-7 + 2 \sigma)/14}
\tilde{l}_{\rm b}^{-3\sigma/14}\Bigg]^{1/2}.
    \label{eq:Uinfcond}
\end{equation}
The coronal temperature in the conduction-dominated limit follows
directly from~(\ref{eq:defcs}) and~(\ref{eq:xcond}).

The second approximate analytic solution is valid in the
expansion-dominated regime, in which the $p\,\d V$ work resulting from
expansion is the dominant sink of internal energy in the
sub-Alfv\'enic region. To leading order, the mass outflow rate in this
parameter regime is given by
\begin{equation}
  \dot{M}^{\rm (exp)}_0 =  \epsilon_\odot \bar{ B} R_\odot^2
  \sqrt{4\upi \rho_\odot} = \frac{P_{\rm AW}(r_{\rm b})}{v_{\rm esc}^2},
  \label{eq:Mdotexp0}
\end{equation} 
where
$P_{\rm AW}(r_{\rm b})$ is the AW power (\ref{eq:defLAW}) evaluated
at the coronal base.  The asymptotic wind
speed in the expansion-dominated regime is to leading order given by
\begin{equation}
    U_{\infty, 0}^{\rm (exp)} = v_{\rm esc},
    \label{eq:Uinf0expdom}
\end{equation}
and the coronal temperature in the expansion-dominated regime is to
leading order
\begin{equation}
  T = \frac{m_{\rm p} v_{\rm esc}^2}{k_{\rm B} \ln(\epsilon_\odot^{-3}
    \tilde{\rho}_{\odot}^{-3/2} \tilde{l}_{\rm b}^{-1} \eta_{\rm b}
    B_\ast)}.
  \label{eq:T0expdom} 
\end{equation}
Equations~(\ref{eq:Mdotexp0}) and (\ref{eq:Uinf0expdom}) reproduce the
approximate scalings of the simplified calculation presented
in \S~\ref{sec:heuristic}. Equation~(\ref{eq:T0expdom}) matches
the right-hand side of~(\ref{eq:Theuristic}) to within five percent
for Sun-like parameters, the difference arising because in the model
developed in this section, $\delta v_{\rm b}$ has a weak dependence on
the coronal temperature via~(\ref{eq:dvbdvsun}) and~(\ref{eq:rhob2})
that is not accounted for in \S~\ref{sec:heuristic}.
Appendix~\ref{ap:analytic} presents higher-order corrections
to~(\ref{eq:Mdotexp0}), (\ref{eq:Uinf0expdom}),
and~(\ref{eq:T0expdom}) that account for conductive losses, wave
momentum deposition inside the wave-modified sonic critical point, and
the fact that only part of~$P_{\rm AW}(r_{\rm b})$ is dissipated
within the sub-Alfv\'enic region. Second- and fourth-order
approximations to~$\dot{M}$ and~$U_\infty$ are shown in
figure~\ref{fig:Mdot_U_fmax} below.

Analytic estimates for the ranges of~$\epsilon_{\odot}$ values
corresponding to the conduction-dominated and expansion-dominated
limits are given in Appendix~\ref{ap:analytic}. There are three
constraints on~$\epsilon_\odot$ in the conduction-dominated limit, the
most stringent of which is that the wave-energy term dominates over the
internal-energy term in the Bernoulli integral at the wave-modified
sonic critical point. The resulting range of~$\epsilon_\odot$ values
is much smaller than the solar value, as illustrated in
figure~\ref{fig:Mdot_asymptotic}. The expansion-dominated limit
corresponds to a finite range of~$\epsilon_\odot$ values that is
sufficiently large that~$p\,\d V$ work dominates over conduction as the
primary internal-energy sink within the sub-Alfv\'enic region, and
sufficiently small that the sound speed makes the dominant contribution
to the outflow velocity at the wave-modified sonic point
in~(\ref{eq:defUc}). This range of $\epsilon_\odot$ values is relevant
to the solar case, as shown in the next section.

\section{Numerical examples}
\label{sec:numerical}

This section presents several numerical solutions and approximate
analytic solutions to the equations of the model developed in
\S~\ref{sec:analytic}.  The numerical solutions are obtained by
solving (\ref{eq:IEbal}), (\ref{eq:crit_point_cond}),
and~(\ref{eq:Bernoulli2}) for $y_{\rm b}$, $y_{\rm c}$, and $x$ using
Newton's method.\footnote{In practice, rather than
  evaluating~$q_{\rm b}$ in~(\ref{eq:IEbal}) using the Lambert~$W$
  function in~(\ref{eq:lambertqb}), I evaluate~$q_{\rm b}$
  using~(\ref{eq:defw}), treat~$w_{\rm b}$ as a fourth unknown, and
  include~(\ref{eq:wb}) as a fourth simultaneous equation to be solved
  numerically. The resulting value of~$q_{\rm b}$ is identical to the
  value from~(\ref{eq:lambertqb}).} Once $y_{\rm b}$, $y_{\rm c}$, and
$x$ are determined, $\dot{M}$, $\chi_{\rm H}$, $\rho_{\rm b}$,
$v_{\rm Ab}$, $q_{\rm b}$, $r_{\rm c}$, $r_{\rm A}$, and~$U_{\infty}$
are computed from~(\ref{eq:Mdot2}), (\ref{eq:defchiH}),
(\ref{eq:rhob2}), (\ref{eq:vAb}), (\ref{eq:lambertqb}),
(\ref{eq:defrc}), (\ref{eq:rA3}), and~(\ref{eq:Uinfty}), respectively.
The approximate analytic solutions are derived in
Appendix~\ref{ap:analytic}.

\subsection{Magnetic-field model}
\label{sec:mag}

The equations in \S~\ref{sec:analytic} are compatible with any
model for the radial profile of the magnetic-field strength, or,
equivalently, any choice of~$\eta(r)$. On the other hand, the approximate
analytic solutions derived in Appendix~\ref{ap:analytic} assume that
\begin{equation}
  \eta_{\rm c} = \gamma_{B{\rm c}} = 1.
  \label{eq:etacgammaBc}
\end{equation} 
To maintain consistency between the numerical
and analytic solutions, and to avoid introducing additional complexity
and free parameters, I take~(\ref{eq:etacgammaBc}) to hold when computing
the numerical solutions presented in this section. In essence,
equation~(\ref{eq:etacgammaBc}) amounts to assuming that all of the
super-radial expansion of the magnetic field occurs inside the
wave-modified sonic critical point. In measurements from
the FIELDS experiment on the Parker Solar Probe (PSP) \citep{bale16} during PSP's
first few orbits,
$|B_r| \simeq 2.2 \mbox{ nT} (1 \mbox{ a.u.}/r)^2$ (Sam Badman, private
communication), where a.u.\ is the abbreviation for astronomical unit. In order to
match this $B_r$ profile, I set
\begin{equation}
  B_\ast = 0.00856 \hspace{0.3cm} \longleftrightarrow \hspace{0.3cm}
  B_r(\mbox{1 a.u.}) = 2.2 \mbox{ nT}
  \label{eq:Bastvalue}
\end{equation}
when computing the results shown in figures~\ref{fig:PSP_E1}
through~\ref{fig:contour1}.

\subsection{Values of the AW power at the coronal base
  and~$\delta v_{\odot \rm eff}$}
\label{sec:PAWb}

Equations~(\ref{eq:chichr}) and~(\ref{eq:dvsuneff}) and the flux-conservation
relation $A_{\rm b} B_{\rm b} = A_{\odot} B_{\odot}$ imply that
\begin{equation}
  \delta v_{\odot \rm eff} =
  \left(\frac{4\upi}{\rho_{\odot}}\right)^{1/4} \left( \frac{P_{\rm
        AWb}}{B_{\rm b} A_{\rm b}} \right)^{1/2}.
  \label{eq:dvsunemp}
\end{equation}
When the solar wind is powered primarily by AWs, $P_{\rm
  AWb}$ is approximately equal to the value of~$\dot{E}_{\rm m0}$
in~(\ref{eq:sandbaek94b}), i.e.,
\begin{equation}
  P_{\rm AWb} \simeq \frac{1}{2} \dot{M} \left( v_{\rm esc}^2 +
    U_{\infty}^2\right).
  \label{eq:PAWb_emp}
\end{equation} 
 Upon evaluating the right-hand side
of~(\ref{eq:PAWb_emp}) using the data in table~1 of \cite{schwadron08}
for {\em Ulysses}' third northern polar pass (3NPP), one obtains
\begin{equation}
  P_{\rm AWb} = 3.6 \times 10^{27} \mbox{ erg} \mbox{ s}^{-1},
  \label{eq:PAWb_emp2}
\end{equation}
which corresponds to $P_{\rm AWb} / (4 \upi R_{\odot}^2) = 0.59 \times 10^5 \mbox{ erg}
\mbox{ cm}^{-2} \mbox{ s}^{-1}$.  I use the 3NPP data 
because this is the
part of {\em Ulysses}'s first three orbits during which the average
scaled radial magnetic field
$|\langle B_r \rangle| \cdot (r/ 1 \mbox{ a.u.})^2 $ was most consistent
with~(\ref{eq:Bastvalue}) (in the 3NPP~data,
$|\langle B_r \rangle| \cdot (r/ 1 \mbox{ a.u.})^2 = 2.1 \pm 0.08 \mbox{
  nT}$), and because~$\dot{M}$ is strongly correlated with the average
scaled radial magnetic field
\citep{schwadron08}.  Equations (\ref{eq:rhoodotval}),
(\ref{eq:Bastvalue}), (\ref{eq:dvsunemp}), and~(\ref{eq:PAWb_emp2})
and the relation
$B_{\rm b} A_{\rm b} = |B_r(1 \mbox{ a.u.})| 4 \upi (1 \mbox{ a.u.})^2$
imply that
\begin{equation}
  \delta v_{\odot \rm eff} = 0.22 \mbox{ km} \mbox{ s}^{-1}.
  \label{eq:dvsuneffval}
\end{equation}
This value of~$\delta v_{\odot \rm eff}$ is used in
figures~\ref{fig:PSP_E1}, \ref{fig:Mdot_U_fmax},
and~\ref{fig:contour2}.

\subsection{Free parameters}
\label{sec:free}

As discussed in \S~\ref{sec:knowns}, there are two free
parameters in the model:
$l_{\rm b}$  (the AW dissipation length scale at~$r_{\rm b}$)
and~$\sigma$ (the dimensionless coefficient in the turbulent heating
rate).
The solutions to (\ref{eq:IEbal}), 
(\ref{eq:crit_point_cond}), and~(\ref{eq:Bernoulli2}) are not very
sensitive to the value of~$l_{\rm b}$. Throughout the rest of
this section, $l_{\rm b}$ is thus simply fixed at a value that seems
physically reasonable:
\begin{equation}
  l_{\rm b} = 0.3 R_{\odot}.
  \label{eq:lb}
\end{equation}
On the other hand, the solutions depend sensitively on the value
of~$\sigma$. Larger values of~$\sigma$ cause a larger fraction of the
AW power at the coronal base~$P_{\rm AWb}$ to be dissipated
in the quasi-isothermal sub-Alfv\'enic region, which
increases~$\dot{M}$ (see, e.g.,~(\ref{eq:IE_heur1})
and~(\ref{eq:Mdot_heur3})). For fixed~$P_{\rm AWb}$,
increasing~$\dot{M}$ decreases~$U_{\infty}$.  In some of the examples
to follow, $\sigma$ is varied to optimise agreement between the model
and observations, as described further in \S~\ref{sec:PSP_E1} and \S~\ref{sec:wang}.  The super-radial
expansion factor at the coronal base,~$\eta_{\rm b}$, also varies in
the model, but this quantity is in principle observable for individual
flux tubes. The value of~$\eta_{\rm b}$ is thus treated as an input
into the model rather than an adjustable parameter.

\subsection{Fiducial solution matching measurements from Parker Solar
  Probe's first perihelion}
\label{sec:PSP_E1}

\begin{figure}
  \includegraphics[width=7cm]{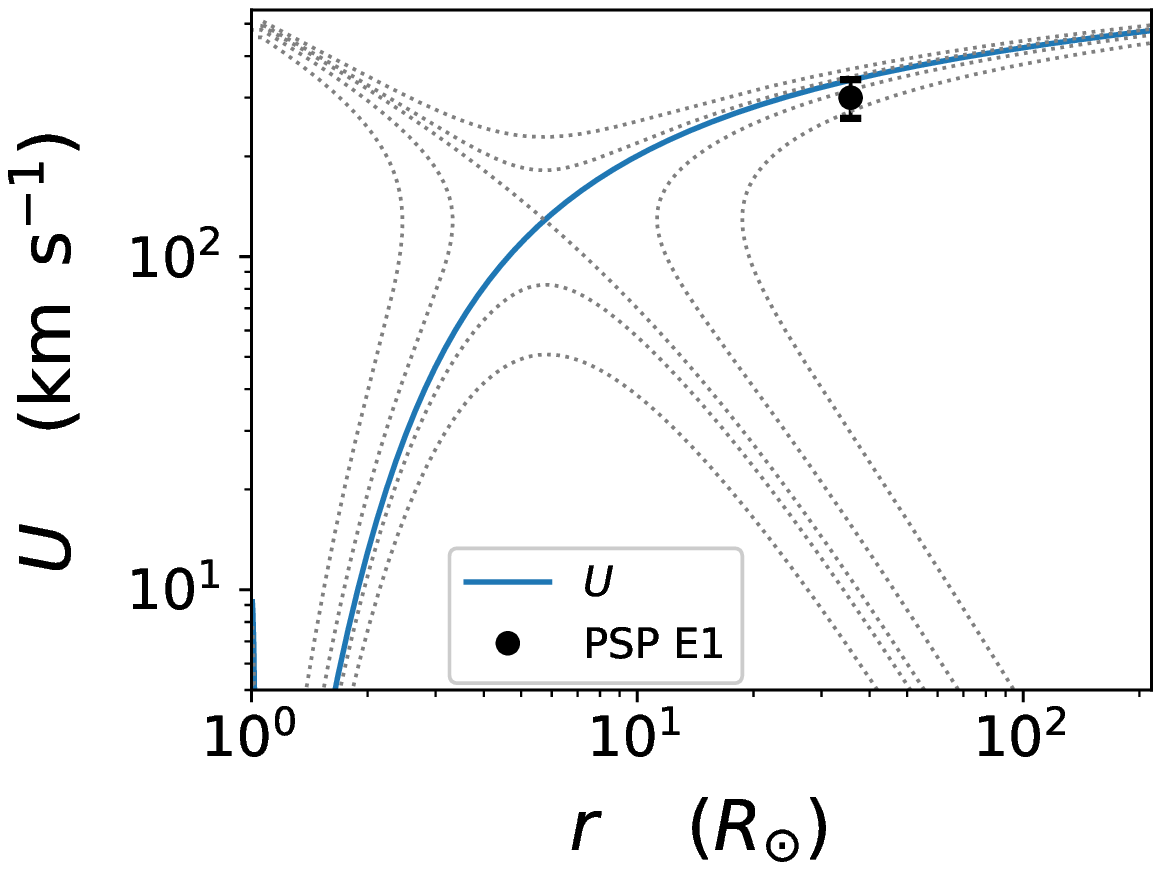}
  \includegraphics[width=7cm]{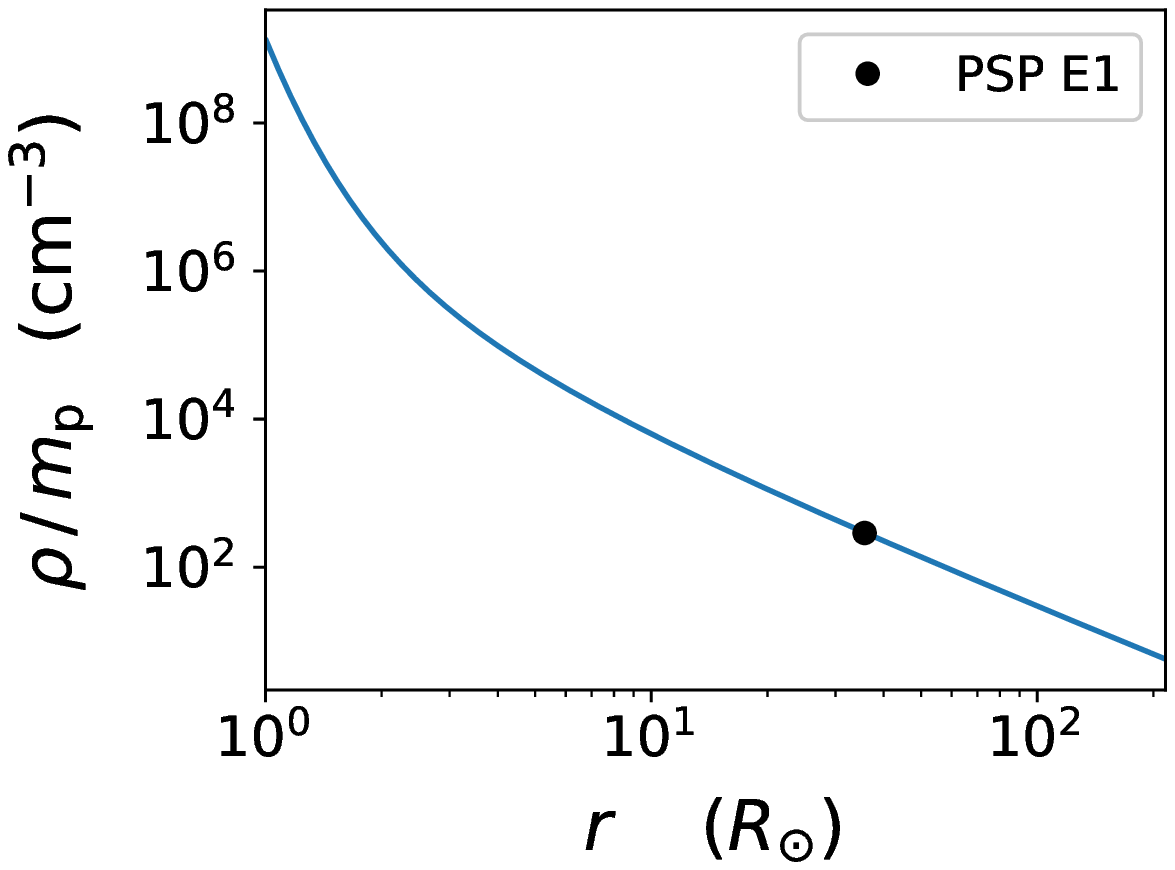}
\\
\includegraphics[width=7cm]{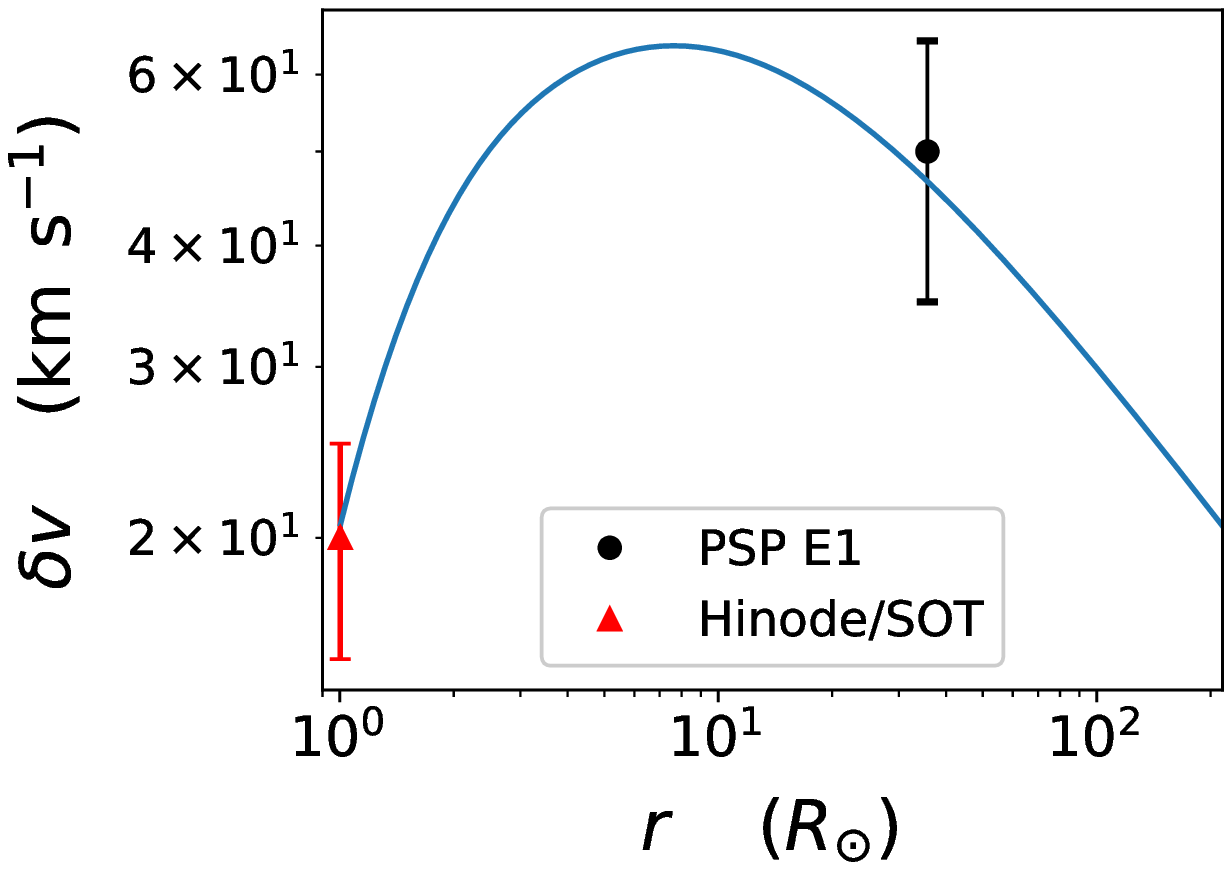}
\includegraphics[width=7cm]{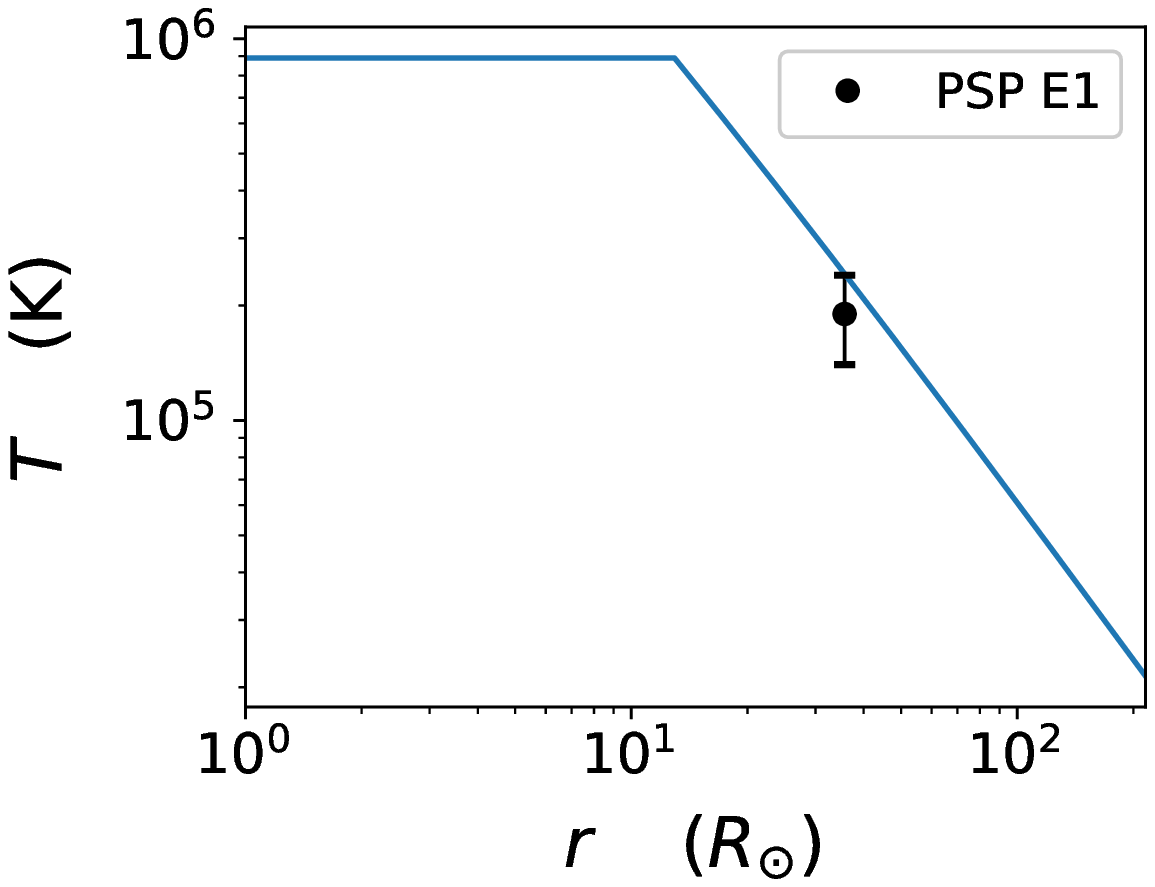}
\caption{The outflow
velocity~$U$, density~$\rho$, r.m.s. amplitude of the 
velocity fluctuation~$\delta v$, and
temperature~$T$ as functions of heliocentric distance~$r$ in a numerical solution to the model equations
with  $\eta_{\rm b} = 100$,  $\sigma =
  0.5$, and the parameter values
  in~(\ref{eq:etacgammaBc}), (\ref{eq:Bastvalue}),
  (\ref{eq:dvsuneffval}), and~(\ref{eq:lb}). The dotted lines in the top-left panel are described in the text.
In the top panels and the lower-right panel, the  circles are
measurements from a 1~h interval containing PSP's first
perihelion on 6~November 2018, from figure~1 of \cite{kasper19}.
The error bars around the PSP data points
in these three panels indicate the approximate
range of values with a relative occurrence rate 
of at least 50\% of
the peak occurrence rate within that 1~h interval. In the top-right panel, the error bars lie
within the data point. In the
lower-left panel, the PSP data point is from \cite{chen20}, and the
error bars around that data point show the approximate range of
measured values near $r=35.7 R_{\odot}$ in figure~7 of \cite{chen20}.
The triangle in the lower-left panel
is the value obtained by \cite{depontieu07} from an analysis of the
motion of filamentary structures in the low solar atmosphere based on
observations from the Solar Optical Telescope on the Hinode satellite.
\label{fig:PSP_E1}}
\end{figure}

Figure~\ref{fig:PSP_E1} illustrates the $r$ dependence of the outflow
velocity, density, fluctuating velocity, and temperature in a
numerical solution to the model equations that is designed to agree
with measurements from PSP's first perihelion encounter (E1) on
6~November 2018. The procedure for computing~$U(r)$ and~$T(r)$ is
described in Appendix~\ref{ap:U_of_r}.  In order to solve for~$U$ at
some radius~$r$, the value of~$\eta$ must be specified at
that~$r$. For figure~\ref{fig:PSP_E1}, I set
$\eta(r) = 1 + (\eta_{\rm b} - 1) \exp(-(r-r_{\rm b})^2/(0.02
R_{\odot} )^2) $, which is effectively unity at $r= r_{\rm c}$,
consistent with~(\ref{eq:etacgammaBc}).  This choice of~$\eta(r)$ is
not realistic for the low corona \citep[see, e.g.,][]{cranmer07}, but
the flow profile in the low corona is not a focus of this
work. It should be noted that while the value of~$\delta v$ at
$r=r_{\rm b}$ (i.e., $\delta v_{\rm b}$)
shown in the lower-left panel
of Figure~\ref{fig:PSP_E1} depends on~$\eta_{\rm b}$
through~(\ref{eq:dvbdvsun}), (\ref{eq:rhob2}), and~(\ref{eq:defxi}),
$\delta v_{\rm b}$ does
not on the way in which $\eta(r)$ decreases from~$\eta_{\rm b }$ to~1.

The solution in Figure~\ref{fig:PSP_E1}  is based upon
the somewhat arbitrary assumption that~$\eta_{\rm b} = 100$ in the
magnetic flux tube encountered by PSP at the time of its first perihelion. The value
of~$\sigma$ is set equal to~0.5 to optimise the agreement between
the model and the data.  Fits of comparable (in some cases superior)
quality can be obtained for different values of $\eta_{\rm b}$.  For
example, the parameter combinations $(\eta_{\rm b}, \sigma) =  (300,
0.47)$ and $(\eta_{\rm b}, \sigma) =  (30,0.55)$ lead to similar agreement with the data.
Modelling of the solar magnetic field during PSP E1 suggests that the
solar-wind stream encountered by PSP at the time of PSP's first
perihelion originated in a small equatorial coronal hole \citep{bale19}.  The fact that all
four quantities plotted in Figure~\ref{fig:PSP_E1} can be matched by
varying the single parameter~$\sigma$ suggests that the model is
reasonably successful at capturing the physical processes that control
the heating and acceleration of coronal-hole outflows.

The dotted lines in the top-left panel of Figure~\ref{fig:PSP_E1} are
solutions of the Bernoulli equation~(\ref{eq:Bernoulli1}) for
different values of the Bernoulli constant~$\Gamma$. One of the dotted
lines intersects the solid line, and that intersection occurs at the wave-modified sonic
critical point, $r=r_{\rm c}$. That dotted line is an accretion-like solution
of~(\ref{eq:Bernoulli1}) with the same value of~$\Gamma$ as in the
model solution but with $\d U / \d r < 0$ at $r=r_{\rm c}$.

\subsection{Illustration of the conduction-dominated and
  expansion-dominated regimes}
\label{sec:asympt}

\begin{figure} \begin{centering}
    \includegraphics[width=7cm]{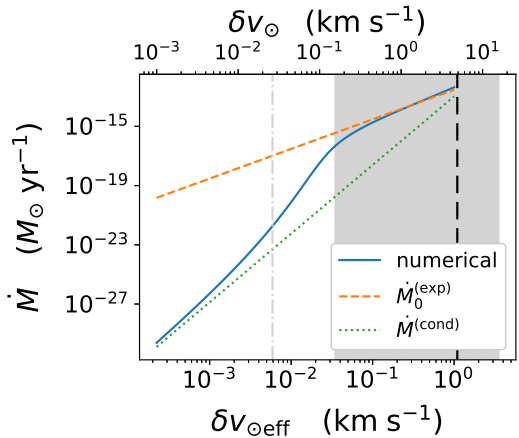} \caption{Mass
      outflow rate as a function of the effective
      fluctuating velocity at the photosphere $\delta v_{\odot \rm
        eff} = f_{\rm chr}^{1/2} \delta v_\odot$ (see
      (\ref{eq:dvsuneff})) for the parameter values
  in~(\ref{eq:etacgammaBc}), (\ref{eq:Bastvalue}),
  (\ref{eq:lb}), $\eta_{\rm b} = 30$, and $\sigma=0.18$, where $f_{\rm chr}$ is the chromospheric/transition-region
      AW transmission coefficient in~(\ref{eq:chichr}). The
corresponding r.m.s. photospheric velocity~$\delta
v_\odot$ is shown at the top of the plot for the case in which~$f_{\rm chr}
= 0.05.$ The solid line plots (\ref{eq:Mdotgeneral})
  using the numerical solution to (\ref{eq:IEbal}), 
(\ref{eq:crit_point_cond}), and~(\ref{eq:Bernoulli2}). The dotted and
short-dashed lines plot the approximate analytic results
from~(\ref{eq:Mdotcond}) and~(\ref{eq:Mdotexp0}), respectively. The
vertical dash-dot line corresponds to $\epsilon_{\odot \rm cond}$
in~(\ref{eq:condlim3}). The left and right edges of the shaded region
correspond to
$\epsilon_{\odot\rm exp, min}$ and $\epsilon_{\odot \rm exp, max}$ in~(\ref{eq:epsexpmin})
and~(\ref{eq:epsexpmax}), respectively. To the right of the vertical long-dashed
line, $r_{\rm c} > r_{\rm A}$, which violates~(\ref{eq:rcrA}) and the
assumptions of the model.
\label{fig:Mdot_asymptotic} }
\end{centering}
\end{figure}

Figure~\ref{fig:Mdot_asymptotic} plots~$\dot{M}$ as a function of~$\delta v_{\odot \rm eff}$ when
$\eta_{\rm b} = 30$ and~$\sigma = 0.18$. This figure
includes  the numerical solution to (\ref{eq:IEbal}),
(\ref{eq:crit_point_cond}), and~(\ref{eq:Bernoulli2})
as well as the approximate analytic results $\dot{M}^{\rm (cond)}$ and
$\dot{M}^{\rm
  (exp)}_0$ from~(\ref{eq:Mdotcond}) and~(\ref{eq:Mdotexp0}).
The vertical dash-dot line in this figure corresponds to
$\epsilon_{\odot \rm cond}$ in~(\ref{eq:condlim3}). The
conduction-dominated approximation that gives rise to~$\dot{M}^{\rm
  (cond)}$ in~(\ref{eq:Mdotcond}) assumes that~$\epsilon_\odot \ll
\epsilon_{\odot \rm cond}$.\footnote{It
should be noted, however, that the solution for~$U_\infty^{\rm
  (cond)}$ in~(\ref{eq:Uinfcond}) exceeds the speed of light when
$\delta v_{\odot \rm eff} \lesssim 2\times 10^{-3} \mbox{ km} \mbox{
  s}^{-1}$, and a relativistic treatment would be needed to model the
outflow correctly in this limit.}
The shaded region corresponds to $\epsilon_{\odot\rm exp, min}<
\epsilon_\odot < \epsilon_{\odot \rm exp, max}$, where
$\epsilon_{\odot\rm exp, min}$ 
and $\epsilon_{\odot\rm exp, max}$ are defined
in~(\ref{eq:epsexpmin}) and~(\ref{eq:epsexpmax}). The expansion-dominated approximation
in \S~\ref{sec:expansion} assumes that $\epsilon_{\odot\rm exp, min}\ll
\epsilon_\odot \ll \epsilon_{\odot \rm exp, max}$. The vertical
long-dashed line corresponds to the condition~$r_{\rm c} = r_{\rm
  A}$. To the right of this line, $\dot{M}$ and the solar-wind density
become so large that~$r_{\rm A} < r_{\rm c}$,
violating~(\ref{eq:rcrA}) and the assumptions underlying
the model. In conjunction with~(\ref{eq:dvsuneffval}),
Figure~\ref{fig:Mdot_asymptotic}  illustrates that the
expansion-dominated regime is relevant to the solar wind
and that the conduction-dominated
regime corresponds to values of~$\delta v_{\odot \rm eff}$ much
smaller than the solar value.

\subsection{Anti-correlation between
  the coronal super-radial expansion factor and  $U_{\infty}$}
\label{sec:wang}

\cite{wang90} showed that the outflow velocity in a solar-wind stream
at $r=1 \mbox{ a.u.}$ is anti-correlated with the
super-radial expansion factor in the coronal magnetic flux tube from
which the solar-wind stream originated. To compare their
results with the model
developed in \S~\ref{sec:analytic}, I follow \cite{cranmer07} by
defining the \cite{wang90} coronal super-radial expansion factor
$f_{\rm max, WS}$ to be $\eta(1.04 R_{\odot})/ \eta(2.5 R_{\odot})$. I
then compute $\eta(1.04 R_{\odot})$ and $\eta(2.5 R_{\odot})$ using a
magnetic-field model similar to that used by~\cite{cranmer07}. In
particular, I employ the global solar-magnetic-field model
of~\cite{banaszkiewicz98} with the parameter values listed below their
equation~(2) ($K=1$, $M=1.789$, $a_1^{\rm (B98)} = 1.538$, and
$Q=1.5$). I supplement this global magnetic field with the
low-solar-atmosphere magnetic-field model of \cite{hackenberg00} using
the same parameter values listed in their figure~1
($L = 30 \mbox{ Mm}$, $d = 0.34 \mbox{ Mm}$, $B_0 = 11.8 \mbox{ G}$,
and $B_{\rm max} = 1.5 \mbox{ kG}$). I then integrate out along a
magnetic-field line at the edge of the polar corona hole in the
\cite{banaszkiewicz98} model, compute~$\eta(r)$, and obtain
\begin{equation}
  f_{\rm max, WS} = \frac{\eta_{\rm b}}{5.75}.
  \label{eq:fmaxWS}
\end{equation}
I take~(\ref{eq:fmaxWS}) to hold even as~$\eta_{\rm b}$ is varied in
Figures~\ref{fig:Mdot_U_fmax},~\ref{fig:contour1}, and~\ref{fig:contour2}. This is
equivalent to assuming that if local magnetic structures on the Sun
cause $\eta_{\rm b}$ to increase or decrease by some factor relative
to the Hackenberg/Banaszkiewicz model just described, then
$\eta(1.04 R_{\odot})$ increases or decreases by the same factor, but
$\eta(2.5 R_{\odot})$ does not change.

If $\sigma$ is fixed while $\eta_{\rm b}$ is varied, the model of
\S~\ref{sec:analytic} does not agree with results
of~\cite{wang90} shown in the top-right panel of
Figure~\ref{fig:Mdot_U_fmax}.  On the other hand, the model becomes
consistent with those results
if $\sigma$ is taken to have a
power-law dependence on~$\eta_{\rm b}$ of the form
\begin{equation}
  \sigma = 5.7\times 10^{-2} \, \eta_{\rm b}^{0.34}.
  \label{eq:sigmascaling}
\end{equation}
Equation~(\ref{eq:sigmascaling}) is used
to determine~$\sigma$ in figures~\ref{fig:Mdot_U_fmax}
through~\ref{fig:contour2}.  The implication 
of~(\ref{eq:sigmascaling}) that $\sigma$
is an increasing function of~$\eta_{\rm b}$ is plausible because
increasing $\eta_{\rm b}$ increases $v_{\rm Ab}$, as can be seen
from~(\ref{eq:vAb}). This, in turn, increases the number
of Alfv\'en-speed scale heights between the coronal base and the
Alfv\'en critical point
($\int_{r_{\rm b}}^{r_{\rm A}} (1/v_{\rm A}) | \d v_{\rm A}/ \d r| \d
r$), which increases the fraction of~$P_{\rm AWb}$ that is dissipated
at~$r<r_{\rm A}$ \citep{chandran09c}.  Further work, however, is
needed to clarify how the structure of the coronal magnetic field
influences the rate of turbulent dissipation along different magnetic
flux tubes with different values of~$\eta_{\rm b}$ and to test the
extent to which (\ref{eq:zplusy}) and (\ref{eq:sigmascaling}) are 
consistent with more rigorous treatments of AW turbulence in the
sub-Alfv\'enic region of the solar wind.

\begin{figure}
  \includegraphics[width=7cm]{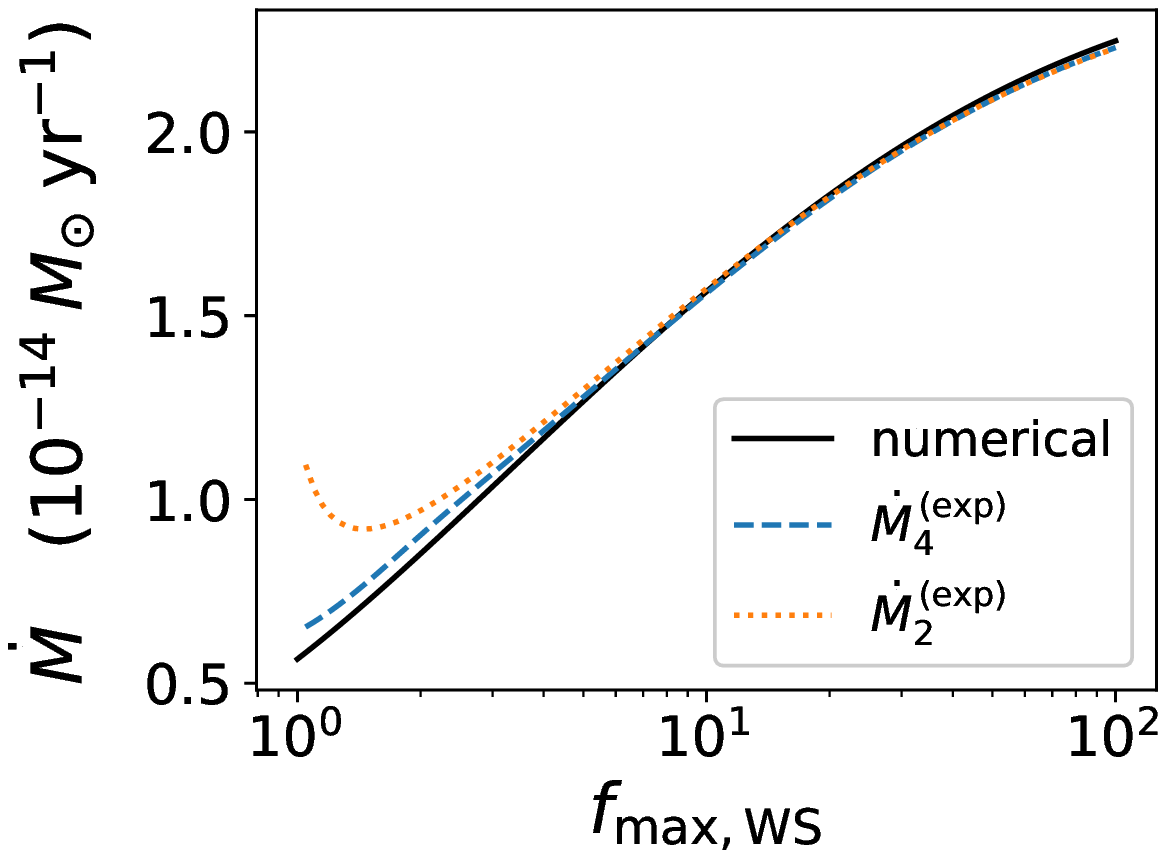}
  \includegraphics[width=7cm]{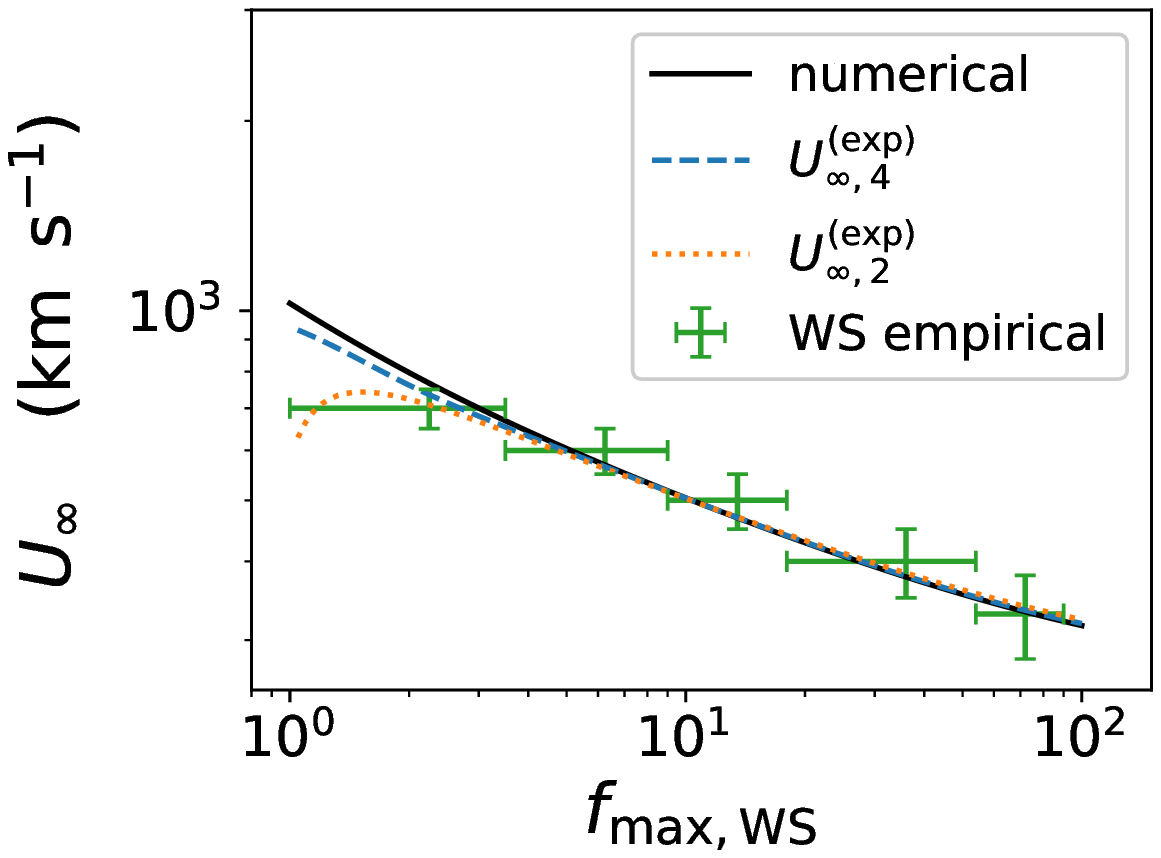}
\\
\includegraphics[width=7cm]{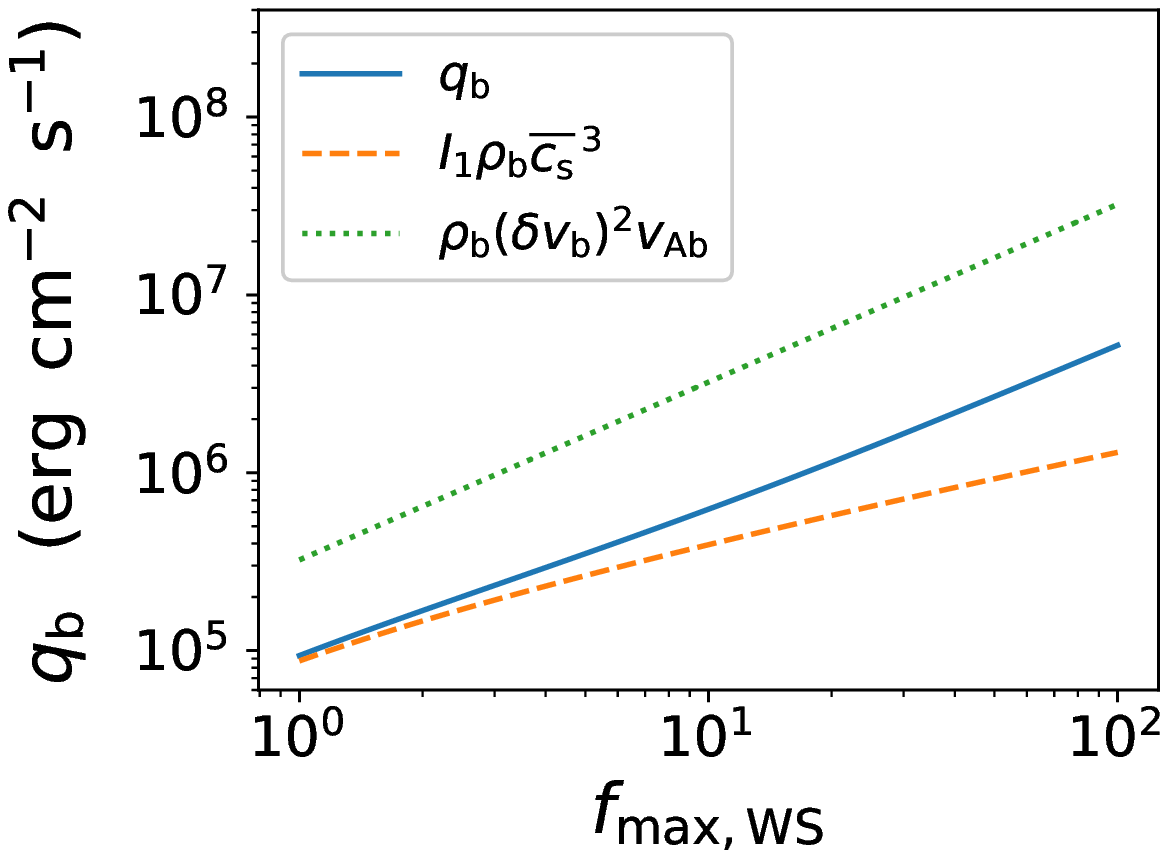}
\includegraphics[width=7cm]{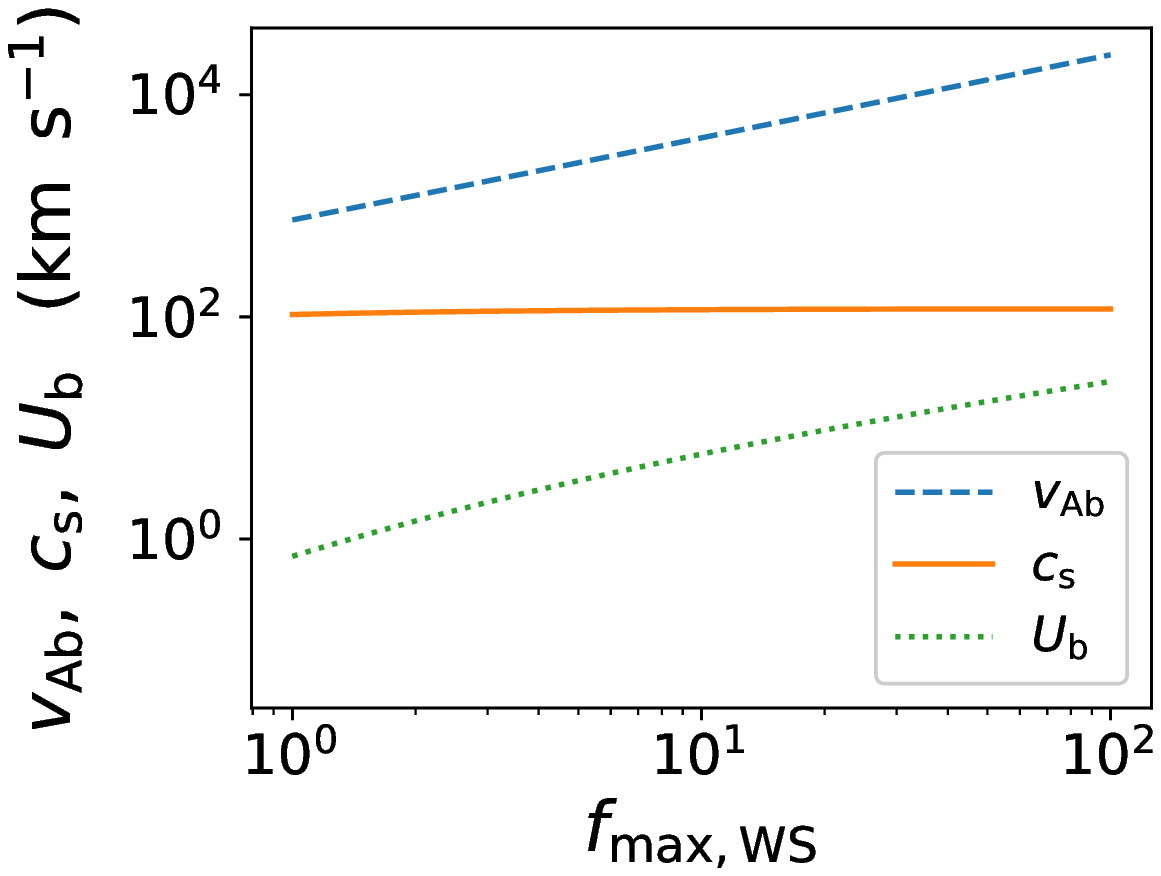}
\\
\includegraphics[width=7cm]{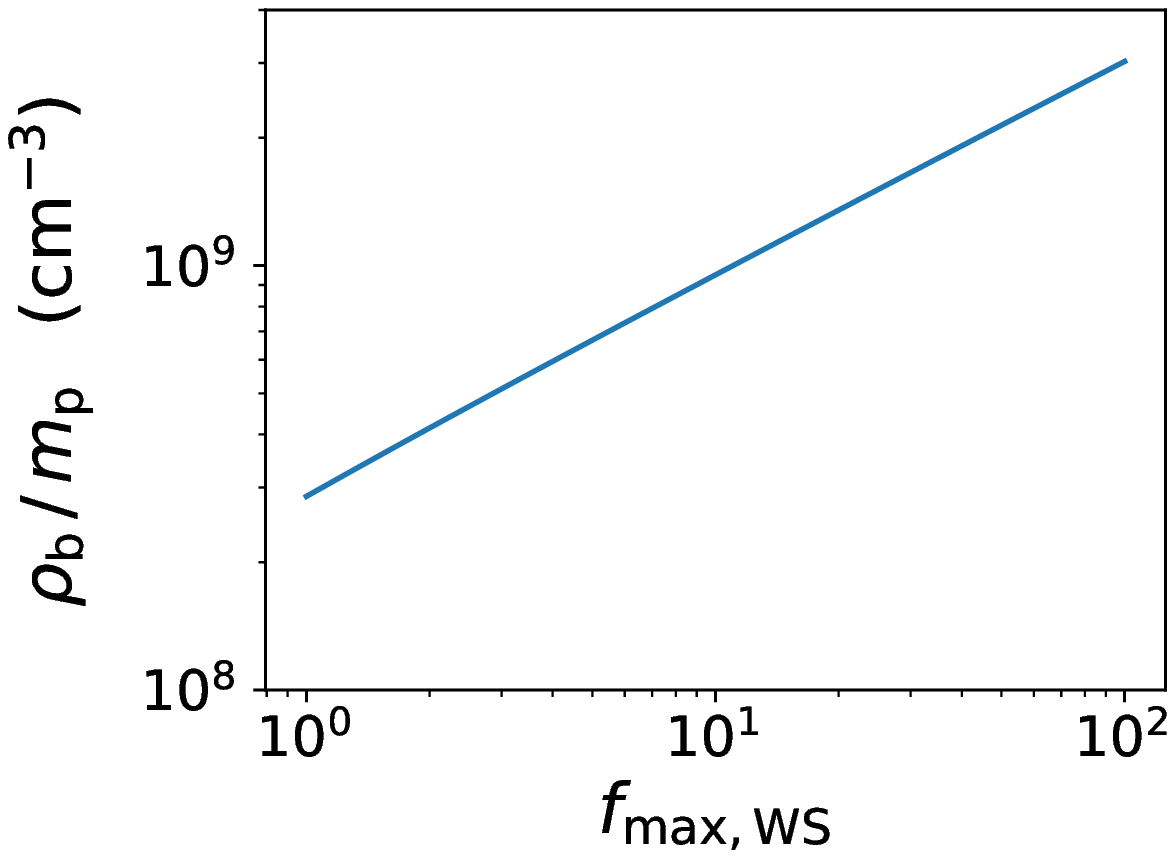}
\includegraphics[width=7cm]{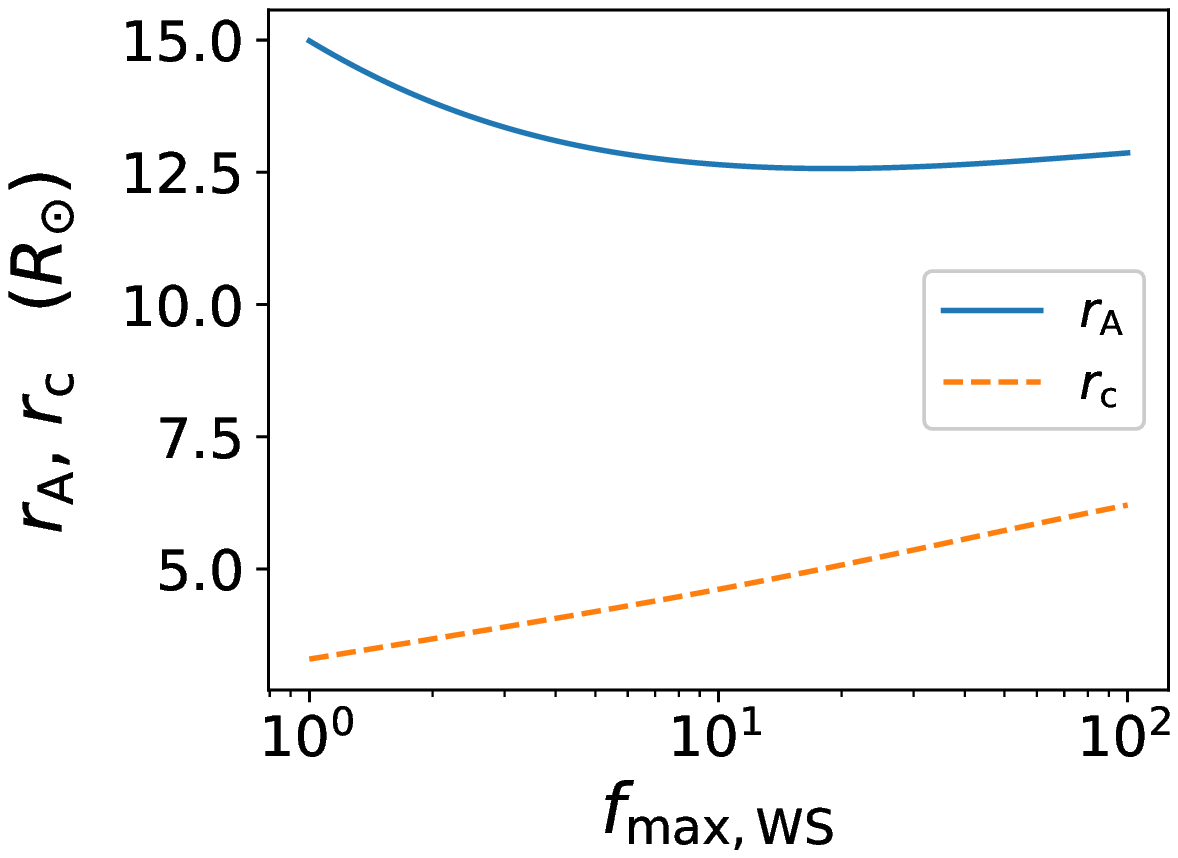}
\caption{The dependence of various flow properties on the Wang-Sheeley
  super-radial expansion factor~$f_{\rm max, WS}$~(\ref{eq:fmaxWS})
  for the parameter values
  in~(\ref{eq:etacgammaBc}), (\ref{eq:Bastvalue}), and
  (\ref{eq:dvsuneffval})--(\ref{eq:sigmascaling}).
All quantities are evaluated using numerical solutions to the model
equations, with the exception of~$\dot{M}_2^{\rm (exp)}$,
$\dot{M}_4^{\rm (exp)}$,
$U_{\infty, 2}^{\rm
    (exp)}$, and~$U_{\infty, 4}^{\rm
    (exp)}$, which are defined in~(\ref{eq:Mdotexpn})
  and~(\ref{eq:Uinftyexp}), and the data points labeled `WS
empirical', which are taken
from table~2 of~\cite{wang90}. The horizontal error bars on these data
points convey the half
widths of the $f_{\rm max, WS}$ data bins in that table. The
vertical error bars correspond to
one-half of the~$ 100 \mbox{ km} \mbox{ s}^{-1}$ increment between the
discretised~$U_{\infty}$ values that define four of the five data
bins.
The quantity~$I_1 \rho_{\rm b} \overline{
    c_{\rm s}}^3$ in the left panel of the second row is the
  low-Mach-number approximation to~$q_{\rm b}$ given in~(\ref{eq:qb_approx}).
  \label{fig:Mdot_U_fmax} }
\end{figure}

The top-left panel of Figure~\ref{fig:Mdot_U_fmax} shows that
as~$f_{\rm max, WS}$ ranges from~$\simeq 3$ to~$\simeq 30$, $\dot{M}$
varies from $10^{-14} M_{\odot} \mbox{ yr}^{-1}$ to
$2\times 10^{-14} M_{\odot} \mbox{ yr}^{-1}$, values that are similar
to the solar-mass-loss rates inferred from {\em Ulysses} and PSP measurements
\citep{mccomas00,schwadron08, kasper19}. The fourth-order analytic
approximation~$\dot{M}_4^{\rm (exp)}$ from~(\ref{eq:Mdotexpn})
reproduces the numerical solution to the model equations
reasonably well. The second-order analytic approximation
~$\dot{M}_2^{\rm (exp)}$ is also reasonably accurate at $f_{\rm max, WS} \gtrsim 3$, but deviates markedly from the
numerical solution at $f_{\rm max, WS} \lesssim 2$. A similar comment applies
to the top-right panel of figure~\ref{fig:Mdot_U_fmax}, which also shows
that the model agrees fairly well
with observational constraints on~$U_{\infty}(f_{\rm max, WS})$ when
(\ref{eq:sigmascaling}) holds. The left panel of the
second row of figure~\ref{fig:Mdot_U_fmax} shows that the
low-Mach-number approximation to~$q_{\rm b}$ given
in~(\ref{eq:qb_approx}) is
quite accurate at small~$f_{\rm max, WS}$, where
$U_{\rm b}/ \overline{ c_{\rm s}} \ll 1$ (see the right panel in the
second row of this figure). However, as~$f_{\rm max, WS}$ increases,
$U_{\rm b}$ increases, because the outflow is concentrated into a
smaller cross-sectional area at the coronal base. This increase
in~$U_{\rm b}$ leads to larger losses of internal energy within the
transition region from $p\,\d V$ work and advection, which, in turn, causes $q_{\rm b}$ to
increase above the low-Mach-number scaling so that conductive heating
within the transition region can balance the additional non-radiative
cooling. This same panel also shows that $q_{\rm b}$ is significantly
smaller than the AW energy flux at the coronal base
($\simeq \rho_{\rm b} (\delta v_{\rm b})^2 v_{\rm Ab} =
\rho_{\odot}^{1/2} (\delta v_{\odot \rm eff})^2 B_{\rm b} /\sqrt{4\upi}$), which
increases with $f_{\rm max, WS}$ because increasing~$f_{\rm max, WS}$
increases $\eta_{\rm b}$ and hence~$B_{\rm b}$.  The lower-left panel
of figure~\ref{fig:Mdot_U_fmax} shows that~$\rho_{\rm b}$ increases by
a factor of~$\simeq 10$ as~$f_{\rm max, WS}$ increases from~1 to~100,
consistent with the $\rho_{\rm b} \propto \eta_{\rm b}^{1/2} x^{1/4}$
scaling in~(\ref{eq:rhob2}) and~(\ref{eq:defxi}), given
that~$x = \overline{ c_{\rm s}}^2/v_{\rm esc}^2$ is approximately
constant over this range of~$f_{\rm max}$. The lower-right panel shows
that $r_{\rm c}$ is typically several solar radii, whereas~$r_{\rm A}$
ranges from~$12 R_{\odot}$ to~$15 R_{\odot}$ for this set of
parameters.

\subsection{Dependence of the flow properties on the 
  AW power, super-radial expansion factor, and strength of
  the interplanetary magnetic field}
\label{sec:contour_plots}

The contour plots in figure~\ref{fig:contour1} show how several
quantities vary as functions of~$f_{\rm max, WS}$
and~$\delta v_{\odot \rm eff}$ in numerical solutions to the model
equations based on the parameter values in~(\ref{eq:etacgammaBc}),
(\ref{eq:Bastvalue}), (\ref{eq:lb}),
and~(\ref{eq:sigmascaling}). The top panels of
figure~\ref{fig:contour1} show that~$\dot{M}$ is an increasing
function of both $f_{\rm max, WS}$ and $\delta v_{\odot \rm eff}$,
whereas $U_\infty$ is a decreasing function of both $f_{\rm max, WS}$
and $\delta v_{\odot \rm eff}$.  The second row shows that~$r_{\rm A}$
is a strongly decreasing function of~$\delta v_{\odot \rm eff}$ and
only weakly dependent on~$f_{\rm max, WS}$ for values
of~$\delta v_{\odot \rm eff}$ comparable to the Sun-like value
in~(\ref{eq:dvsuneffval}).  The right panel of this row shows that
$r_{\rm c}$ is an increasing function of~$f_{\rm max, WS}$ and, for
the most part, a decreasing function of~$\delta v_{\odot \rm eff}$.
The lower-left panel of figure~\ref{fig:contour1} shows that
$\overline{ c_{\rm s}}$ varies only weakly with $f_{\rm max, WS}$ and
$\delta v_{\odot \rm eff}$. As shown in the lower-right panel,
$\chi_{\rm H}$ varies by a factor of~$\simeq 2$ over the parameter
range shown. If $\delta v_{\odot \rm eff}$ is set equal to the value
in~(\ref{eq:dvsuneffval})
and $f_{\rm max}$ is restricted to the
interval~$(2, 8)$ so that $U_{\infty}$ in the upper-right panel takes
on fast-wind-like values of 600--800~$\mbox{ km} \mbox{ s}^{-1}$,
then $\chi_{\rm H} \simeq 0.5 - 0.7$, consistent with direct numerical
simulations of reflection-driven AW turbulence~\citep{perez21}.

\begin{figure}
  \includegraphics[width=6.5cm]{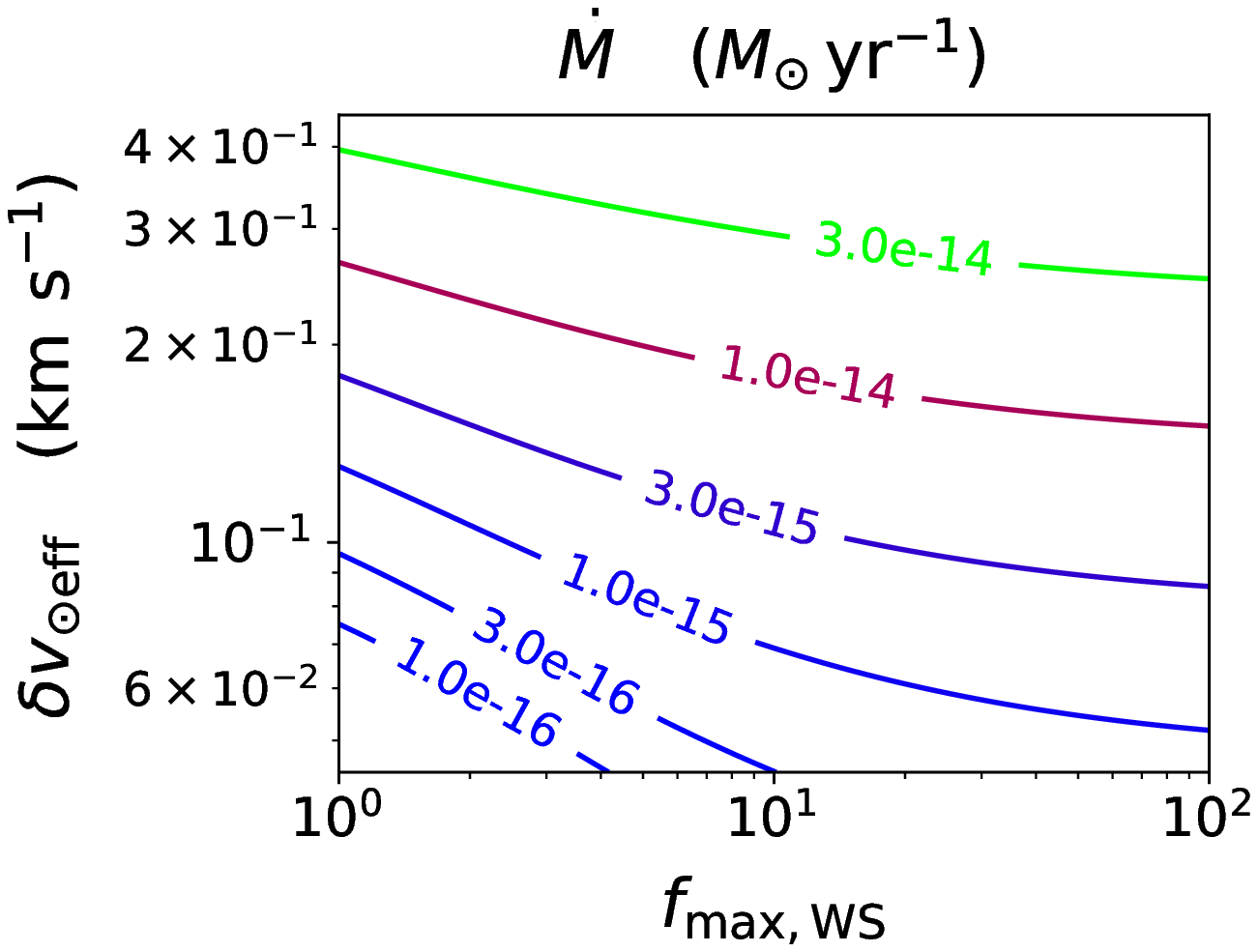}
  \includegraphics[width=6.5cm]{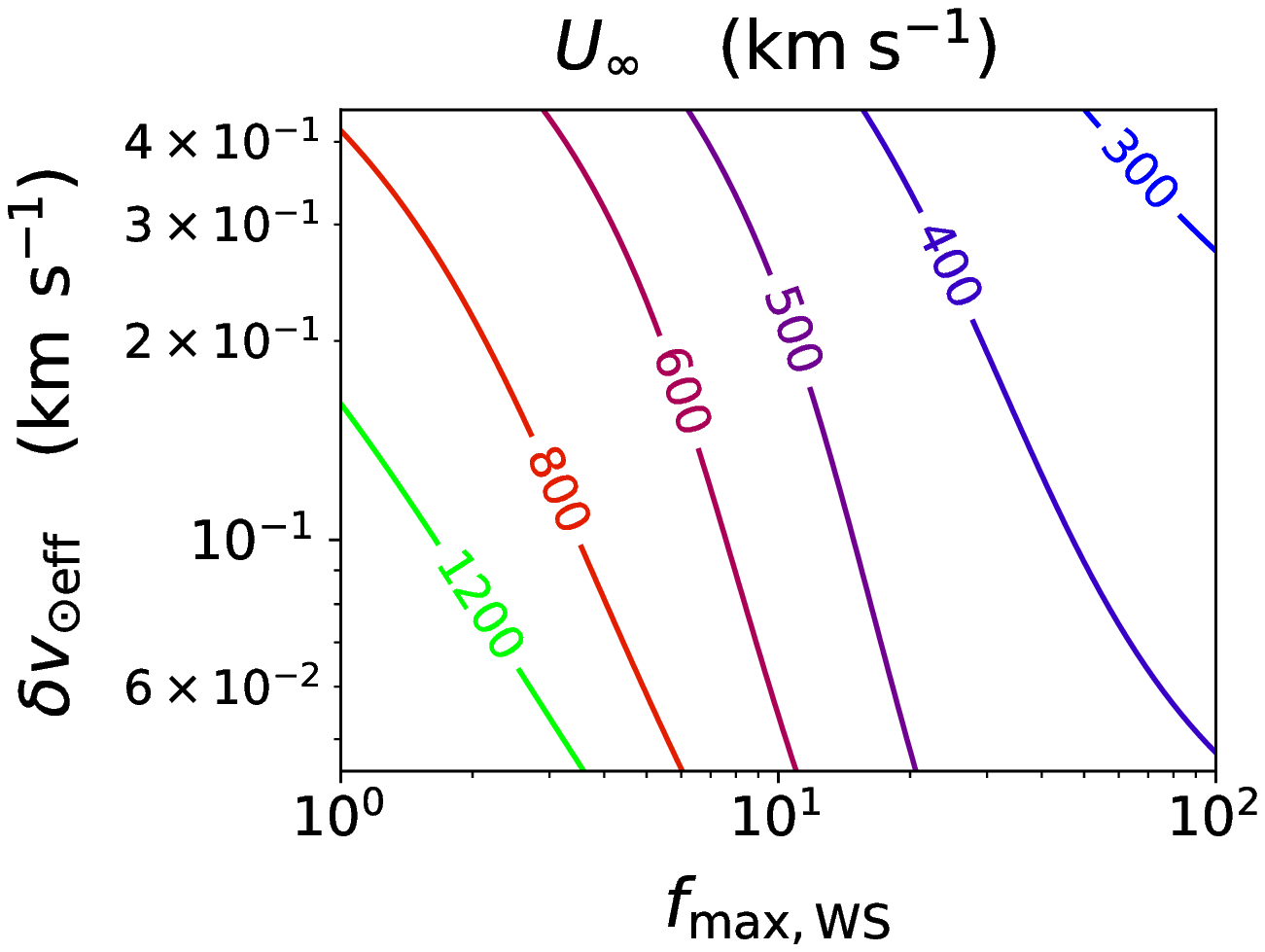}
  \vspace{0.2cm} 
  \\
  \includegraphics[width=6.5cm]{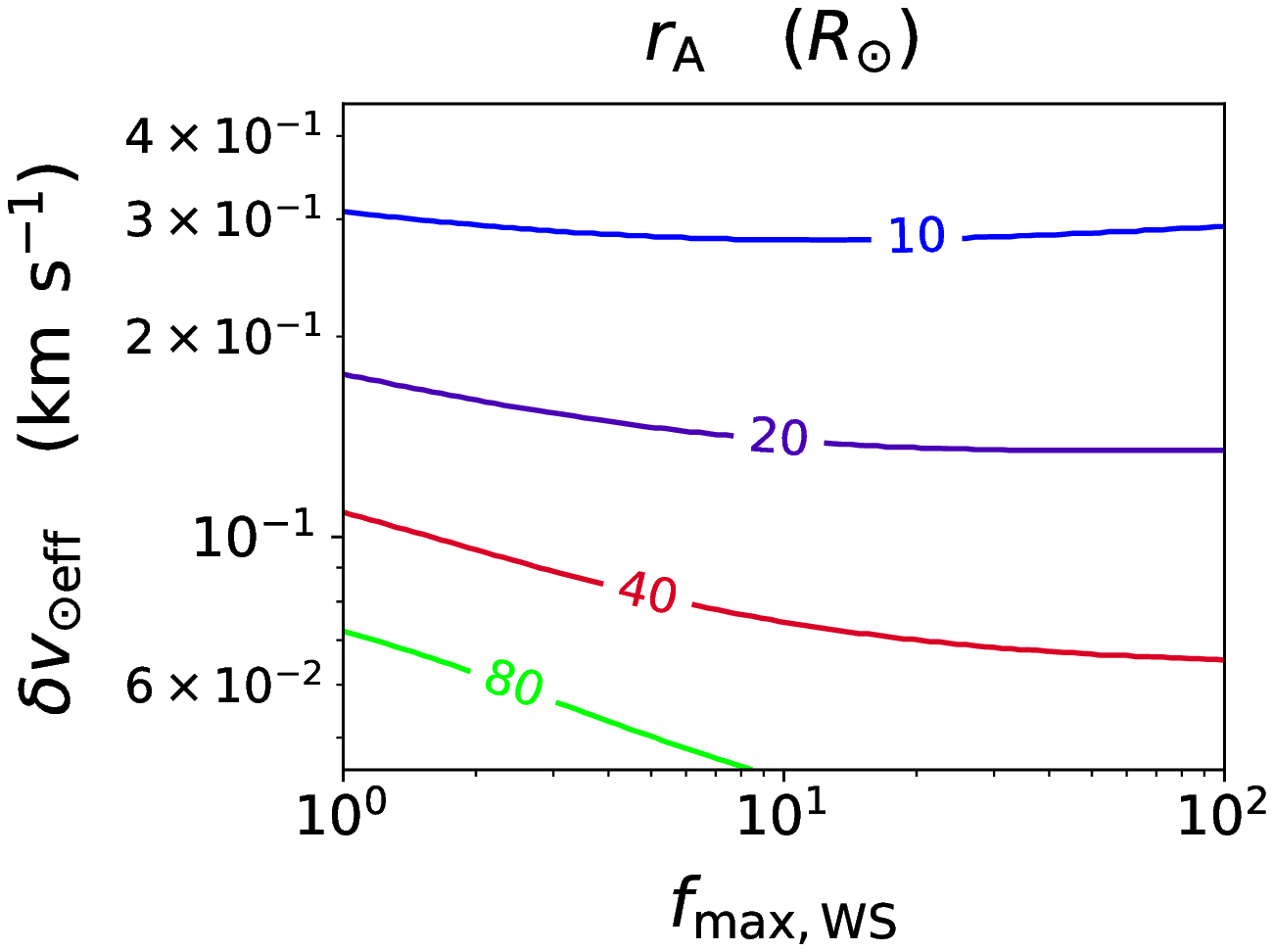}
  \includegraphics[width=6.5cm]{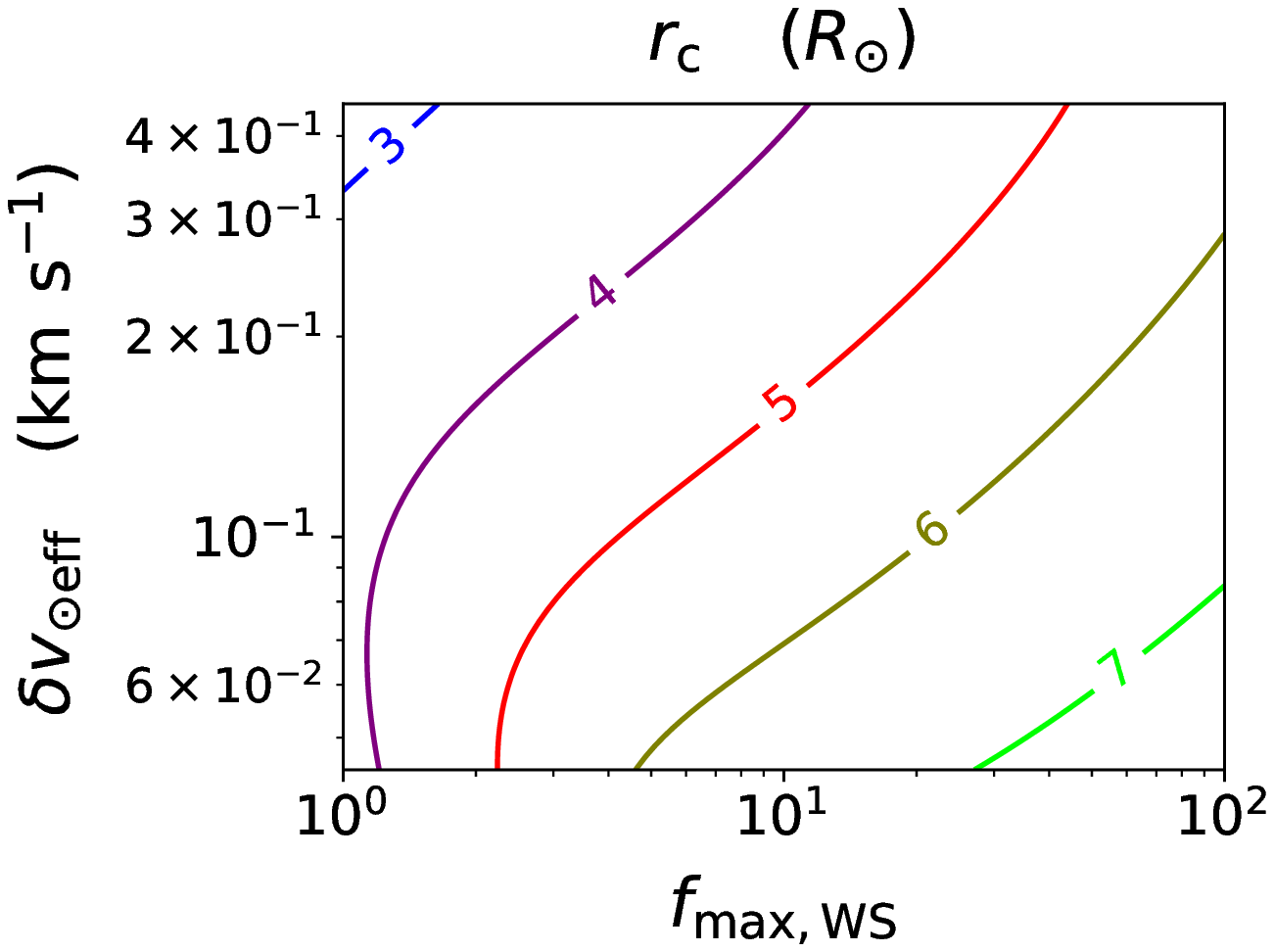}
  \vspace{0.2cm} 
  \\
  \includegraphics[width=6.5cm]{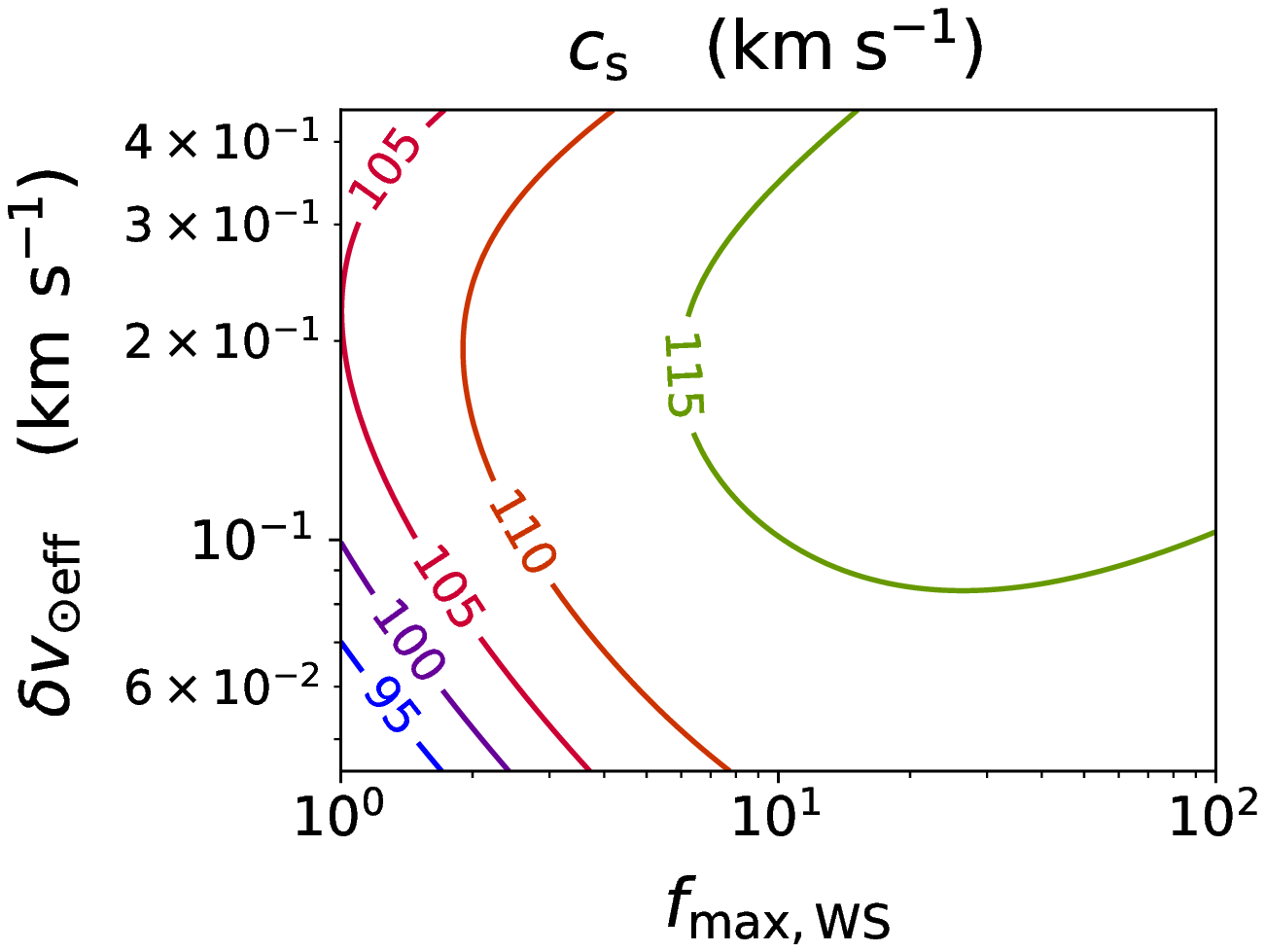}
  \includegraphics[width=6.5cm]{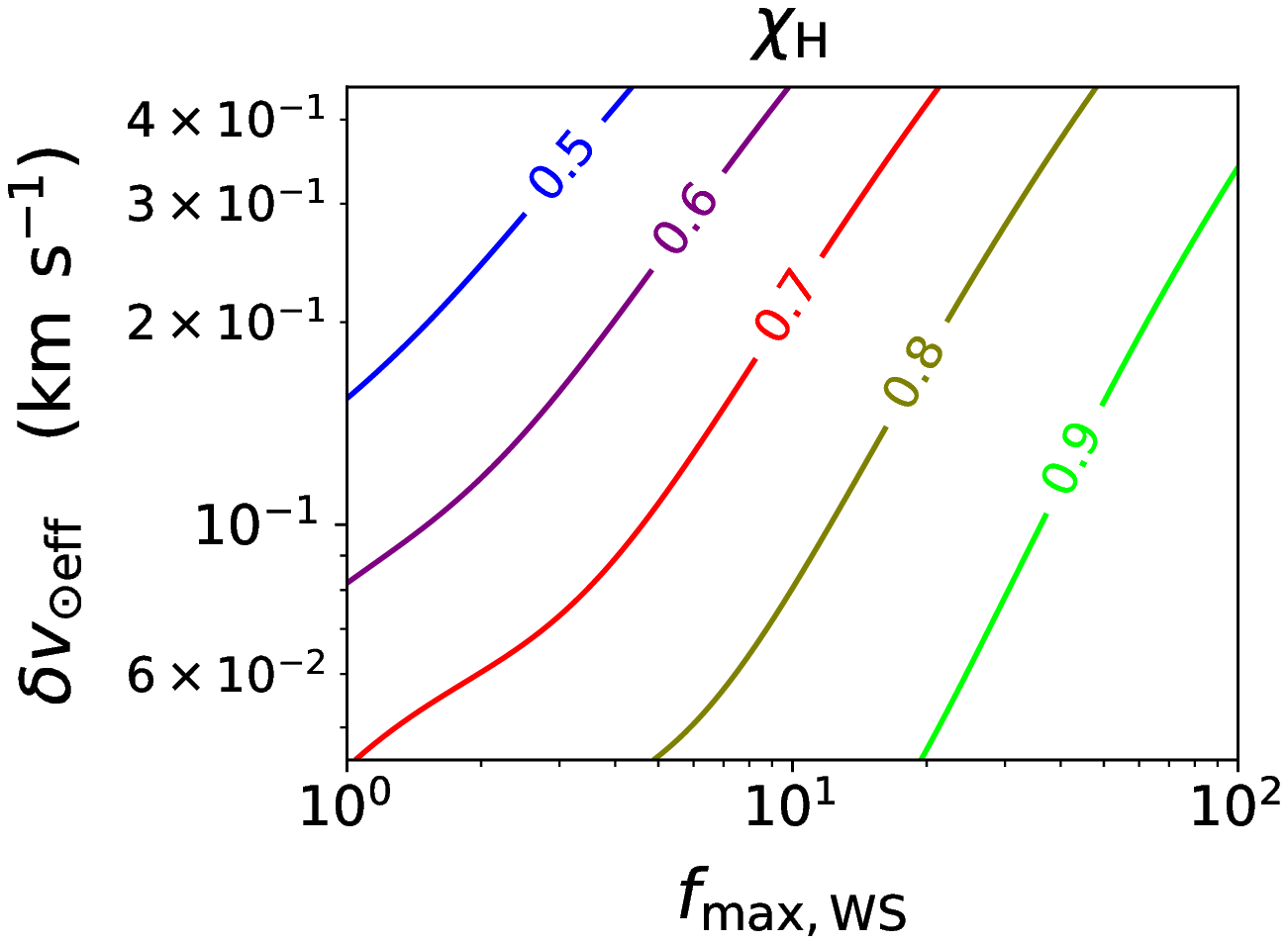}	
  \caption{The dependence of various flow properties on the Wang-Sheeley
    super-radial expansion factor~$f_{\rm max,WS}$~(\ref{eq:fmaxWS}) and the effective
    fluctuating velocity at the photosphere~$\delta v_{\odot\rm eff}$
    defined in~(\ref{eq:dvsuneff}). 
\label{fig:contour1} }
\end{figure}

Figure~\ref{fig:contour2} displays contour plots of the same
quantities as in figure~\ref{fig:contour1}, but this time as functions
of~$f_{\rm max, WS}$ and~$B_\ast$, or, equivalently,
$B_r(\mbox{1 a.u.})$, using the parameters in~(\ref{eq:etacgammaBc}),
(\ref{eq:dvsuneffval}), (\ref{eq:lb}),  and~(\ref{eq:sigmascaling}).  The
top panels show that $\dot{M}$ increases approximately linearly
with~$B_r(\mbox{1 a.u.})$ at fixed~$f_{\rm max, WS}$, and that
$U_\infty$ varies only weakly with~$B_r(\mbox{1 a.u.})$, consistent with
measurements from the {\em Ulysses} spacecraft~\citep{schwadron08, riley10}.
The left panel of the middle row shows that~$r_{\rm A}$ depends more
strongly on~$B_r(\mbox{1 a.u.})$ than on~$f_{\rm max, WS}$, whereas the
reverse is true for~$r_{\rm c}$. As in figure~\ref{fig:contour1},
$\overline{ c_{\rm s}}$ varies very weakly across the entire parameter
range observed. The lower-right panel shows that,
when (\ref{eq:sigmascaling}) holds, $\chi_{\rm H}$
depends more strongly on~$f_{\rm max, WS}$ than on~$B_r(\mbox{1 a.u.})$.

\begin{figure}
  \includegraphics[width=6.5cm]{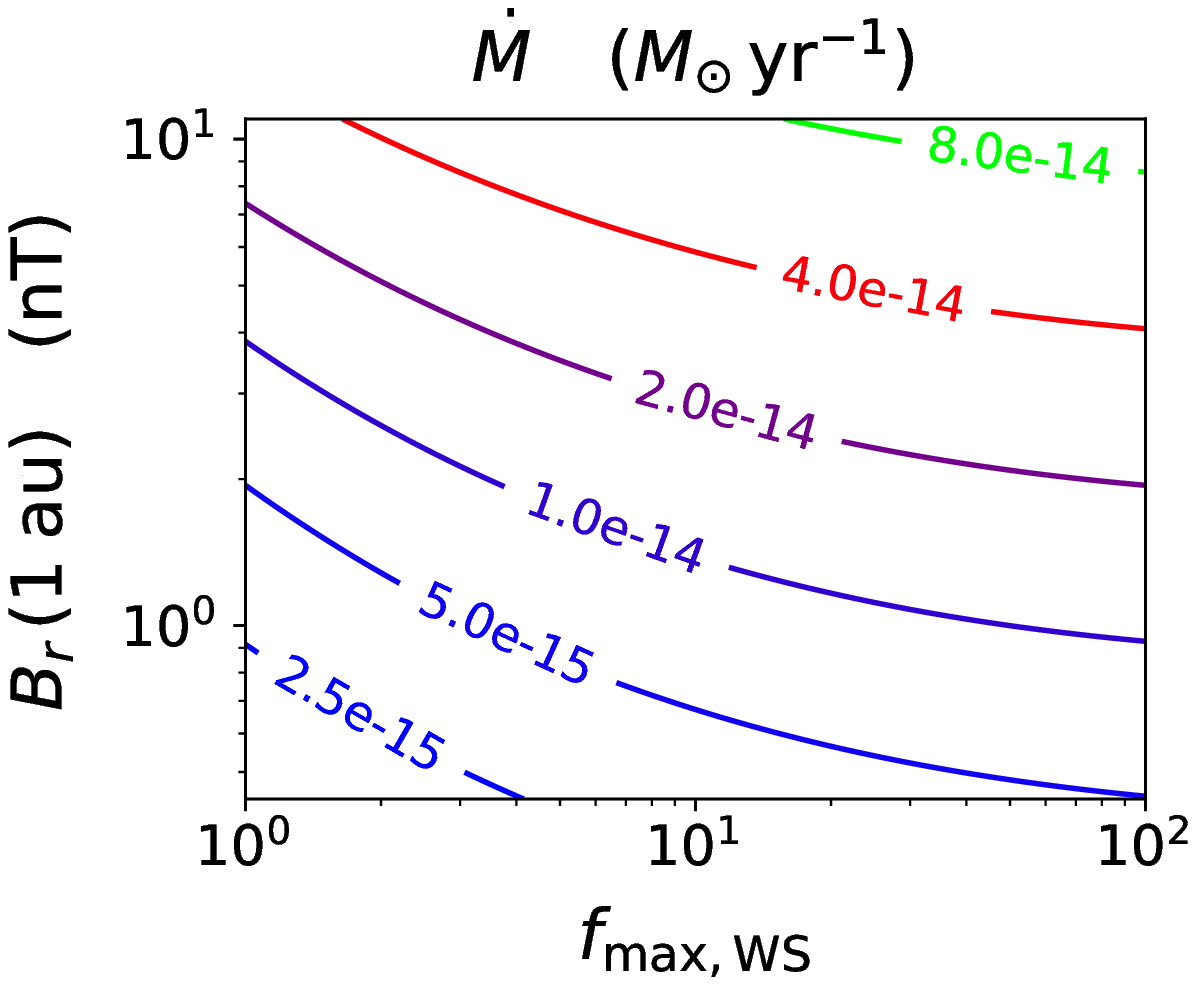}
  \includegraphics[width=6.5cm]{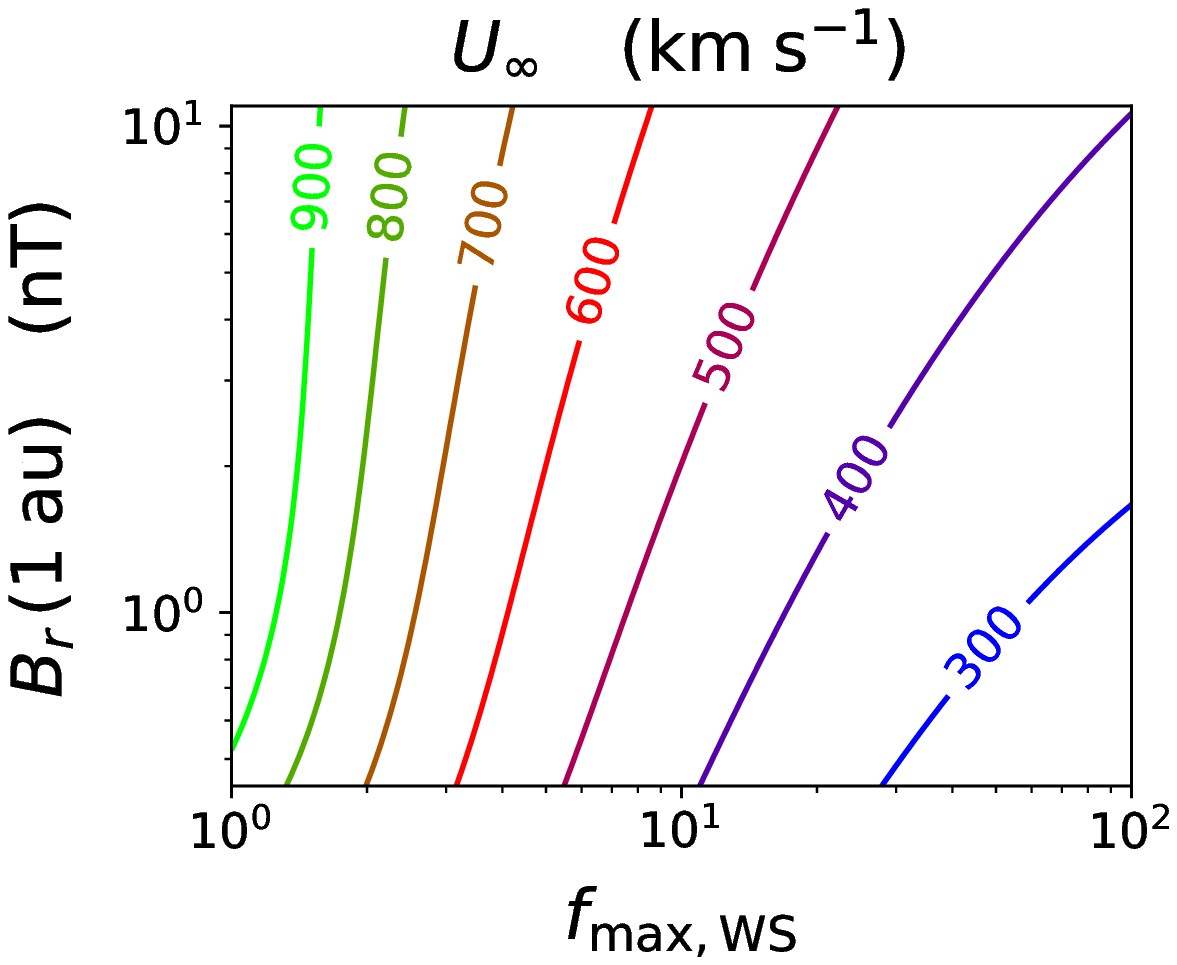}
  \vspace{0.2cm} 
  \\
  \includegraphics[width=6.5cm]{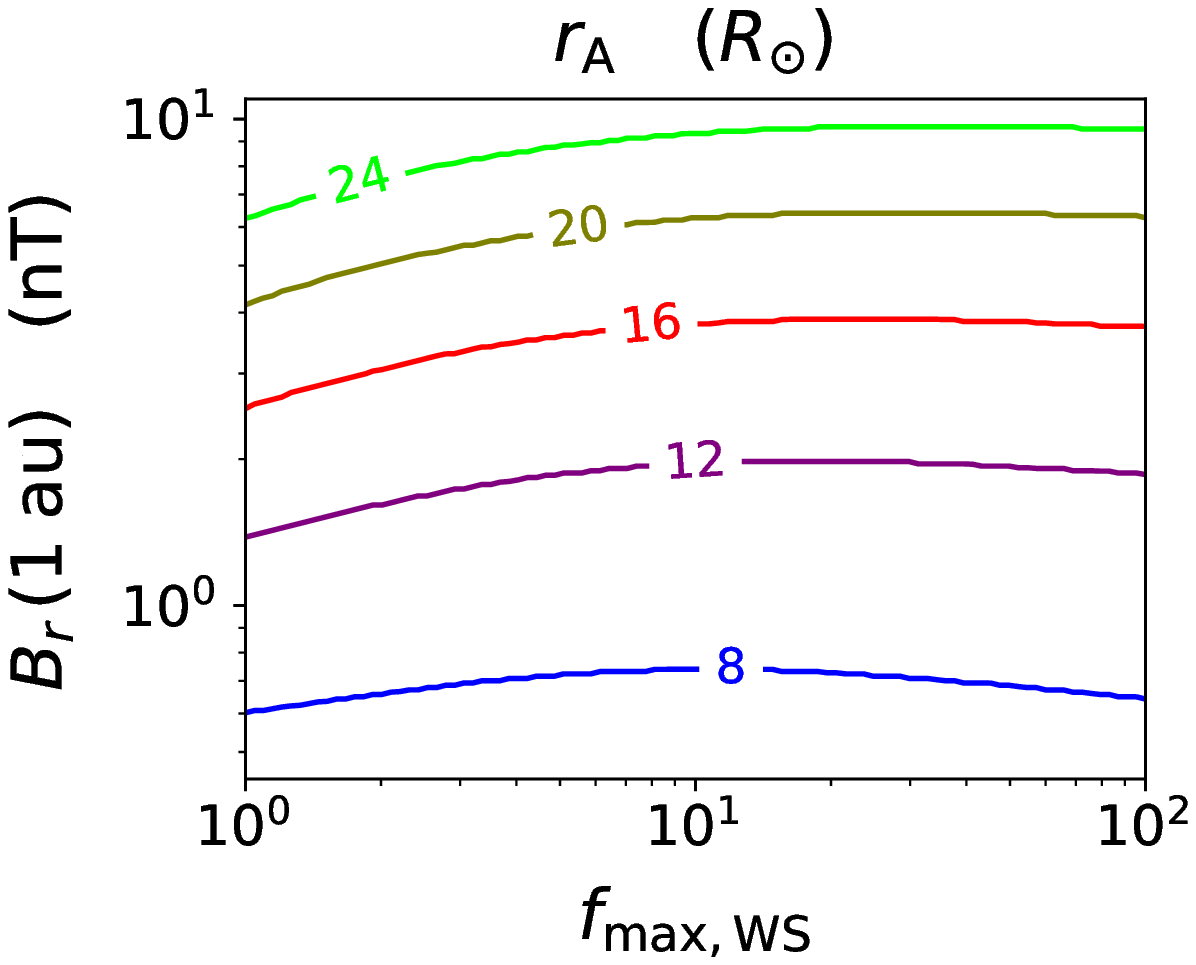}
  \includegraphics[width=6.5cm]{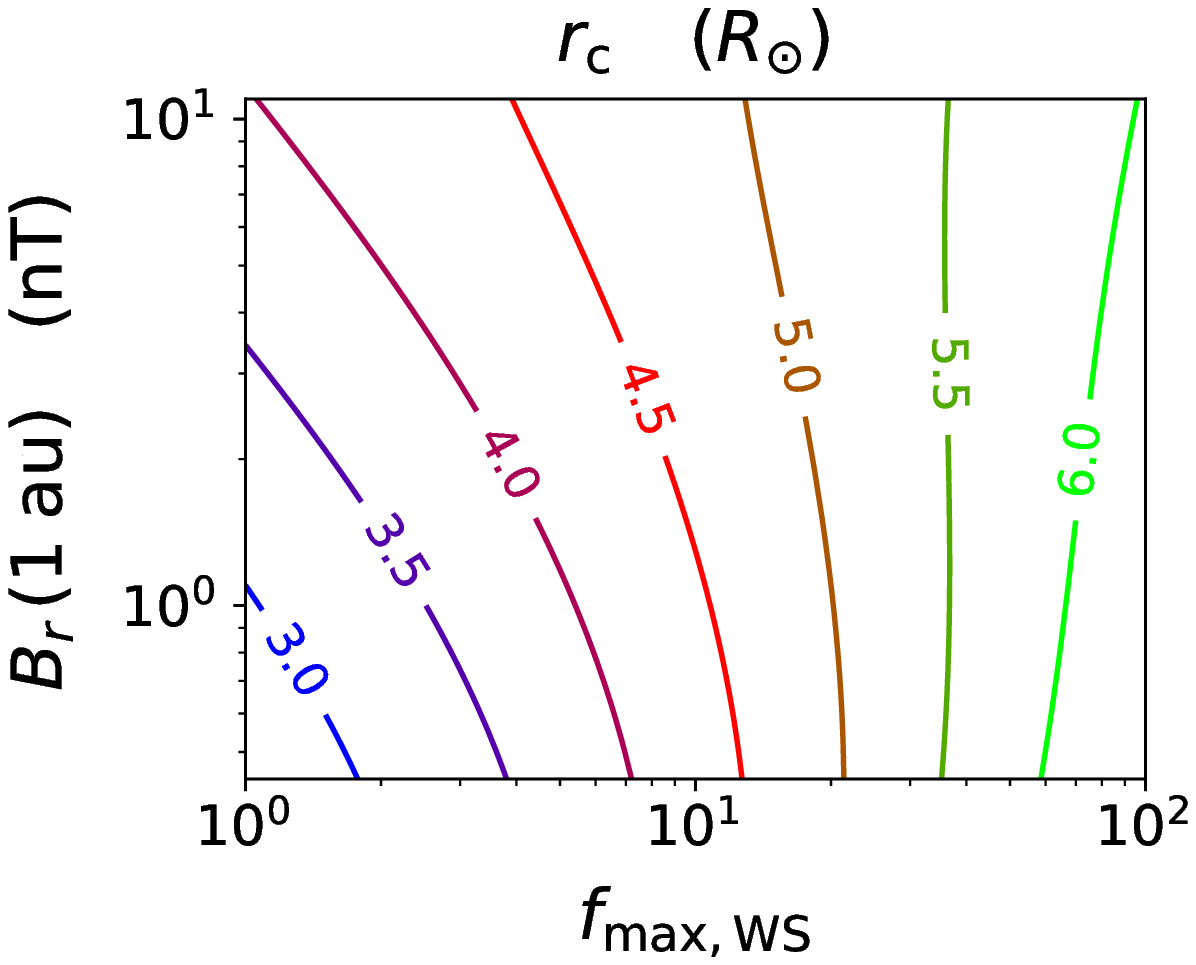}
  \vspace{0.2cm} 
  \\
  \includegraphics[width=6.5cm]{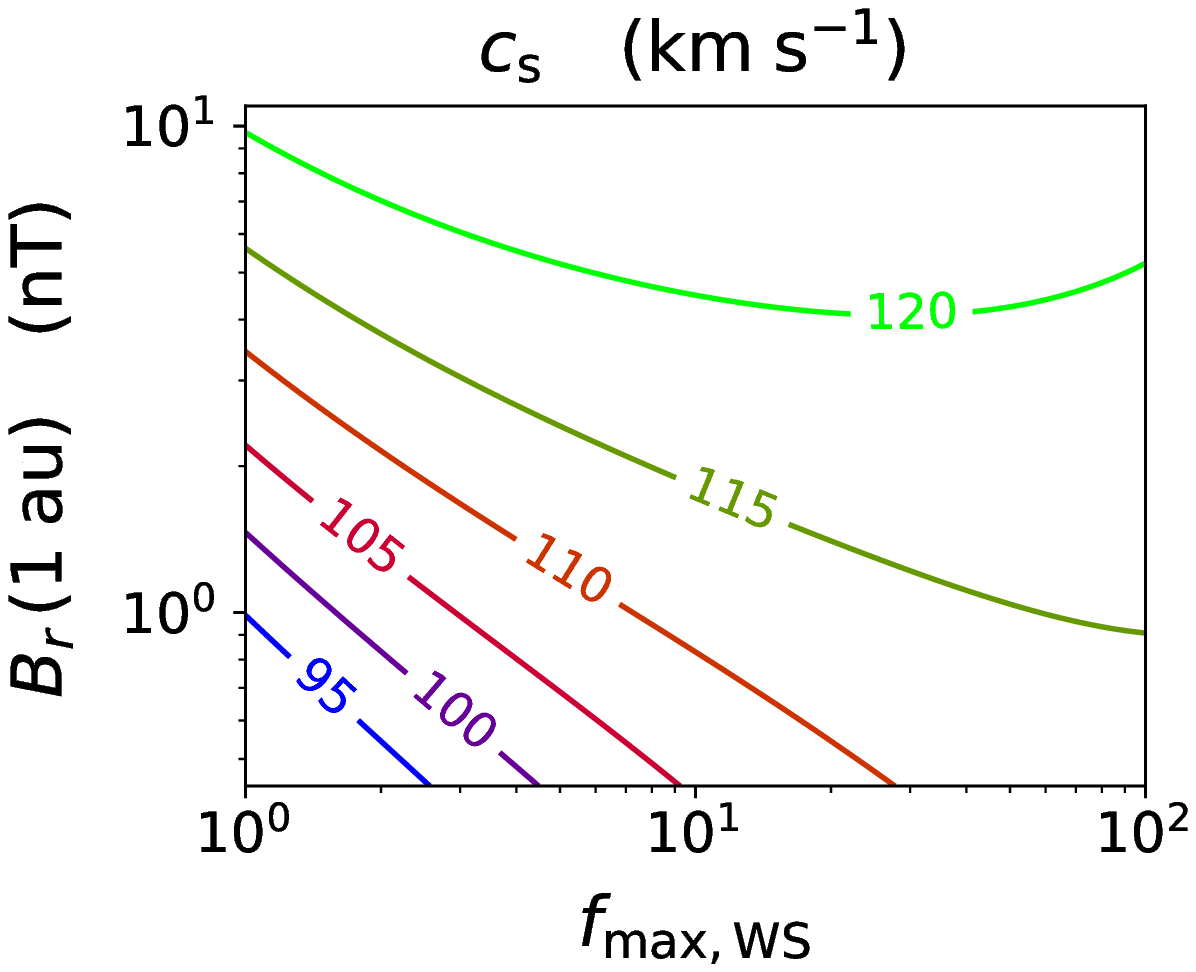}
  \includegraphics[width=6.5cm]{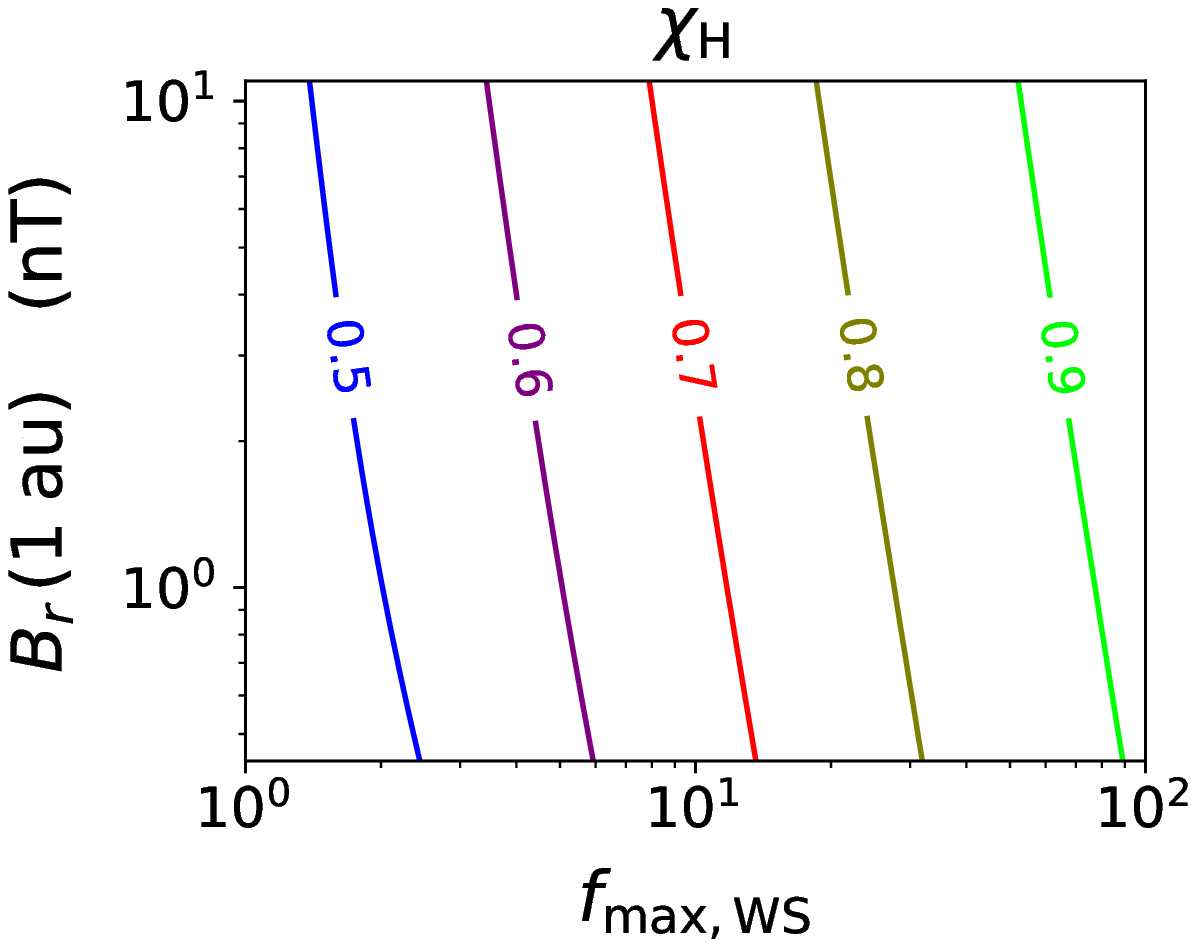}	
  \caption{The dependence of various flow properties on the Wang-Sheeley
    super-radial expansion factor~$f_{\rm max,WS}$~(\ref{eq:fmaxWS}) and the strength
    of the radial magnetic field at $r= 1 \mbox{ a.u.}$
\label{fig:contour2} }
\end{figure}

\section{Discussion and conclusion}
\label{sec:conclusion}

The main goal of this paper is to obtain an approximate analytic solution
to the coupled problems of coronal heating and solar-wind acceleration
under the assumption that the solar wind is powered primarily by an AW
energy flux.  Section~\ref{sec:heuristic} presents a first step toward
this goal, namely, a simplified calculation of~$\dot{M}$,
$U_{\infty}$, and the coronal temperature in a spherically symmetric,
steady-state solar wind in the absence of solar rotation. This
calculation is based upon: (1) the assumption that $p\,\d V$ work rather
than thermal conduction is the primary sink of internal energy in the
sub-Alfv\'enic region as a whole; (2) the result of
\S~\ref{sec:isothermal} that a solar or stellar wind heated
primarily by AW turbulence becomes approximately isothermal between
the coronal base~$r_{\rm b}$ and Alfv\'en critical point~$r_{\rm A}$;
and (3) the finding in recent direct numerical simulations that most of
the AW power at the coronal base~$P_{\rm AW}(r_{\rm b})$ is dissipated
between~$r=r_{\rm b}$ and~$r=r_{\rm A}$~\citep{perez21}. The calculation of \S~\ref{sec:heuristic}
shows that
\begin{equation}
  \dot{M} \simeq \frac{P_{\rm AW}(r_{\rm b})}{v_{\rm esc}^2} \qquad
  U_\infty \simeq v_{\rm esc}
  \qquad
  T \simeq \frac{m_{\rm p} v_{\rm esc}^2}{8 k_{\rm B} \ln(v_{\rm
      esc}/\delta v_{\rm b})}
  .
  \label{eq:summary1}
\end{equation}

Deviations from~(\ref{eq:summary1}) can be caused by a number of
factors, including conductive losses into the transition region, wave
momentum deposition inside the wave-modified sonic-critical point, and
the fact that part of the AW power reaches the super-Alfv\'enic
region, where it enhances~$U_{\infty}$ without contributing to the
heating that helps drive the outflow of mass past the wave-modified
sonic critical point. These factors are accounted
for in the more detailed solar-wind model developed in
\S~\ref{sec:analytic}. It is worth noting that in this model:

\vspace{0.2cm} 
\begin{myitemize} 
\item the plasma density at the coronal base~$\rho(r_{\rm b}) = \rho_{\rm
      b}$ is determined by equating the turbulent heating
    rate~$Q(r_{\rm b})$ and radiative
    cooling rate~$R(r_{\rm b})$; and
    \vspace{0.2cm}
    
\item an analytic solution for the heat flux from
      the corona into the transition region~$q_{\rm b}$ is obtained by balancing,
      within the transition region,
      conductive heating against internal-energy losses from $p\,\d V$
      work, advection, and radiative cooling.
\end{myitemize}
\vspace{0.2cm} 

\noindent The expression (\ref{eq:rhob2}) for~$\rho_{\rm b}$ that
results from setting~$Q(r_{\rm b}) = R(r_{\rm b})$ contains a factor
of~$x^{1/4}$, where $x = \overline{ c_{\rm s}}^2/v_{\rm esc}^2$ is the
dimensionless temperature. Thus,  $x$ must be found before
the exact value of~$\rho_{\rm b}$ can be determined. However,
because~$\dot{M}$ is so sensitive to the coronal temperature,
$x^{1/4}$ can to a good approximation be treated as a constant (see,
e.g., figures~\ref{fig:contour1} and~\ref{fig:contour2}),
and (\ref{eq:rhob2}) with~$x^{1/4} = 0.44$ can be used to obtain
the approximate value of~$\rho_{\rm b}$ without solving the full
model equations.

The equations of the solar-wind model developed in
\S~\ref{sec:analytic} are solved analytically in two different parameter
regimes. One of these is the conduction-dominated limit, in which heat
conduction into the transition region is the dominant mechanism for draining internal
energy from the sub-Alfv\'enic region, wave pressure makes the dominant
contribution to the critical-point velocity~$U_{\rm c}$
in~(\ref{eq:defUc}), and the wave-energy term dominates over the
plasma-internal-energy term in the Bernoulli
equation~(\ref{eq:Bernoulli1}).  This limit corresponds to
photospheric velocities much smaller than those of the Sun. The second
parameter regime is the expansion-dominated limit, in which $p\,\d V$ work
is the dominant sink of internal energy in the sub-Alfv\'enic region,
the sound speed makes the dominant contribution to the critical-point
velocity~$U_{\rm c}$ in~(\ref{eq:defUc}), and the
plasma-internal-energy term in the Bernoulli
integral~(\ref{eq:Bernoulli1}) dominates over the wave-energy
term. As illustrated in Figure~\ref{fig:Mdot_asymptotic}, the
expansion-dominated regime is relevant to the solar wind.
The leading-order solution in the expansion-dominated regime
reproduces~(\ref{eq:summary1}), with a small difference in the coronal
temperature arising from the fact that $\delta v_{\rm b}$ has a weak
dependence on the coronal temperature in the model
of~\S~\ref{sec:analytic}.  Numerical solutions to the model
equations approach the approximate analytic solutions in the
appropriate parameter regimes, match a range of solar-wind
observations, and illustrate how the properties of the solar wind depend
upon the r.m.s. photospheric velocity, super-radial expansion factor,
and interplanetary magnetic-field strength.

\subsection{Top-down causality for determining~$\dot{M}$, $q_{\rm b}$,
  and the pressure, temperature range, and altitude of the transition
  region within the solar atmosphere}
\label{sec:topdown} 

In this paper, as in Parker's original model
\citep{parker58,parker65}, the average rate at which mass flows out
through the lower solar atmosphere is determined in large part by the
outflow condition at the wave-modified sonic critical
point~$r_{\rm c}$, several~$R_\odot$ out from the Sun, at which the
plasma transitions from being gravitationally bound to gravitationally
unbound.  Since the gravitational force weakens with increasing~$r$,
there is no physical mechanism that can prevent plasma at~$r_{\rm c}$
from flowing outward at approximately the wave-enhanced effective
sound
speed,~$c_{\rm s, eff} \equiv \displaystyle [(p + p_{\rm
  wave})/\rho]^{1/2} = [ c_{\rm s}^2 + 0.5 (\delta v)^2]^{1/2}$,
evaluated at $r=r_{\rm c}$, which is comparable to the square root of
the right-hand side of~(\ref{eq:defUc}).  This is why
$\dot{M} \sim A(r_{\rm c}) \rho_{\rm c} c_{\rm s, eff}(r_{\rm c})$.

On the other hand, localised motions near the transition region at
speeds~$\ll v_{\rm esc}$ are gravitationally bound, and the mass flux
that they carry is not determinative of~$\dot{M}$. For example, if at
some time, such motions led to an overall mass outflow rate at
$r_{\rm b}$ exceeding the rate
$\sim A(r_{\rm c}) \rho_{\rm c} c_{\rm s, eff}(r_{\rm c})$ at which mass flows
past the critical point~$r_{\rm c}$, then plasma would build up in the
corona. This would, in turn, weaken the pressure gradient relative to
the gravitational force per unit volume in the vicinity of the transition
region, thereby reducing the amount of plasma flowing up from the
chromosphere.

Nevertheless, the transition region and chromosphere do
affect~$\dot{M}$ in two ways. First, $q_{\rm b}$ reduces the net
heating power within the quasi-isothermal sub-Alfv\'enic
region,~$P_{\rm net}$. As discussed following~(\ref{eq:Mdot_heur3}),
the heating cost per unit mass for plasma to transit the
quasi-isothermal sub-Alfv\'enic region is
$\overline{ c_{\rm s}}^2 \ln(\rho_{\rm b}/\rho(r_{\rm A})) \simeq
v_{\rm esc}^2$, and $\dot{M} \simeq P_{\rm net}/v_{\rm esc}^2$. By
reducing~$P_{\rm net}$, the conduction of heat from the corona into
the transition region reduces~$\dot{M}$. The second way that the lower
solar atmosphere affects~$\dot{M}$ is via the
chromospheric/transition-region transmission
coefficient,~$f_{\rm chr} = P_{\rm AW}(r_{\rm b})/P_{\rm
  AW}(R_{\odot})$, whose value again influences the value
of~$P_{\rm net}$. To summarise this paragraph and the preceding paragraph, the
chromosphere and transition region influence~$\dot{M}$
thermodynamically, but not dynamically.

The regulation of the mass flux at~$r_{\rm b}$ by the critical-point
condition at~$r_{\rm c}$ is an example of `top-down' causality, in
which physical processes at larger~$r$ control the plasma properties
at smaller~$r$. In the model of this paper, top-down causality also
characterises the determination of~$q_{\rm b}$, the pressure within
the transition region, the altitude of the transition region in the
solar atmosphere, and the temperature jump across the transition
region. As mentioned previously, $\rho_{\rm b}$ is determined by the
condition that~$Q(r_{\rm b}) = R(r_{\rm b})$, without reference to
conditions in the chromosphere or the value of~$q_{\rm b}$. The sound
speed at $r=r_{\rm b}$, which is~$\simeq \overline{ c_{\rm s}}$, is
approximately determined by balancing turbulent heating against
internal-energy losses from~$p\,\d V$ work within the corona and
sub-Alfv\'enic solar wind. The outflow
velocity at~$r=r_{\rm b}$, $U_{\rm b}$, follows from the value
of~$\dot{M}$, which is controlled by the critical-point condition, as
described above. The values of~$\rho_{\rm b}$,
$\overline{ c_{\rm s}}$, and~$U_{\rm b}$ jointly determine~$q_{\rm b}$
via~(\ref{eq:lambertqb}), which embodies the requirement that
conductive heating offset internal-energy losses (from radiation,
$p\,\d V$ work, and advection) within the transition region. The values
of~$\rho_{\rm b}$ and~$\overline{ c_{\rm s}}$ are sufficient to
determine the approximate transition-region
pressure,~$p_{\rm tr} \simeq \rho_{\rm b} \overline{ c_{\rm s}}^2$.
The pressure within the upper chromosphere, $p_{\rm chr}(r)$, is an
approximately exponentially decreasing function of altitude. The
altitude of the transition region is determined by setting
$p_{\rm chr}(r)= p_{\rm tr}$. The shape of the radiative loss function
plotted in figure~\ref{fig:Lambda_T} constrains the temperature at the
bottom of the transition region to be approximately~$10^4$~K, so that radiative
cooling within the comparatively dense upper chromosphere can be
balanced by local heating mechanisms in the absence of strong
conductive heating. Given this constraint, the factor by which the
temperature changes across the transition region is determined by the
value of~$T(r_{\rm b})$, which is controlled by the balance between
heating and $p\,\d V$ work in the corona, as discussed previously.

Although the condition $R(r_{\rm b}) = Q(r_{\rm b})$ determines the
density~$\rho_{\rm b}$ at the upper boundary of the transition region
within the corona (subject to the caveats at the end of the paragraph
following~(\ref{eq:summary1})), setting $R(r)=Q(r)$ within the
chromosphere does not determine the density at the lower edge of the
transition region, because small changes in~$T$ within the upper
chromosphere lead to dramatic changes in the radiative loss
function~$\Lambda(T)$, as shown in Figure~\ref{fig:Lambda_T}. The
density at the bottom of the transition region is instead largely
determined by the values of~$\rho_{\rm b}$ and
$\overline{ c_{\rm s}}$, the near constancy of the pressure across the
transition region, and the above-mentioned constraint (arising from
the shape of the radiative loss function) that~$T\sim 10^4$~K in the
upper chromosphere.

\subsection{Limitations and future work}
\label{sec:limitations}

The model of \S~\ref{sec:analytic} has a number of
shortcomings. First, the sub-Alfv\'enic region is not truly
isothermal, and, hence, the quasi-isothermal approximation
in~(\ref{eq:QIT1}) leads to some error.  Second, for the solutions
shown in figure~\ref{fig:Mdot_U_fmax}, the temperature
$m_{\rm p} \overline{ c_{\rm s}}^2/(2k_{\rm B})$ of the sub-Alfv\'enic
region increases from $\simeq 7\times 10^5 \mbox{ K}$ when
$U_\infty \simeq 800 \mbox{ km} \mbox{ s}^{-1}$ to
$\simeq 8.4 \times 10^5 \mbox{ K}$ when
$U_\infty \simeq 400 \mbox{ km} \mbox{ s}^{-1}$.  In contrast, in
measurements from the {\em Ulysses} spacecraft, the coronal freeze-in
temperature increases from $\simeq 9 \times 10^5 \mbox{ K}$
when~$U_{\infty} \simeq 800 \mbox{ km} \mbox{ s}^{-1}$ to
$\simeq 1.35 \times 10^6 \mbox{ K}$ when
$U_\infty \simeq 400 \mbox{ km} \mbox{ s}^{-1}$ \citep{mccomas02,
  schwadron03}. Although the numerical solutions in
figure~\ref{fig:Mdot_U_fmax} reproduce the observed anti-correlation
between coronal temperature and asymptotic wind speed, the
isothermal-sub-Alfv\'enic-region approximation of the present paper is
too crude to be able match the measured freeze-in temperatures in
detail. Another shortcoming is that the dimensionless
coefficient~$\sigma$ that appears in the turbulent heating rate has a
large impact on the solution to the model equations, but is an
adjustable free parameter. Further work is needed to provide a
physical basis for determining~$\sigma$ and
how it varies from one flux tube to another.

In a future study, the model developed in \S~\ref{sec:analytic}
could be used in conjunction with studies that map the magnetic-field
line traversed by PSP back to a source region
on the Sun to rapidly predict flow properties at PSP's location based
on the observed super-radial expansion factor within the source region
\citep[see, e.g.,][]{bale19,kasper19}. In addition, the modelling
results and approaches developed in this paper could be applied to
outflows from other astrophysical objects, such as stars with
differing masses and winds from accretion disks around compact
astrophysical objects.

\acknowledgements I thank Sam Badman, Stuart Bale, Chris Chen, Eugene Churazov,
Joe Hollweg, Justin Kasper, Brian Metzger, Jean Perez, Eliot Quataert,
Alex Schekochihin, and Marco Velli for valuable discussions, and the
two anonymous referees for comments that led to improvements in the
manuscript. This work was supported in part by NASA grant NNN06AA01C
to the Parker Solar Probe FIELDS Experiment and by NASA grants
NNX17AI18G and 80NSSC19K0829.

\appendix

\section{Solving for $U(r)$ and~$r_{\rm A}$}
\label{ap:U_of_r}

Once $y_{\rm b}$, $y_{\rm c}$, and $x$ are determined by solving
(\ref{eq:IEbal}), 
(\ref{eq:crit_point_cond}), and~(\ref{eq:Bernoulli2}),
the
value of~$U(r)$ between $r=r_{\rm b}$ and~$r=r_{\rm A}$ (which is as yet unknown)
can be found by solving the Bernoulli
integral~(\ref{eq:Bernoulli1}) with~$\Gamma$ given
by~(\ref{eq:Gamma}). 
An equation for~$r_{\rm A}$ can be obtained by evaluating the
Bernoulli integral (\ref{eq:Bernoulli1}) at $r=r_{\rm A}$,
setting $U(r_{\rm A}) = v_{\rm A}(r_{\rm A})$ and $y=1$, and
rewriting~$\Gamma$ using~(\ref{eq:Gamma}), which leads to 
\[
  0 = \frac{v_{\rm A}(r_{\rm A})^2}{2} - 2 \overline{ c_{\rm s}}^2 \ln
  y_{\rm b} + \frac{v_{\rm esc}^2}{2} \left(\psi^{1/2} -
    \frac{R_{\odot}}{r_{\rm A}}\right)
\]
\begin{equation}
  - \frac{(\delta v_{\rm
      b})^2}{2(1-\sigma)} \left[ \frac{2^{\sigma
        -2}(3+\sigma)(1+y_{\rm b})^{2-\sigma} -2}{y_{\rm b}}
    - 1 - \sigma\right]
  - \frac{v_{\rm Ab}^2}{2 y_{\rm b}^2}.
  \label{eq:rA1}
\end{equation} 
Upon setting
\begin{equation}
  v_{\rm A}(r_{\rm A}) = v_{\rm Ab} y_{\rm b} \left[ \frac{B(r_{\rm
        A})}{B_{\rm b}}\right] = \frac{v_{\rm Ab} y_{\rm b}
    \eta(r_{\rm A}) R_{\odot}^2}{\eta_{\rm b} r_{\rm A}^2 \psi}
  \label{eq:vAA}
\end{equation} 
in~(\ref{eq:rA1}) and rewriting~$v_{\rm Ab}$ in~(\ref{eq:rA1})
using~(\ref{eq:vAb}), one obtains the following equation
for~$r_{\rm A}$:
\[
0 =  \left[\frac{y_{\rm b}^2 \eta(r_{\rm A})^2}{(x
      \tilde{\rho}_\odot)^{1/4} \xi^2 \psi^{1/2}}\right]
  \left(\frac{R_{\odot}}{r_{\rm A}}\right)^{4} -
  \frac{R_{\odot}}{r_{\rm A}} - 4 x \ln y_{\rm b} + \psi^{1/2} -
  \frac{\psi^{3/2} \eta_{\rm b}^2}{(x \tilde{\rho}_\odot)^{1/4} \xi^2
    y_{\rm b}^2}
\]
\begin{equation}
    + \frac{\epsilon}{1-\sigma} \left[\frac{2 - 2^{\sigma - 2}(3+\sigma)(1+y_{\rm
        b})^{2-\sigma}}{y_{\rm b}} + 1+\sigma \right].
  \label{eq:rA3}
\end{equation}

At $r> r_{\rm A}$, the outflow velocity~$U(r)$ cannot be determined
from the Bernoulli integral, because the quasi-isothermal
approximation does not apply. In addition, equation~(\ref{eq:zminus2}) yields a
poor approximation for~$z_-$ at $r>r_{\rm A}$, because $1+y$
approaches a constant when $r \gg r_{\rm A}$. A better
approximation for~$z_-$ in the super-Alfv\'enic region can be obtained
from~(\ref{eq:zminusCH09}) and the simplifying approximation that
$|(\d / \d r) \ln(v_{\rm A})| = -(\d /\d r) \ln y$, which holds when
$B\propto r^{-2}$ and $\rho \propto r^{-2}$ (the latter scalings being
fairly accurate for $10 R_{\odot} \lesssim r \lesssim 60 R_{\odot}$, a
region in which the field lines are approximately radial and the flow
speed is approximately constant).  In this case, equation~(\ref{eq:zminusCH09})
becomes
\begin{equation}
  \frac{z_-}{L_\perp} = - (U+v_{\rm A}) \diff{}{r} \ln y.
  \label{eq:zminussuperAlfvenic}
\end{equation}
Integrating~(\ref{eq:dgdr}) from~$r=r_{\rm A}$ out to larger~$r$ then
gives $z_+^2 = [z_+(r_{\rm A})]^2 4y^2 / (1+y)^2$, where the value
of~$z_+(r_{\rm A})$ is obtained by setting~$y=1$ in~(\ref{eq:zplusy}).
The approximate value of~$c_{\rm
  s}^2(r)$ can then be found in terms of~$\rho$ by solving the internal-energy
equation~(\ref{eq:internal_energy}) with~$Q$ given by~(\ref{eq:defQ}) and neglecting radiative cooling
and thermal conduction. This leads to
\begin{equation}
  c_{\rm s}^2(r) =     \left[\frac{\rho(r)}{\rho_{\rm
        A}}\right]^{\gamma-1}
  \left(\overline{ c_{\rm s}}^2
+ K \int_y^1
    \frac{y_1^{3-2\gamma} \d y_1}{1+y_1} \right)
    \label{eq:cs(r)}
  \end{equation}
at $r>r_{\rm A}$,  where
  \begin{equation}
    K = \frac{(\gamma-1) z_{+ \rm b}^2}{y_{\rm b}}\left( \frac{1+y_{\rm
        b}}{2}\right)^{2-\sigma}.
    \label{eq:defK}
  \end{equation}
Total-energy conservation implies
that the mechanical luminosity
\begin{equation}
  L_{\rm mech}(r) = \dot{M} \left[ \frac{U^2}{2} +
    \frac{5 c_{\rm s}^2(r)}{2} - \frac{v_{\rm esc}^2
      R_{\odot}}{2r} + \frac{z_+^2}{4}\left(\frac{3}{2} + y\right) +
    \frac{q_{\rm r}}{\rho U}\right]
  \label{eq:defLmech}
\end{equation}
is independent of~$r$, where the ratio of specific heats~$\gamma$ has
been set equal to~$5/3$. The value of~$U(r)$ at $r>r_{\rm A}$ can be
obtained by setting
\begin{equation}
  L_{\rm mech}(r) = L_{\rm mech}(r_{\rm A}),
  \label{eq:LrLA}
\end{equation}
and rewriting~$\rho(r)$ in terms of~$U(r)$ using~(\ref{eq:cont1});
i.e., $\rho(r) U(r)/B(r) = \rho_{\rm b} U_{\rm b}/ B_{\rm b}$. 
As $U(r_{\rm A}) = v_{\rm A}(r_{\rm A})$ and $y(r_{\rm A}) = 1$,
(\ref{eq:defLmech}) implies that 
\begin{equation}
  L_{\rm mech}(r_{\rm A}) = \dot{M}\left[\frac{v_{\rm A}(r_{\rm A})^2}{2}
    + \frac{5  \overline{ c_{\rm s}}^2}{2} - \frac{v_{\rm esc}^2
      R_{\odot}}{2 r_{\rm A}} + \frac{5(\delta v_{\rm b})^2 (1+y_{\rm
        b})^{2-\sigma}}
    {2^{3-\sigma} y_{\rm b}} 
  \right].
  \label{eq:LmechA}
\end{equation}

Determining~$U(r)$ at $r>r_{\rm A}$ via~(\ref{eq:defLmech})
through~(\ref{eq:LmechA}) requires evaluating~$r_{\rm A}$
and~$v_{\rm A}(r_{\rm A})$. An alternative method that avoids this
requirement results from  noting that
\begin{equation}
  L_{\rm mech}(r_{\rm b})= \dot{M}\left\{ \frac{v_{\rm Ab}^2}{2y_{\rm b}^2} +
    \frac{5 \overline{ c_{\rm s}^2}}{2} - \frac{v_{\rm esc}^2
      \psi^{1/2}}{2} + (\delta v_{\rm b})^2 \left[ \frac{1}{2} +
      (1+y_{\rm b})(1 - \chi_{\rm H})\right] + 2 \overline{ c_{\rm
        s}}^2 \ln y_{\rm b}\right\},
  \label{eq:Lmechb}
\end{equation}
where~(\ref{eq:Eq3}) has
been used to eliminate~$q_{\rm b}$.
Replacing $\chi_{\rm H}$ in~(\ref{eq:Lmechb}) with the expression
on the right-hand side of~(\ref{eq:defchiH}) 
leads to the consistency check that
\begin{equation}
  L_{\rm mech}(r_{\rm A}) = L_{\rm mech}(r_{\rm b}).
  \label{eq:LALb}
\end{equation}
Combining (\ref{eq:LrLA}) and~(\ref{eq:LALb}), one can find~$U(r)$ at $r>r_{\rm A}$ by 
setting
\begin{equation}
  L_{\rm mech}(r) = L_{\rm mech}(r_{\rm b}).
  \label{eq:LrLb}
\end{equation} 
Equation~(\ref{eq:LrLb}) leads to a simple expression for the
asymptotic wind velocity~$U_{\infty}$, i.e., $U(r)$ as $r\rightarrow \infty$. As
$r\rightarrow \infty$, the kinetic-energy flux dominates the total
energy flux, and $L_{\rm mech}(r) \rightarrow \dot{M}
U_{\infty}^2 / 2$. Setting $\dot{M} U_{\infty}^2/2 = L_{\rm
  mech}(r_{\rm b})$ yields
\begin{equation}
  U_{\infty} = \left[\frac{v_{\rm Ab}^2}{y_{\rm b}^2} + 5 \overline{
    c_{\rm s}}^2 - v_{\rm esc}^2 \psi^{1/2} + 2 (\delta v_{\rm b})^2
  \left(
    \frac{3}{2} + y_{\rm b}\right) - \frac{2 q_{\rm b} y_{\rm
      b}}{\rho_{\rm b} v_{\rm Ab}}\right]^{1/2}.
\label{eq:Uinfty}
\end{equation} 

\section{Approximate analytic solutions}
\label{ap:analytic} 

As discussed in \S~\ref{sec:knowns}, the core of the solar-wind
model developed in \S~\ref{sec:analytic}  is a set of three
simultaneous equations,
(\ref{eq:IEbal}), (\ref{eq:crit_point_cond}),
and~(\ref{eq:Bernoulli2}), for the three
unknowns~$y_{\rm b}$, $y_{\rm c}$, and $x$. In this
section, two different approximate analytic solutions to these
equations are obtained in two 
different parameter regimes. Both solutions rely on the 
simplifying approximations in~(\ref{eq:approxs0}), which are repeated here:
\begin{equation}
 y_{\rm b} \gg 1 \qquad \psi = 1 \qquad \epsilon \ll 1
  \qquad \eta_{\rm c} = \gamma_{B \rm c} = 1.
  \label{eq:approxs}
\end{equation}
In particular, $y_{\rm b}$ is taken to be sufficiently large that: 
(1) the $U_{\rm b}^2  = v_{\rm Ab}^2/y_{\rm b}^2$ term in~(\ref{eq:Gamma})
can be dropped, which amounts to dropping the second-to-last term on
the left-hand side of~(\ref{eq:Bernoulli2}); and (2)
$O(y_{\rm b}^{-1})$ terms can be dropped in~(\ref{eq:defchiH}), so that
\begin{equation}
  \chi_{\rm H} = 1 - \zeta y_{\rm b}^{-\sigma},
  \label{eq:chiH_approx}
\end{equation}
where
\begin{equation}
  \zeta = 2^{\sigma-1}\left(\frac{2-\sigma}{1-\sigma}\right).
  \label{eq:defzeta}
\end{equation}
The $y_{\rm b}^{-\sigma}$ term in~(\ref{eq:chiH_approx}) is retained,
even though terms of order~$y_{\rm b}^{-1}$ are discarded,
on the working assumption that $\sigma \sim$ 0.1--\,0.5, as is the case in the numerical
examples in \S~\ref{sec:numerical}.
As mentioned in \S~\ref{sec:knowns},
the last equality in~(\ref{eq:approxs}) amounts to taking all of the
super-radial expansion of the field lines to occur inside the
wave-modified sonic critical point and the field lines to be completely radial at
$r\geq r_{\rm c}$.
With these approximations, 
(\ref{eq:IEbal}), (\ref{eq:crit_point_cond}), 
and~(\ref{eq:Bernoulli2}) become, respectively,
\begin{equation}
  \frac{\epsilon_\odot \tilde{\rho}_{\odot}^{3/8} \chi_{\rm
      H}}{B_\ast \xi x^{1/8}} - \frac{2 x \ln y_{\rm b}}{y_{\rm b}}
  - \frac{q_{\rm b}}{\rho_{\rm b} v_{\rm Ab} v_{\rm esc}^2} = 0,
  \label{eq:IEbal2}
\end{equation}
\begin{equation}
    \left(\frac{y_{\rm c}^2 x^{1/8} \xi \tilde{\rho}_{\odot}^{1/8}}{16
        y_{\rm b}}\right)^{2/3} - x
    - \frac{\epsilon_\odot \tilde{\rho}_\odot^{3/8} y_{\rm
        b}^{1-\sigma}[y_{\rm c}^2(1+\sigma) + 3y_{\rm c}]}{4 B_\ast
      \xi x^{1/8}(1+y_{\rm c})^{3-\sigma}}= 0,
    \label{eq:crit_point2}
\end{equation}
and
\begin{equation}
  1  - x\left[4 \ln \left(\frac{y_{\rm b}}{y_{\rm c}}\right) + 3\right]
  - \frac{\epsilon_\odot \tilde{\rho}_\odot^{3/8} y_{\rm
      b}^{1-\sigma}[y_{\rm c}^2(1+\sigma)(7-3\sigma) + y_{\rm c}(21 -
    5\sigma) + 8]}{4(1-\sigma)B_\ast \xi x^{1/8} (1+y_{\rm
      c})^{3-\sigma}} = 0.
  \label{eq:Bernoulli3}
\end{equation}

\subsection{Conduction-dominated limit}
\label{sec:conduction}

When $\epsilon_\odot$ is sufficiently small, an approximate solution
to (\ref{eq:IEbal2}), (\ref{eq:crit_point2}),
and~(\ref{eq:Bernoulli3}) can be obtained through the method of
dominant balance~\citep{bender78}, in which two of the three terms in each equation are
taken to be dominant, and the third term is taken to be much smaller
in magnitude. Neglecting the smaller term in each equation yields the
leading-order solution, with the smaller term producing higher-order
corrections.  In the present case, it is the second term on the
left-hand side of each equation that can be neglected to leading
order. In~(\ref{eq:IEbal2}), this corresponds to balancing turbulent
heating of the sub-Alfv\'enic region against the heat that is
conducted from the corona into the transition region.  In other words,
conduction into the transition region rather than $p\,\d V$ work is the
dominant sink of internal energy in the sub-Alfv\'enic region as a
whole.  Neglecting the second term in~(\ref{eq:crit_point2}) amounts to
assuming that the fluctuating velocity makes the dominant contribution
to~$U_{\rm c}$ in~(\ref{eq:defUc}), which is equivalent to
taking~$\delta v(r_{\rm c}) \gg \overline{ c_{\rm s}}$. The latter equality
appears paradoxical, because the small-$\epsilon_\odot$ limit
corresponds to small values of the fluctuating velocity at the coronal
base. However, as $\epsilon_\odot \rightarrow 0$, the coronal
temperature drops, the density scale height in the corona decreases, and
the wave amplitude at the critical point grows as the AWs attempt to
conserve wave action, which would lead
to~$\delta v \propto \rho^{-1/4}$ when $U\ll v_{\rm A}$ in the absence
of reflection and dissipation (see (\ref{eq:constgsq})
and~(\ref{eq:defg})).  Neglecting the second term
in~(\ref{eq:Bernoulli3}) amounts to taking the wave pressure force to
have a larger cumulative effect than the thermal pressure force on the
acceleration of the outflowing plasma between~$r_{\rm b}$
and~$r_{\rm c}$, which can again be understood as a consequence of the
wave amplitudes growing rapidly with increasing~$r$ when the density
scale height in the corona is small.

As already noted in (\ref{eq:Mdotcond}) and illustrated in
Figure~\ref{fig:Mdot_asymptotic}, $\dot{M}$ is exceedingly
small in the conduction-dominated limit, and as a consequence~$U_{\rm
  b} \ll \overline{ c_{\rm s}}$.  The heat flux at the coronal base
is thus approximately given by~(\ref{eq:qb_approx}),
and~(\ref{eq:IEbal2}) can be rewritten as
\begin{equation}
  \frac{\epsilon_\odot \tilde{\rho}_{\odot}^{3/8} \chi_{\rm
      H}}{B_\ast \xi x^{1/8}} - \frac{2 x \ln y_{\rm b}}{y_{\rm b}}
  - \frac{I_1 x^{13/8} \xi \tilde{\rho}_\odot^{1/8}}{\eta_{\rm b}} = 0.
  \label{eq:IEbal2_low_Mach_number}
\end{equation}
Balancing the first and third terms in~(\ref{eq:IEbal2_low_Mach_number}) and dropping
the~$y_{\rm b}^{-\sigma}$ term in~(\ref{eq:chiH_approx}) yields the
leading-order solution for the dimensionless temperature in the
sub-Alfv\'enic region:
\begin{equation}
  x = I_1^{-4/7} \tilde{\rho}_\odot^{1/7} (\epsilon_\odot \eta_{\rm b}
  B_\ast \tilde{l}_{\rm b})^{2/7}.
  \label{eq:xcond}
\end{equation}
Anticipating the solution for~$y_{\rm c}$, it is useful to predict at
the outset (as
will shortly be confirmed by (\ref{eq:ybyccond}) and~(\ref{eq:ybcond}))
that
\begin{equation}
y_{\rm c} \gg 1.
\label{eq:yclim} 
\end{equation} 
Balancing the first and third terms in~(\ref{eq:Bernoulli3}) then
gives
\begin{equation}
\frac{y_{\rm b}}{y_{\rm c}} = \left[
\frac{4(1-\sigma) B_\ast^{2/7} \eta_{\rm b}^{2/7}}
{I_1^{1/14}(1+\sigma)(7-3\sigma)\epsilon_\odot^{5/7}\tilde{l}_{\rm b}^{3/14}\tilde{\rho}_\odot^{5/14}}
\right]^{1/(1-\sigma)}.
\label{eq:ybyccond} 
\end{equation} 
Balancing the first and third terms in~(\ref{eq:crit_point2}) and
making use of~(\ref{eq:yclim}) and~(\ref{eq:ybyccond}), one arrives at
the leading-order solution for~$y_{\rm b}$:
\begin{equation} 
y_{\rm b} = I_2 \left[
  \epsilon_\odot^{-(12-2\sigma)/7} B_\ast^{(9-5\sigma)/7} 
\eta_{\rm b}^{2(1+\sigma)/7} \tilde{l}_{\rm b}^{-3(1+\sigma)/14} 
\tilde{\rho}_\odot^{-(6-\sigma)/7}
\right]^{1/(1-\sigma)},
\label{eq:ybcond} 
\end{equation} 
where
\begin{equation}
  I_2 = 16 I_1^{-(1+\sigma)/(14-14\sigma)} \left(\frac{4}{1+\sigma}\right)^{2/(1-\sigma)}
\left(\frac{1-\sigma}{7-3\sigma}\right)^{(7-3\sigma)/(2-2\sigma)} .
  \label{eq:defI2}
\end{equation}
Equations~(\ref{eq:Mdot2}), (\ref{eq:rhob2}), (\ref{eq:xcond}),
and~(\ref{eq:ybcond}) then yield the leading-order mass outflow rate
in the conduction-dominated regime,~$\dot{M}^{\rm (cond)}$, given
in~(\ref{eq:Mdotcond}).

The asymptotic wind velocity $U_\infty$ is obtained by setting
$\dot{M} U_\infty^2 /2 $ equal to the mechanical luminosity at the
coronal base $L_{\rm mech}(r_{\rm b})$ as described in
Appendix~\ref{ap:U_of_r}, where $L_{\rm mech}$ is defined
in~(\ref{eq:defLmech}).  When this procedure is carried out
using~(\ref{eq:xcond}) and~(\ref{eq:ybcond}), the two leading-order
terms in~$L_{\rm mech}(r_{\rm b})$ cancel. To obtain the leading-order
non-vanishing term in~$U_{\infty}^2$, one must account for the next largest
term in~(\ref{eq:IEbal2_low_Mach_number}), which results from the $\chi_{\rm H}$
correction; i.e., the second term on the right-hand
side of~(\ref{eq:chiH_approx}). When this term is retained, one
obtains the leading-order asymptotic wind speed in the
conduction-dominated regime,~$U_{\infty}^{\rm (cond)}$, given
in~(\ref{eq:Uinfcond}).

The range of~$\epsilon_\odot$ values for which (\ref{eq:Mdotcond}),
(\ref{eq:Uinfcond}),
(\ref{eq:xcond}), (\ref{eq:ybyccond}), and~(\ref{eq:ybcond}) are 
approximately valid can be determined by requiring that
the neglected terms in (\ref{eq:crit_point2}), 
(\ref{eq:Bernoulli3}), and~(\ref{eq:IEbal2_low_Mach_number}) be small compared to the other terms
when $y_{\rm b}$, $y_{\rm c}$, and~$x$ are given by~(\ref{eq:xcond}),
(\ref{eq:ybyccond}), and~(\ref{eq:ybcond}).  The most stringent
condition on~$\epsilon_{\odot}$ arises from carrying out this
procedure for~(\ref{eq:Bernoulli3}), which results in the requirement
that 
\begin{equation}
  I_3 \epsilon_\odot^{2/7}\left[ 4 \ln I_4 + 3 -
    \frac{20}{7(1-\sigma)} \ln \epsilon_\odot\right] \ll 1,
\label{eq:condlim2} 
\end{equation}
where $I_3   = I_1^{-4/7} \tilde{\rho}_{\odot}^{1/7} (\eta_{\rm b}
B_\ast \tilde{l}_{\rm b})^{2/7}$, and $I_4$ equals the right-hand side
of~(\ref{eq:ybyccond}) without the~$\epsilon_{\odot}$ term; i.e., $I_4
= \epsilon_\odot^{5/(7-7\sigma)} y_{\rm b}/y_{\rm c}$, with $y_{\rm
  b}/y_{\rm c}$ given by the right-hand side of~(\ref{eq:ybyccond}).
Upon neglecting the quantity $4\ln I_4 + 3$ on the left-hand side
of~(\ref{eq:condlim2}), which is smaller in magnitude than the
remaining term when~$\epsilon_\odot$ is in the conduction-dominated
regime but other
parameters take on Sun-like values, one finds that the
conduction-dominated regime corresponds to
\begin{equation}
  \epsilon_\odot \ll \epsilon_{\odot \rm cond} \equiv
  \exp\left(\frac{7}{2} W_{-1}\left(\frac{\sigma-1}{10
        I_3}\right)\right),
  \label{eq:condlim3}
\end{equation}
where $W_{-1}$ is the lower branch of the Lambert $W$ function.

\subsection{Expansion-dominated limit}
\label{sec:expansion} 

In the expansion-dominated limit, the first two terms on the left-hand
sides of~(\ref{eq:IEbal2}), (\ref{eq:crit_point2}),
and~(\ref{eq:Bernoulli3}) are treated as dominant. For this case, it
is useful to define the variables
\begin{equation}
  u = \frac{A_1 y_{\rm b}}{x^{1/8}} \qquad p = \frac{y_{\rm b}}{y_{\rm c}},
  \label{eq:defu}
\end{equation}
where
\begin{equation}
  A_1 = \frac{\epsilon_{\odot}^{3/4} \tilde{\rho}_\odot^{3/8} \tilde{l}_{\rm
      b}^{1/4}}{(\eta_{\rm b} B_\ast)^{1/4}}.
    \label{eq:defA1}
  \end{equation}
Rewriting~(\ref{eq:IEbal2}), (\ref{eq:crit_point2}),
and~(\ref{eq:Bernoulli3}) in terms of $u$, $p$, and~$x$ rather
than~$y_{\rm b}$, $y_{\rm c}$, and~$x$, one obtains
\begin{equation}
  1 + \frac{2x\ln A_1}{u} = a,
  \label{eq:IEbal3}
\end{equation} 
\begin{equation}
    \left(\frac{u x^{1/4} \xi \tilde{\rho}_\odot^{1/8}}{16 A_1
        p^2}\right)^{2/3} - x = b,
    \label{eq:crit_point3}
\end{equation}
and
\begin{equation}
    1 - x(4 \ln p + 3) = c.
    \label{eq:Bernoulli4}
\end{equation}
The quantities $a$, $b$, and $c$, which are treated as small,
are functions of $u$, $p$,
and~$x$. In order to see how (\ref{eq:IEbal3}),
(\ref{eq:crit_point3}), and (\ref{eq:Bernoulli4}) follow from (\ref{eq:IEbal2}),
(\ref{eq:crit_point2}),
and~(\ref{eq:Bernoulli3}), it is helpful to leave terms containing~$y_{\rm b}$ and~$y_{\rm c}$
in the expressions for $a$, $b$, and~$c$,
with the understanding
that $y_{\rm b} = x^{1/8} u / A_1$ and $y_{\rm c} = x^{1/8} u / (A_1
p)$:
\begin{equation}
  a \equiv \frac{2x}{u} \ln(u x^{1/8}) + \zeta y_{\rm b}^{-\sigma} +
  \frac{ x^{1/8} q_{\rm b }}{A_1 \rho_{\rm b} v_{\rm Ab} v_{\rm esc}^2},
  \label{eq:defa}
\end{equation}
\begin{equation}
b \equiv \frac{\epsilon_\odot \tilde{\rho}_\odot^{3/8} y_{\rm
    b}^{1-\sigma}[y_{\rm c}^2(1+\sigma) + 3 y_{\rm c}]}{4 B_\ast \xi
  x^{1/8} (1+y_{\rm c})^{3-\sigma}} ,
  \label{eq:defb}
\end{equation}
and
\begin{equation}
c \equiv \frac{\epsilon_\odot \tilde{\rho}_\odot^{3/8} y_{\rm
    b}^{1-\sigma}[y_{\rm c}^2(1+\sigma)(7-3\sigma) + y_{\rm
    c}(21-5\sigma) + 8]}{4 (1-\sigma) B_\ast \xi x^{1/8} (1+y_{\rm c})^{3-\sigma}}.
  \label{eq:defc}
\end{equation}

From~(\ref{eq:IEbal3}), it follows that
\begin{equation}
  x = \frac{(1-a)u}{-2\ln A_1}.
  \label{eq:xsolveexp}
\end{equation}
Equation~(\ref{eq:Bernoulli4}) implies that
\begin{equation}
  p = \exp\left(\frac{1-c}{4x} - \frac{3}{4}\right).
  \label{eq:psolveexp}
\end{equation}
After substituting~(\ref{eq:xsolveexp}) and~(\ref{eq:psolveexp})
into~(\ref{eq:crit_point3}), one finds that
\begin{equation}
  \frac{1}{u} - 1 = S(u, a, b, c),
  \label{eq:usolveexp}
\end{equation}
where
\[
  S(u, a, b, c) = \frac{c}{u} - a - \frac{(1-a)}{\ln A_1}\left[   \frac{5}{4} \ln u
\right.
\]
\begin{equation}
  \left.
 + \frac{\ln (1-a)}{4} - \frac{\ln(-2\ln A_1)}{4}
    - \ln A_2 + \frac{3}{2} - \frac{3}{2} \ln \left(b -
      \frac{(1-a)u}{2\ln A_1}\right)\right],
  \label{eq:defS}
\end{equation}
and
\begin{equation}
  A_2 = \frac{16}{\xi \tilde{\rho}_\odot^{1/8}},
  \label{eq:defA2} 
\end{equation}
which is~$\simeq 1$ for typical solar parameters.
The quantities $a$, $b$, and~$c$ are themselves functions of~$u$ by
virtue of~(\ref{eq:defu}) and
(\ref{eq:defa}) through~(\ref{eq:psolveexp}).

In the expansion-dominated regime
\begin{equation}
  u \sim O(1) \qquad |S(u, a, b, c)| \ll 1.
  \label{eq:expineq} 
\end{equation}
The latter inequality is achieved when $-1 / \ln A_1$, $|a|$, $b$,
and~$c$ are much smaller than~$1$.  When (\ref{eq:expineq}) is satisfied,
equations~(\ref{eq:xsolveexp}), (\ref{eq:psolveexp}), and~(\ref{eq:usolveexp})
can be solved perturbatively through the recursion relations
\begin{eqnarray}
  \frac{1}{u_n} - 1  & = & \left\{\begin{array}{cc} 0 & \mbox{ if
                                                        $n=0$} \\
                                    S(u_{n-1}, a_{n-1}, b_{n-1},
                                    c_{n-1}) & \mbox{ if $n\geq
                                               1$} \end{array} \right.  \label{eq:usolvepert} \\
  x_n & = &\frac{(1-a_{n-1})u_n}{-2\ln A_1} \\
  p_n & = &\exp\left(\frac{1-c_{n-1}}{4x_n} - \frac{3}{4}\right)  ,
\end{eqnarray} 
where $n= 0, 1, 2, \dots$,  and $a_{-1} = c_{-1} = 0$. The values
of~$a_n$, $b_n$, and~$c_n$ are obtained by replacing~$(a,b,c,u,y_{\rm b},
y_{\rm c}, x)$ with $(a_n, b_n, c_n, u_n,y_{{\rm b},n}, y_{{\rm c}, n}, x_n)$
in~(\ref{eq:defa}), (\ref{eq:defb}), and~(\ref{eq:defc}). For
reference below, the values
of $\rho_{\rm b}$, $v_{\rm Ab}$, and $q_{\rm b}$ that result from this
substitution (via~(\ref{eq:rhob2}), (\ref{eq:vAb}),
and~(\ref{eq:lambertqb})) are denoted $\rho_{{\rm b}, n}$, $v_{{\rm Ab},
  n}$, and~$q_{{\rm b},n}$. The values
of~$y_{{\rm b},n}$ and $y_{{\rm c},n}$ are obtained by replacing
$(y_{\rm b}, y_{\rm c}, x, u, p)$ with $(y_{{\rm b},n}, y_{{\rm c},n}, x_n,
u_n, p_n)$ in~(\ref{eq:defu}).

The $n^{\rm th}$-order approximation for~$\dot{M}$ in the expansion-dominated
regime, denoted $\dot{M}^{\rm (exp)}_n$, can be found by
setting $\dot{M} = \dot{M}^{\rm (exp)}_n$, $x=x_{\rm n}$, $y_{\rm b} = y_{{\rm
  b},n}$ and $\psi = 1$ in~(\ref{eq:Mdotgeneral}):
\begin{equation}
  \dot{M}^{\rm (exp)}_n= \frac{R_\odot^2 \bar{ B}^2}{v_{\rm
      esc}} y_{{\rm b},n}^{-1} (x_n \tilde{\rho}_\odot)^{1/8} \xi.
  \label{eq:Mdotexpn}
\end{equation} 
For $n\geq 1$, I define the $n^{\rm th}$-order approximation
for~$U_\infty$, denoted $U_{\infty, n}^{\rm (exp)}$,
to be the value of~$U_{\infty}$ in~(\ref{eq:Uinfty}) when
$(x, y_{\rm b}, \psi, q_{\rm b}, \rho_{\rm b}, v_{\rm Ab}) = (x_n,
y_{{\rm b},n}, 1, q_{{\rm b},n}, \rho_{{\rm b},n}, v_{{\rm Ab},n})$: 
\begin{equation}
U_{\infty, n}^{\rm (exp)} = v_{\rm esc} \left[
\frac{v_{{\rm Ab}, n}^2}{y_{{\rm b}, n}^2 v_{\rm esc}^2} + 5 x_n - 1 + \frac{2y_{{\rm b},n} \epsilon_\odot \tilde{\rho}_\odot^{3/8}
}{B_\ast \xi x_n^{1/8}}  - \frac{2 q_{{\rm b},n} y_{{\rm
      b},n}}{\rho_{{\rm b},n} v_{{\rm Ab},n} v_{\rm esc}^2}
\right]^{1/2},
\label{eq:Uinftyexp} 
\end{equation} 
where I have invoked~(\ref{eq:approxs}) to approximate $3/2 + y_{{\rm b},
  n}$ as~$y_{{\rm b},n}$.
The leading-order approximation for~$U_\infty$ in the
expansion-dominated regime, denoted $U_{\infty, 0}^{\rm (exp)}$, is
obtained from~(\ref{eq:Uinftyexp}) with~$n=0$ after dropping terms
that were neglected in the calculation of~$u_0$ --- in particular, the
first, second, and last terms inside the brackets on the right-hand
side of~(\ref{eq:Uinftyexp}). This yields
\begin{equation}
  U_{\infty, 0}^{\rm (exp)} = v_{\rm esc}.
  \label{eq:Uinftyexpn0}
\end{equation}

As in the conduction-dominated limit,
the range of~$\epsilon_\odot$ values for which (\ref{eq:usolvepert}), (\ref{eq:Mdotexpn}),
and~(\ref{eq:Uinftyexpn0}) are approximately valid
can be determined by imposing the constraint that
the neglected terms in (\ref{eq:IEbal2}), (\ref{eq:crit_point2}), and
(\ref{eq:Bernoulli3}) be small compared to the terms that are kept
when $y_{\rm b}$, $y_{\rm c}$, and~$x$ are given by~(\ref{eq:defu}),
(\ref{eq:defA1}), (\ref{eq:xsolveexp}), (\ref{eq:psolveexp}), and~$u=1$.
Carrying out this procedure for~(\ref{eq:IEbal2}) and making the
simplifying approximations that~$a$ is dominated by the last term on
the right-hand side of~(\ref{eq:defa}), that $q_{\rm b}$ is
given by~(\ref{eq:qb_approx}), and that $\ln A_1 \simeq \ln
\left(\epsilon_\odot^{3/4}\right)$, one obtains the 
requirement that
\begin{equation}
  \epsilon_\odot \gg \epsilon_{\odot \rm exp,min} \equiv \exp\left(
    \frac{7}{2} W_{-1}\left( - \frac{4 I_1^{1/14}}{21
        \tilde{\rho}_{\odot}^{1/7}
        (\tilde{l}_{\rm b} \eta_{\rm b} B_\ast)^{2/7}}\right)\right),
  \label{eq:epsexpmin}
\end{equation}
where $W_{-1}$ is, as above, the lower branch of the Lambert $W$
function. Equation~(\ref{eq:epsexpmin}) corresponds to the requirement
that the conductive losses from the sub-Alfv\'enic region into the
transition region be negligible compared to~$p\,\d V$ work in this region.
Carrying out the above procedure for~(\ref{eq:crit_point2}) and again
making the simplifying approximation that $\ln A_1 \simeq \ln\left( \epsilon_\odot^{3/4}\right)$,
one obtains
\begin{equation}
  \epsilon_\odot \ll \epsilon_{\odot \rm exp,max} \equiv \exp\left(
    \frac{3}{(1+\sigma)}
    W_{-1}\left( - \frac{2^{25/9}  (1+\sigma)^{1/9} e^{2(1-\sigma)/3}
        }{9}\left[\frac{\eta_{\rm b}B_\ast}{\tilde{l}_{\rm
            b} \tilde{\rho}_{\odot}^{3/2}}\right]^{(1+\sigma)/9}
    \right)
    \right).
    \label{eq:epsexpmax}
\end{equation}
Equation~(\ref{eq:epsexpmax}) corresponds to the requirement that the
sound speed make the dominant contribution to the outflow
velocity~$U_{\rm c}$ at the critical point in~(\ref{eq:defUc}).  As
illustrated by the shaded gray rectangle in
figure~(\ref{fig:Mdot_asymptotic}), for Sun-like parameters (and in
particular, for~$\eta_{\rm b} = 30$), $\epsilon_{\odot \rm exp,max}$
is approximately four orders of magnitude larger than
$\epsilon_{\odot \rm exp, min}$. There is thus a finite range
of~$\epsilon_\odot$ values that satisfy both~(\ref{eq:epsexpmin})
and~(\ref{eq:epsexpmax}). However, it should be noted that the
approximations~$\ln A_1 \simeq \ln (\epsilon_\odot^{3/4})$ and
$  q_{\rm b} = I_1 \rho_{\rm b} \overline{ c_{\rm s}}^3$
cause~(\ref{eq:epsexpmin}) to underestimate the lower bound on
$\epsilon_\odot$ in the expansion-dominated regime. Also, for Sun-like
parameters, the assumption $r_{\rm c} < r_{\rm A}$ that underlies the
model of \S~\ref{sec:analytic} breaks down at values
of~$\epsilon_{\odot}$ smaller than~$\epsilon_{\odot \rm exp, max}$, as
illustrated in Figure~\ref{fig:Mdot_asymptotic}.

\bibliography{articles}

\bibliographystyle{jpp}

\end{document}